\documentclass[acmsmall]{acmart}

\usepackage{amsmath,amsfonts}
\usepackage{color}
\usepackage{colortbl}
\usepackage{multirow}
\usepackage{subcaption}
\usepackage[ruled,noend,vlined,linesnumbered]{algorithm2e}
\usepackage[noend]{algcompatible}
\usepackage[flushleft]{threeparttable}
\usepackage{wasysym}
\usepackage{makecell}
\usepackage{tabularx}
\usepackage[export]{adjustbox}

\newcommand{\hl}[1] {\textcolor{red}{#1}}

\newcommand{\bheading}[1]{{\vspace{2pt}\noindent{\textbf{#1}}\hspace{2pt}}}

\makeatletter
\newcommand{\removelatexerror}{\let\@latex@error\@gobble}
\makeatother

\definecolor{dkgreen}{rgb}{0,0.6,0}
\definecolor{gray}{rgb}{0.5,0.5,0.5}
\definecolor{mauve}{rgb}{0.58,0,0.82}

\begin{document}

\title{A Survey of Microarchitectural Side-channel Vulnerabilities, Attacks and Defenses in Cryptography}

\author{Xiaoxuan Lou}
\affiliation{%
   \institution{Nanyang Technological University}
   \country{Singapore}}
\email{XIAOXUAN001@e.ntu.edu.sg}
\author{Tianwei Zhang}
\affiliation{%
   \institution{Nanyang Technological University}
   \country{Singapore}}
\email{tianwei.zhang@ntu.edu.sg}
\author{Jun Jiang}
\affiliation{%
   \institution{Two Sigma Investments, LP}
   \country{USA}}
\email{jiangcj@pathsec.org}
\author{Yinqian Zhang}
\affiliation{%
   \institution{Southern University of Science and Technology}
   \country{China}}
\email{yinqianz@acm.org}



\begin{abstract}
Side-channel attacks have become a severe threat to the confidentiality of computer 
applications and systems. One popular type of such attacks is the microarchitectural attack,
where the adversary exploits the hardware features to break the protection enforced
by the operating system and steal the secrets from the program. 
In this paper, we systematize microarchitectural side channels with a focus on attacks and defenses in cryptographic applications. We make three contributions.
(1) We survey past research literature to categorize microarchitectural
side-channel attacks. Since these are hardware attacks targeting software, we summarize the
vulnerable implementations in software, as well as flawed designs in hardware.
(2) We identify common strategies to mitigate microarchitectural attacks, from the application, OS and hardware levels. (3) We conduct a large-scale evaluation on popular
cryptographic applications in the real world, and analyze the severity, practicality
and impact of side-channel vulnerabilities.
This survey is expected to inspire side-channel research community
to discover new attacks, and more importantly, propose new defense solutions against them.
\end{abstract}

\ccsdesc[500]{Security and privacy~Side-channel analysis and countermeasures}

\keywords{Microarchitecture, Cryptography, Vulnerability Analysis}

\maketitle

\section{Introduction}
\label{sec:intro}



The history of side-channel attacks dates back to the year of 1996, when Kocher \cite{Ko:96}
demonstrated that the data leaked from timing channels was sufficient for an attacker to
recover the entire secret key. To generalize, vulnerable implementations of cryptographic operations
can exhibit secret-dependent non-functional behaviors during the time of execution, which an adversary
can observe and utilize to fully or partially recover sensitive information. Since then, numerous types of side channels (e.g., execution timing \cite{Be:05, BoMi:06}, acoustic emission \cite{GeShTr:14}, electromagnetic radiation \cite{GePiTr:15} and power consumption \cite{Co:99}) have been discovered and exploited to defeat modern cryptographic schemes, allowing adversaries to break strong ciphers in a short period of time with very few trials.



%
Among these side-channel threats, microarchitectural attacks are particularly dangerous and prevalent. A fundamental cause of such attacks is the conflict between \emph{performance} and \emph{security}. During the evolution of computer architecture, various strategies were introduced to speed up the execution, which may bring side channels that leak the information of applications running on the system. One example is Simultaneous Multithreading (SMT), where multiple threads execute in parallel and share the same CPU core and functional units. This brings not only high performance, but also side channels due to contention for the shared hardware components. Another example is caching: a small hardware component is introduced (e.g., CPU caches, Translation Look-aside Buffer, DRAM row buffer) to store the previously accessed data, which is usually expected to be used again soon due to the principle of locality. Fetching data directly from this component is much faster. However, such timing differences can reveal the victim program's access traces \cite{Pe:05, OsShTr:06, GrRaBo:18}.

It is obviously infeasible to disable those features for side-channel mitigation, which can incur tremendous performance overhead. Therefore, effective elimination of side-channel vulnerabilities has been a long-standing goal. Although security-aware cryptographic applications, systems and architectures were designed to mitigate side-channel attacks, it is however still very challenging to remove all side-channel vulnerabilities from the software implementations and hardware designs. As such, the arms race between side-channel attacks and defenses remains heated.

This paper provides a comprehensive survey of microarchitectural side-channel attacks and defenses in cryptographic applications. Since we focus on hardware attacks on software, it is necessary to study the vulnerabilities and defense opportunities in both hardware and software levels. We are particularly interested in three questions: (1) \emph{What are the common and distinct features of software vulnerabilities and hardware flaws that lead to side-channel attacks?} (2) \emph{What are the typical mitigation strategies for applications, operating systems and hardware, respectively?} (3) \emph{What is the status quo of cryptograhpic applications in terms of side-channel vulnerabilities?}

\noindent\textbf{Existing surveys.}
Past efforts summarized side-channel studies from different perspectives and fail to answer the above questions. First, some works mainly focused on the physical attacks \cite{Na:16,HuAlZa:16,SpMoKo:18}, networking attacks \cite{ZaArBr:07,UlZsFa:17} or fault injection attacks with integrity breach \cite{el2015survey}, which have different characteristics or requirements from microarchitectural side-channel attacks. Second, a few surveys \cite{BiGhNa:17,GeYaCo:18,Sz:18} only considered the hardware flaws that result in side channels, while ignoring the software vulnerabilities. Third, several efforts focused on vulnerabilities and countermeasures in one certain cryptosystem (e.g., Elliptic Curve Cryptography \cite{Av:05, FaGuDe:10, FaVe:12}, Pairing-based cryptography \cite{el2015survey}). These summaries are outdated due to a large quantity of newly discovered vulnerabilities and implementation improvements afterwards. Fourth, some works only considered specific platforms (e.g., Trusted Execution Environments \cite{schwarz2020trusted}, smart card \cite{Tu:17}, cloud \cite{UlZsFa:17, BeWeMu:17}) or target applications (e.g., key logging \cite{Mo:18, HuAlZa:16}), which did not provide comprehensive conclusions.

\noindent\textbf{Our contributions.}
Our survey has three significant contributions. First, we characterize microarchitectural side-channel attacks comprehensively. We summarize the attack vectors in both hardware designs (Section \ref{sec:hardware}) and software implementations (Section \ref{sec:crypto}). Second, we identify and abstract the key defense strategies, which are categorized into application, system and hardware, respectively (Section \ref{sec:insights}). Third, we conduct a large-scale evaluation of mainstream cryptographic applications. We analyze the side-channel vulnerabilities and the corresponding patches in various libraries and products, and evaluate the severity and impact from a practical perspective (Section \ref{sec:history}). We hope this work can help researchers, developers and users better understand the current status and the future direction of side-channel research and countermeasure development.

\section{Background}
\label{sec:bg}


\subsection{Basics of Side-channel Attacks}
In a microarchitectural side-channel attack, the adversary steals secrets by exploiting observable information from the microarchitectural components. Given a secret input $\mathcal{S}$, the target application exhibits certain runtime behaviors $\mathcal{R}$ (e.g., memory access patterns) and causes the underlying host system to reveal some characteristics $\mathcal{I}$. By identifying the correlation $\mathcal{I} \sim \mathcal{R} \sim \mathcal{S}$, the adversary is able to capture the microarchitectural characteristics as the side-channel information and infer the secret input. The success of microarchitectural side-channel attacks relies on vectors from both software and hardware levels. 
 
\bheading{Software vectors.}
One necessary condition for microarchitectural attacks is that \emph{application's runtime behaviors need to be correlated with the secrets}: $\mathcal{R} \sim \mathcal{S}$. Generally there are two sources of leakage. (1) \emph{Secret-dependent control flow}: when the secret $\mathcal{S}$ changes, the application executes another code path. 
(2) \emph{Secret-dependent data flow}: the application may rely on the secret $\mathcal{S}$ to determine the data access location. They yield different behaviors distinguishable by the attacker. 

\bheading{Hardware vectors.}
The key factor is that \emph{application's behaviors can be reflected by the microarchitectural characteristics}: $\mathcal{I} \sim \mathcal{R}$. Two kinds of techniques exist to capture useful microarchitectural characteristics. (1) An adversary can directly check the states of the hardware component altered by the execution of the application. 
In this case, the attacker program needs to share the same component with the victim. (2) An adversary can measure the application's execution time to indirectly infer its microarchitectural characteristics.
In this case, the attack can be performed without the need to co-locate with the victim, but is only able to capture coarser-grained side-channel information. Thus, a large quantity of sessions are needed to statistically infer useful information.

\subsection{Multi-core Architecture}
\begin{figure}[h]
  \centering
  \vspace{-10pt}
  \includegraphics[width=0.8\linewidth]{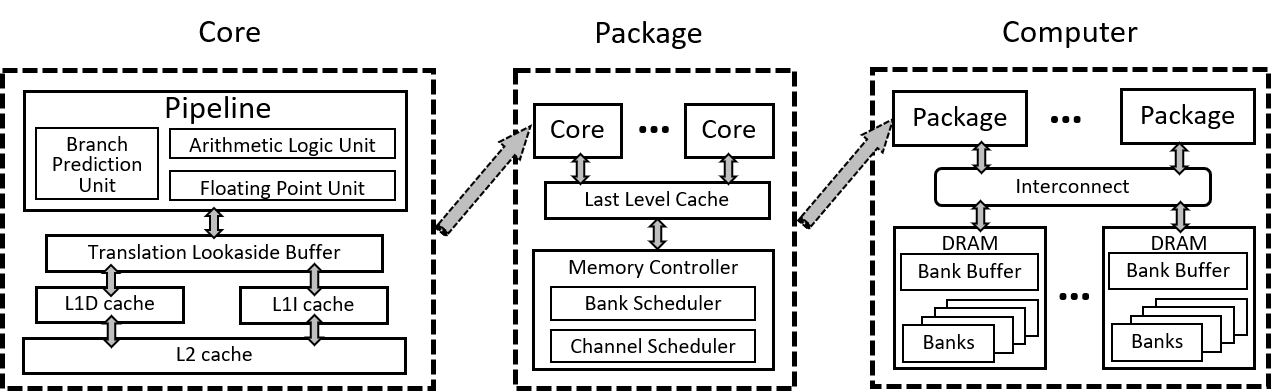}
  \caption{A Multi-core System}
  \vspace{-10pt}
  \label{fig:microarch}
\end{figure}

Figure \ref{fig:microarch} shows the overview of a multi-core system in a hierarchic structure. Basically, a computer has multiple \emph{CPU packages} and \emph{DRAM chips}, interconnected by memory buses (right). Each package is comprised of multiple \emph{CPU cores}, Last Level Caches and a memory controller (middle). Each CPU core consists of a pipeline, Translation Lookaside Buffer and two levels of caches (left).  

\bheading{CPU Core.}
A key feature in modern processors is the multi-stage pipeline, which allows executing instructions in a continuous and parallel manner.
The pipeline involves various functional units. For instance, the Branch Prediction Unit predicts the next branch to follow using recently executed branch targets held in the Branch Target Buffer (BTB). The Arithmetic Logic Unit is responsible for the arithmetic and bit-wise operations of integers, while the Floating Point Unit performs computation on floating point numbers. Modern pipeline designs support Simultaneous Multithreading (SMT), where multiple threads can execute concurrently in the pipeline. This feature can facilitate side-channel attacks in two ways: (1) the pipeline and functional units are shared among all active threads on the core, and such contention can expose side-channel information. (2) The attacker can measure the victim states concurrently at the same core without interrupting the execution of the victim, and obtain finer-grained information than in the case without SMT\footnote{For remote timing attacks, the adversary does not need to launch spy programs on the victim machine, hence SMT does not affect the attack results.}.

Processes use virtual addresses for data access, but the memory system uses physical addresses to store the data. Thus, the processor must perform a translation from virtual to physical addresses, based on the page table maintained by the operating system. To accelerate the translation, a hardware component named Translation Lookaside Buffer (TLB) is introduced to store recent translations, which can be retrieved later at a much higher speed than walking the page table.

CPU caches store recently accessed data for the processor to reuse in the near future and avoid time-consuming main memory access. A cache system is hierarchical and typically consists of three levels. Level 1 (L1) and Level 2 (L2) caches are on-core, while Last Level Caches (LLCs) are off-core. Caches closer to the processor are faster to access. There are separate data cache and instruction cache in L1, while L2 and LLC both have mixed data and instruction caches.
The smallest storage unit in a cache is a cache line that stores aligned adjacent bytes, which means the processor has to fetch or evict the cache line in its entirety. Modern caches commonly employ the $n$-way set-associative design, where a cache is divided into multiple sets, each containing $n$ cache lines. A data block is mapped to one cache set determined by its memory address. It can be stored in an arbitrary cache line within this set, selected by a replacement policy. For instance, the Least Recently Used (LRU) policy selects the cache line that is least recently accessed to hold the new block when this set is full. Particularly, a cache that has only one way in every set (i.e., $n=1$) is a direct-mapped cache, while a cache that has only one set is called fully-associative. 

\bheading{Package.}
Each package has one LLC that caches data from applications running on all cores. 
If a data access request cannot be fulfilled by the LLC, the memory controller will be involved. The memory controller buffers the requests in multiple queues, schedules them for high performance and fairness, and directs them to the DRAM chips. 
Cores, the LLC and the memory controller are interconnected by the memory buses with very high bandwidth. 

\bheading{Computer.}
A computer consists of several packages and DRAM chips. A DRAM chip has several banks. Each bank can be viewed as a two-dimensional array with multiple rows and columns, and has a row buffer to hold the most recently used row to speed up DRAM accesses. A memory access to a DRAM bank may either be served by the row buffer (buffer hit), which is fast, or in the bank itself (buffer miss), which is slow. Packages and DRAM chips are interconnected in a Non-Uniform Memory Architecture (NUMA): each DRAM is associated with a package, and each package can access all DRAM chips, but it's faster for the package to access its own local DRAM.

\noindent\textbf{Trusted Execution Environment (TEE).} This feature protects the security of unprivileged programs from the malicious OS through isolated execution and memory encryption. It has been implemented in ARM TrustZone \cite{alves2004trustzone} and Intel SGX \cite{costan2016intel}. However, as the design of TEE does not consider side-channel attacks, it is possible to use conventional techniques to steal secrets from the protected application. If the attacker is the malicious OS, she can obtain fine-grained information in an easier way by manipulating the OS interrupt (e.g., SGX-Step \cite{van2017sgx}). If the attacker is a normal user, she can use enclaves to hide malicious behaviors \cite{schwarz2017malware}.


\subsection{Cryptography}


\bheading{Asymmetric cryptography.}
Also known as public key cryptography, it adopts two keys: the user keeps a private key to herself and distributes a public key to the world. This design can provide confidentiality protection: anyone can use the public key to encrypt a message, which can only be decrypted by the user using the private key. It can also provide digital signature functionality: given a message, the user can use her private key to generate a signature, which can be verified by anyone using the public key and cannot be forged without the private key. Various algorithms were designed for asymmetric cryptography. 

The most famous algorithm is RSA \cite{RiShAd:78}. The key pair is generated using two large prime numbers that are kept secret, and the public key includes their product. The security of RSA relies on the practical difficulty of prime factorization of large numbers. ElGamal \cite{El:85} is another public-key cryptosystem, defined over any cyclic group, such as the multiplicative group of integers modulo $n$. Its security is supported by the difficulty of solving the Discrete Logarithm Problem. Yet another approach is Elliptic Curve Cryptography (ECC) \cite{Mi:85} that is based on the algebraic structure of elliptic curves over finite fields. Its security depends on the difficulty of solving the Elliptic Curve Discrete Logarithm Problem. Schemes based on ECC are designed for digital signature (ECDSA) and key exchange (ECDH).

\bheading{Symmetric cryptography.}
It uses the same key for both encryption and decryption, which is shared between two participants and cannot be distributed to the general public.
There are generally two types of symmetric-key algorithms. (1) In stream ciphers, each digit of the plaintext is encrypted at a time by a digit from a key stream to produce the ciphertext stream. One common component in stream ciphers is digital shift registers, which generate the key stream from a random seed value. (2) In block ciphers, fixed-length blocks of plaintext bits are blended with the key blocks to generate the ciphertext blocks. The encryption process usually adopts the Substitution-Perturbation Network (SPN), which takes a block of the plaintext and the key as the input, and applies multiple alternating rounds of substitution and permutation. AES is the most widely adopted block cipher, which is implemented as a multi-round SPN.



\bheading{Post-quantum cryptography.}
The advent of quantum computers in the near future can break the security of classical asymmetric cryptography. 
As such, post-quantum cryptography, a family of asymmetric ciphers, is proposed to survive attacks by a quantum computer. One popular scheme is lattice-based cryptography. For instance, NTRU \cite{HoPiSi:98} utilizes simple polynomial multiplication in the ring of truncated polynomials. 
Bimodal Lattice Signature Scheme (BLISS) \cite{DuDuLe:13} provides the digital signature function secure against quantum computers. Other algorithms were proposed based on the Ring Learning With Errors (RLWE) hard problem \cite{LyPeRe:10}.

\bheading{Cryptographic protocol.}
SSL/TLS allows a server and a client to use the handshake protocol to exchange a symmetric key $K$
for later secure communications. Specifically, the client first sends a list of its supported cipher suites and the server responds with a list of its supported cipher suites and the server certificate.
Then the client picks a cipher (e.g., RSA) supported by both parties, and generates a random secret string $K$ as the master key. The client generates a random non-zero padding string \texttt{pad} that is at least 8 bytes, creates a block $0x00||0x02||\texttt{pad}||0x00||K$, encrypts it using the server's public key and sends the ciphertext to the server. The server decrypts and accepts the message only when the format is valid. Finally, the server sends a ``finished'' message to the client, and the client replies with a ``finished'' message, marking the completion of the key exchange. 

After the key is established, the server and client adopt CBC-MAC to encrypt messages. The plaintext $P$ is created by concatenating the message $m$, its Message Authentication Code (MAC) and a padding string chosen to make the byte length of $P$ a multiple of the block size. Then $P$ is divided into blocks of $b$ bytes, each block encrypted with key $K$. The final message is the concatenation of a header and all encrypted blocks. The receiver decrypts the ciphertext in CBC mode and validates the padding format and the MAC. If both are correct, she accepts the original intact message $m$.

\subsection{Threat Model}
\bheading{What we cover.}
The target of our surveyed works is microarchitectural side-channel attacks. Microarchitecture is defined as the hardware implementation of an Instruction Set Architecture (ISA). We mainly focus on the x86 ISA (e.g., Intel and AMD) due to its wide adoption in modern PCs and servers, although some techniques can also be extended to the ARM processors \cite{green2017autolock,lipp2016armageddon}. Some works may need the processor to have additional hardware features such as Intel SGX \cite{lee2017inferring,brasser2017software,gotzfried2017cache,hahnel2017high, moghimi2017cachezoom, dall2018cachequote,schwarz2017malware,moghimi2019memjam,xu2015controlled,shinde2016preventing,wang2017leaky,van2017telling}, Intel TSX \cite{JaLeKi:16, disselkoen2017prime+} and AMD's cache-way predictor \cite{lipp2020take}. We will mention the requirements when discussing these works.

We consider the attacker as a normal user in the target system without root privileges. She can launch a malicious program on the same machine as the victim program, but cannot control the scheduling of the attacking process or the victim. One exception is the TEE scenario, where the attacker can be the OS that has the privilege to schedule all processes, but cannot introspect into the victim's protected memory. In remote timing attacks, the attacker can only query the victim cryptographic program remotely without launching the malicious program on the host machine.

\noindent\textbf{What we do not cover.} The following attacks and scenarios are out of the scope of this paper:

Physical side-channel attacks: these require the attacker to be physically local to the target system in order to collect the physical signals (e.g., power consumption \cite{Co:99}, electromagnetic radiation \cite{GePiTr:15}, acoustic emission \cite{GeShTr:14}) during the execution.

Network side-channel attacks: an adversary can exploit the network application features (e.g., response messages, packet pattern and size) as side channels to attack the network services protected by the cryptographic protocols, including RSA padding oracle attacks \cite{Bl:98} and CBC-MAC padding oracle attacks \cite{Va:02}. These network attacks have fundamentally different causes from microarchitectural attacks, and hence are not summarized in this paper. Note that we still consider the timing attacks which analyze the information leaked from the microarchitectural states of a remote machine.

Transient execution attacks: Meltdown \cite{MoMiDa:18} and Spectre
\cite{KoHoFo:19} attacks were demonstrated to bypass the protection schemes
in OSes, followed by many variants \cite{JoMaOf:18, ChChXi:19, TrLuMa:18, SaAhId:19, ClJoMi:18, schwarz2019netspectre}. Although side channel techniques are used in such attacks as a tool to leak secrets, these attacks target all data in the protected memory region instead of only cryptographic secrets.

Invasive attacks: following the most conventional microarchitectural side-channel attacks, we assume the attacker can only passively spy the behaviors of the victim, rather than actively compromising the integrity of the victim data. For instance, Rowhammer \cite{kim2014flipping}, an inherent vulnerability in modern high-density memory modules, can induce bit flips in the adjacent rows by frequently accessing a memory row. Fault attacks can also be achieved via physical means (e.g., laser injection) \cite{el2015survey}. Although such active attacks can break cryptographic ciphers (e.g., RSA \cite{bhattacharya2016curious}, AES \cite{zhang2018persistent}), we do not elaborate relevant works about Rowhammer \cite{frigo2020trrespass,mutlu2019rowhammer,kim2020revisiting,lou2019understanding} and fault attacks \cite{hou2019fully,roy2019safari,srivastava2020solomon,keerthi2020feds} in this paper. Note that Rambleed \cite{kwong2020rambleed} is an exception as it does not interfere with the victim data. 

Attacks against non-cryptographic applications: at the application level, attacks exist to identify keystrokes \cite{SoWaTi:01} and application states/activities \cite{SpPaMa:18}. At the system level, adversaries may infer host configurations \cite{schwarz2019javascript} and memory layout information \cite{HuWiHo:13}. We do not systematize these attacks.

\section{Characterization of Hardware Attack Vectors}
\label{sec:hardware}

We characterize the attack vectors of side-channel techniques from the level in the computer system and the category of side-channel information, as summarized in Table \ref{table:hardware-vector}.



\subsection{Instruction Level}
We first consider the instruction level attacks, which aim to identify when and what instructions are issued by the victim program. Based on the instruction trace, the adversary can infer the cryptographic secrets. Modern processors normally contain numerous arithmetic or logical functional units to perform designated computation. 
To launch an instruction level attack, the adversary must share the same CPU core and the target functional units with the victim process. The contention on these units can leak information of issued instructions from the victim to the adversary.

\bheading{Multiply instruction.} Multiplication is a fundamental operation in cryptographic applications. Hardware multiplier units are implemented in the CPU core to accelerate the computation. Wang et al. \cite{wang2006covert} demonstrated that processes running on the same core can interfere with the multiplier units, and the adversarial process is able to identify the multiply instruction of the victim based on the timing difference. Aciicmez et al. \cite{aciicmez2007cheap} designed a side-channel attack against the RSA implementation in OpenSSL by distinguishing the multiplications from square operations.

\begin{table*}[t]
\centering
\resizebox{\linewidth}{!}{
\begin{threeparttable}
\begin{tabular}{lllll}
  \hline
  \textbf{Level} & \textbf{Category} & \textbf{Sharing} & \textbf{Attacks} & \textbf{Requirements} \\
  \hline
  \hline

  \multirow{4}{*}{Instruction} & Multiply & $\blacksquare$ & Multiplier unit contention \cite{wang2006covert,aciicmez2007cheap} & \\  \cline{2-5}
  & Floating point & $\blacksquare$ & FPU contention \cite{andrysco2015subnormal} & \\ \cline{2-5}
  & Branch & $\blacksquare$ & BTB contention \cite{aciiccmez2007predicting,lee2017inferring} & \cite{lee2017inferring} requires Intel SGX\\ \cline{2-5}
  & Micro-operation & $\blacksquare$ & Port contention \cite{AlBrHa:19} & \\ \cline{2-2}\cline{4-4}
  \hline

  \multirow{5}{*}{Cache} & Cache set & \makecell[l]{$\blacksquare$: L1 \& L2,\\$\CIRCLE$: LLC} & \makecell[l]{\textsc{Prime-Probe} \cite{OsShTr:06,neve2006advances,OsShTr:06,Pe:05,aciiccmez2007yet,ZhJuRe:12} \\\cite{LiYaGe:15,InGuIr:16,irazoqui2015s,brasser2017software,gotzfried2017cache,hahnel2017high,briongos2019cache,aciiccmez2010new,kayaalp2016high,genkin2018drive} \\\cite{brasser2017software,gotzfried2017cache,hahnel2017high, moghimi2017cachezoom, dall2018cachequote,schwarz2017malware}\\\textsc{Evict-Time} \cite{OsShTr:06}, \textsc{Prime-Abort} \cite{disselkoen2017prime+}} & \makecell[l]{\cite{brasser2017software,gotzfried2017cache,hahnel2017high, moghimi2017cachezoom, dall2018cachequote,schwarz2017malware} \\ requires Intel SGX\\\cite{disselkoen2017prime+} requires Intel TSX} \\ \cline{2-5}
  & Cache line & \makecell[l]{$\blacksquare$: L1 \& L2,\\$\CIRCLE$: LLC} & \makecell[l]{\textsc{Flush-Reload} \cite{GuBaKr:11,YaFa:14,VaSmYa:15,BeVaSm:14,gruss2015cache}\\\textsc{Flush-Flush} \cite{gruss2016flush+}, \textsc{Reload+Refresh} \cite{briongos2020reload+} \\\textsc{Collide-Probe}, \textsc{Load-Reload}\cite{lipp2020take} \\LRU state leaking\cite{xiong2020leaking}} & \makecell[l]{Requires KSM \\\cite{lipp2020take} requires AMD predictor }\\ \cline{2-5}
  & Cache bank & $\blacksquare$ & Bank contention \cite{YaGeHe:17}, MemJam \cite{moghimi2019memjam} & \cite{moghimi2019memjam} requires Intel SGX \\ \hline
  
  \multirow{3}{*}{Memory Page} & \multirow{2}{*}{Page} & $\blacksquare$ & TLB contention \cite{GrRaBo:18,canella2019fallout} & \\ \cline{3-5}
  & & $\blacktriangle$ & Page Fault/Table Entry \cite{xu2015controlled,shinde2016preventing,wang2017leaky,van2017telling} & Requires Intel SGX \\  \cline{2-5}
  & DRAM bank row & $\blacktriangle$ & \makecell[l]{Row buffer contention \cite{pessl2016drama} \\ Rambleed \cite{kwong2020rambleed}} & \\ 
  \hline
  \end{tabular}
  \end{threeparttable}}
\caption[Caption for LOF]{Side-channel attack vectors in hardware. $\blacksquare$: sharing the same CPU core\protect\footnotemark; $\CIRCLE$: sharing the same package. $\blacktriangle$: sharing the same computer.}
\vspace{-25pt}
\label{table:hardware-vector}
\end{table*}
\footnotetext{If the SMT is enabled, the attacker and victim programs only need to share the physical core, instead of the logical core. An attacker in a different logical core from the victim but the same physical core can monitor the victim concurrently without interrupting it. This setting improves the success rate of side-channel attacks, and is commonly adopted by these works.}

\bheading{Floating point instruction.}
Another type of arithmetic operations is computation on floating point numbers. Such operations usually have large internal states, and are accelerated by the Floating Point Unit (FPU). Thus, FPU context switch can cause longer computation time.
Additionally, floating point instructions with different operands also have distinguishable execution times, which can leak sensitive information \cite{andrysco2015subnormal}. However, this technique is limited to applications with floating point instructions for critical operations, which are relatively rare in cryptographic applications.

\bheading{Branch instruction.}
Given that branch operations widely exist in many applications, speculative execution is introduced to accelerate such operations. The basic idea is to guess a branch path and execute the code in that path. Correct branch prediction saves the wait time for branch condition calculation and can significantly improve the performance, dominating the small overhead due to a misprediction. The speculation is implemented by hardware units, such as Branch Target Buffer (BTB) which records the target addresses of multiple previous branches. The adversary can observe the reduced execution time of the victim thanks to this technique and deduce the corresponding operations. Aciicmez et al. \cite{aciiccmez2007predicting} demonstrated such an attack against RSA in OpenSSL by selectively evicting entries from the BTB. Similar attacks were realized in the Intel SGX platform \cite{lee2017inferring}.
Evtyushkin et al. \cite{EvRiAb:18} further exploited the directional branch predictor as a new attack vector to steal secret from an SGX enclave.

\bheading{Micro-operation.}
The execution of an instruction can be divided into multiple micro-operations in the CPU pipeline. Contention on the corresponding functional units can also reveal the traces of micro-operations. Aldaya et al. \cite{AlBrHa:19} demonstrated a novel side-channel vector exploiting the port contention in the Execution Engine, a built-in component of modern processors with Intel Hyper-Threading technology. The adversary can capture side-channel information derived from port contention with very fine spatial granularity.


\subsection{Cache Level}
The cache system has become one of the most popular microarchitectural side channels due to its large channel capacity and low attack requirement. According to the granularity of leaked information, these attacks can be further divided into three categories. Below we briefly discuss the attack techniques and the literatures. Detailed modeling of these attacks can be found in \cite{zhang2018analyzing}.

\bheading{Cache set.} This type of attack aims at identifying the cache set trace of the victim process, with the limitation that different memory accesses mapped to the same cache set cannot be distinguished. There are multiple techniques to achieve this goal. 
The most common technique is \textsc{Prime-Probe} \cite{OsShTr:06}. The adversary first fills the critical cache set with its own memory lines (\textsc{Prime}). Then the victim executes for a period of time and potentially touches the set. After that, the adversary can measure the access time to those previously loaded memory lines (\textsc{Probe}). A longer access time indicates that the corresponding cache set has been used by the victim. While it is normally observed through cache hits, \cite{briongos2019cache} proposed that the adversary can use cache miss information for better attack efficiency.

\textsc{Prime-Probe} was first adopted to attack the AES encryption on the L1 data cache \cite{neve2006advances,OsShTr:06,Pe:05}. Then Aciicmez et al. \cite{aciiccmez2007yet} applied it to L1 instruction cache to check whether certain instructions are executed by the victim. This attack was enhanced in \cite{aciiccmez2010new}, which combines vector quantization and hidden Markov models to monitor each instruction cache set individually. Zhang et al. \cite{ZhJuRe:12} further explored the attack in the cloud, and demonstrated the practicality to steal information across VMs using the \textsc{Prime-Probe} technique.


Researchers shifted the interest from L1 cache to LLC as the adversary and victim do not need to share the same CPU core. Liu et al. \cite{LiYaGe:15} proposed the first \textsc{Prime-Probe} attack on LLC by reverse engineering the cache slice mapping and attacking specific cache sets. Following this work, Kayaalp et al. \cite{kayaalp2016high} further relaxed the attack assumptions and achieved higher resolution. Besides that, Inci et al. \cite{InGuIr:16} conducted the \textsc{Prime-Probe} attack on Amazon EC2 and retrieved the RSA key from the co-located instance. Irazoqui et al. \cite{irazoqui2015s} used the technique to monitor cache set traces of LLC in both Xen and VMware ESXi hypervisors, recovering the AES key in just a few minutes. This attack technique can also be mounted from a browser with the portable code, e.g., JavaScript, as demonstrated in \cite{genkin2018drive}. 


\textsc{Prime-Probe} attacks were also applied to the Intel SGX platform, enabling a malicious OS to retrieve secret information from the enclave applications \cite{brasser2017software,gotzfried2017cache,hahnel2017high, moghimi2017cachezoom, dall2018cachequote}. Since the OS is responsible for process scheduling and interruption, it can easily conduct \textsc{Prime-Probe} side-channel attacks on different levels of caches either synchronously or asynchronously. Besides, the attacker can also use SGX to conceal the cache attacks \cite{schwarz2017malware}.

Another technique to monitor the cache set access is \textsc{Evict-Time} \cite{OsShTr:06}. At the \textsc{Evict} stage, the adversary fills up one cache set and evicts the victim's memory lines out of the cache. Then at the \textsc{Time} stage, the victim executes certain blocks of code (e.g., encryption of one plaintext) and the corresponding execution time is measured. A long execution time means that the victim has accessed the critical cache sets during the execution and competed for the cache with the adversary.

In addition to timing attacks, Disselkoen et al. \cite{disselkoen2017prime+} proposed the \textsc{Prime-Abort} attack on the Intel Transactional Memory (TSX) processors, where the occurrence of aborts is used to infer the victim's access. At the \textsc{Prime} stage, the adversary initiates a TSX transaction for its memory blocks and fills up the target cache sets. When the victim evicts the adversary’s block out of the cache, the adversary observes an abort and detects the victim’s access.


\bheading{Cache line.} We next consider the attacks that can retrieve information at the granularity of one cache line, typically realized by the \textsc{Flush-Reload} technique.
This requires the adversary to share the same memory line with the victim, e.g., via memory deduplication. The adversary first evicts the critical memory lines out of the cache using dedicated instructions (e.g., \emph{clflush}). After a period of time, she reloads these lines into the cache and measures the access time. A shorter time indicates that the memory lines were accessed by the victim and betrays the access trace to the adversary. This attack was first mounted by Gullasch et al. \cite{GuBaKr:11} against the AES implementation on the L1 cache. Then Yarom and Falkner \cite{YaFa:14} adopted this technique on the LLC to monitor the square and multiply operations and steal keys from the RSA implementation. This method was further used to attack other ciphers such as ECDSA \cite{VaSmYa:15,BeVaSm:14}. Gruss et al. \cite{gruss2015cache} proposed a cache template attack, which leverages \textsc{Flush-Reload} to automatically build templates and attack critical applications.

A variant of \textsc{Flush-Reload} is \textsc{Flush-Flush} \cite{gruss2016flush+}, where the \textsc{Reload} operation is replaced by \textsc{Flush} at the second stage. This technique works as the execution time of \textsc{Flush} can also reflect whether the memory line is in the cache or not. This technique can reduce the activity on the cache and achieve better stealthiness, but has higher error rates due to the noise in the observation.

Cache line states with the replacement policy can also leak side-channel information. Lipp et al. \cite{lipp2020take} exploited the cache way predictor in the AMD processor to identify the victim's memory accesses with two new techniques: \textsc{Collide-Probe} and \textsc{Load-Reload}. Briongos et al. \cite{briongos2020reload+} reverse engineered the cache replacement policies of the Intel processors and then proposed the \textsc{Reload+Refresh} technique to monitor memory accesses in a cache set without evicting the victim's data. Xiong et al. \cite{xiong2020leaking} also presented that the LRU states of cache lines can leak information, and demonstrated the attacks on both Intel and AMD processors. Bhattacharya et al. \cite{bhattacharya2012hardware} discovered that the prefetching state of the cache lines can result in non-constant time encryption, which leaks timing information for the attacker to reveal the key from CLEFIA.



\bheading{Cache bank.}
The adversary can even get finer-grained side-channel information than the cache line. A cache line is divided into multiple cache banks. Concurrent requests to the same line but different banks can be served in parallel. However, requests to the same bank would cause a conflict, resulting in observable execution delay. This cache bank conflict can reveal the access pattern of the secret within one cache line. Yarom et al. \cite{YaGeHe:17} demonstrated such a side-channel attack on L1 cache targeting RSA in OpenSSL. Moghimi et al. \cite{moghimi2019memjam} designed a cache attack in the SGX platform, which is based on the false dependency of memory read-after-write (i.e., 4K Aliasing). This creates a new timing channel, enabling the adversary to observe the memory accesses in the same cache line with different offsets.


\subsection{Memory Page Level}
The memory page is the smallest unit for memory management in the OS and computer architecture. It is a contiguous and aligned memory block with a specific size, e.g., 4KB. The microarchitectural components responsible for manipulating memory pages can leak side-channel information at the granularity of the page size, which is coarser than that of instruction level or cache level attacks, but still allows the adversary to steal secrets from certain applications.


\bheading{Page.}
The TLB is an address translation cache, which is similar to CPU caches in terms of timing channels. Gras et al. \cite{GrRaBo:18} introduced a TLB-based side-channel attack, where interferences with the TLB are exploited to infer the victim's memory page trace. Canella et al. \cite{canella2019fallout} identified a new attack, which exploits the interactions with the store buffer to steal information of store addresses.

Page faults can also be used as side-channel information to capture the memory accesses \cite{xu2015controlled,shinde2016preventing}. A malicious OS can allocate a restricted number of physical pages to the victim application. When the application needs to access pages not available in the memory, a page fault is triggered and reported by the CPU. The OS is thus able to observe the memory pages the application tries to access. This technique, however, can induce huge performance overhead due to the large number of page faults. Researchers then proposed more advanced attacks \cite{wang2017leaky,van2017telling}, where the adversary can infer the accessed pages based on the flags in the page table entries, without the need to raise page faults. Moghimi et al. \cite{moghimi2020copycat} combined the SGX-Step mechanism \cite{van2017sgx} with the page-fault based technique to count the number of instructions issued within one page. This can reveal more information (instruction-level) about the victim program inside SGX enclaves for cryptanalysis.


\bheading{DRAM bank row.}
Each DRAM bank has a row buffer that caches the recently used DRAM row which normally contains multiple pages. It accelerates the memory access, but also introduces a timing channel. Pessl et al. \cite{pessl2016drama} designed a DRAM-based attack by reverse engineering the DRAM addressing schemes. This attack is less practical as it can only recover very coarse-grained information. However, Kwong et al. \cite{kwong2020rambleed} recently exploited the data-dependent bit flips induced by the Rowhammer \cite{kim2014flipping} to reveal RSA private key stored in the adjacent pages bit by bit.

\section{Characterization of software Attack Vectors}
\label{sec:crypto}

We systematically characterize side-channel vulnerabilities from past works based on
different operations in different cryptographic algorithms and protocols. Table
\ref{table:summary_crypto} summarizes the vulnerabilities covered in this paper.
For each vulnerability, we present the vulnerable operations, causes and the corresponding
attack techniques.



\begin{table*}[ht]
\centering
\resizebox{0.9\linewidth}{!}{
\begin{threeparttable}
\begin{tabular}{lllllll}
  \hline
  \textbf{Category} & \textbf{Operation} & \textbf{Implementation} & \textbf{Application} & \textbf{Cause} & \textbf{Attack Technique} & \textbf{Reference} \\
  \hline
  \hline
  \multirow{22}{*}{\makecell[l]{Asymmetric \\Cryptography}} & \makecell[l]{Modular \\Multiplication} & \makecell[l]{Basic and Karatsuba\\ multiplication} & RSA & $\blacksquare$ & Remote timing & \cite{BrBo:05}\\ \cline{2-7}
  & \multirow{18}{*}{\makecell[l]{Modular \\Exponentiation \\ \& Scalar\\ Multiplication}} & \multirow{3}{*}{\makecell[l]{Square-and-multiply \\\& Double-and-add}} & RSA & $\blacksquare$ & Cache \textsc{Flush-Reload} & \cite{YaFa:14} \\ \cline{4-7}
  & & & ElGamal & $\blacksquare$ & Cache \textsc{Prime-Probe} & \cite{ZhJuRe:12, LiYaGe:15} \\
  \cline{4-7}
  & & & EdDSA & $\blacksquare$ & TLB & \cite{GrRaBo:18} \\
  \cline{3-7}

  & & \multirow{2}{*}{\makecell[l]{Square-and-Multiply-always \\\& `Double-and-Add-always}} & RSA & $\blacksquare$ & Branch & \cite{DoKo:17} \\
  \cline{4-7}
   & & & RSA & $\blacksquare$ & TLB & \cite{GrRaBo:18} \\
  \cline{3-7}
   & & \multirow{7}{*}{Sliding window} & RSA & $\blacksquare$ & Cache \textsc{Prime-Probe} & \cite{Pe:05} \\
  \cline{4-7}
   & & & RSA & $\blacksquare$ & Cache \textsc{Flush-Reload} & \cite{BeBrGe:17} \\
  \cline{4-7}
  & & & ECDSA & $\blacksquare$ & Cache \textsc{Flush-Reload} & \cite{BeVaSm:14, VaSmYa:15, FaWaCh:16, AlBrFa:16} \\
  \cline{4-7}
  & & & ECDSA & $\blacksquare$ & Execution Port & \cite{AlBrHa:19} \\
  \cline{4-7}
  &  & & RSA & $\square$ & Cache \textsc{Prime-Probe} & \cite{InGuIr:16} \\
  \cline{4-7}
  &  & & ElGamal & $\square$ & Cache \textsc{Prime-Probe} & \cite{LiYaGe:15} \\
  \cline{4-7}
  & & & ECDSA & $\square$ & Cache \textsc{Prime-Probe} & \cite{BrHa:09} \\
  \cline{3-7}

  & & Fixed window & RSA & $\square$  & Cache bank & \cite{YaGeHe:17} \\
  \cline{3-7}

  & & \multirow{2}{*}{Montgomery ladder} & ECDSA & $\blacksquare$ & Cache \textsc{Flush-Reload} & \cite{YaBe:14} \\
  \cline{4-7}
  & &  & ECDSA & $\blacksquare$ & Remote timing & \cite{BrTu:11} \\
  \cline{3-7}

  & & \multirow{3}{*}{\makecell[l]{Branchless montgomery \\ladder}}  & ECDH & $\square$ & Cache \textsc{Flush-Reload} & \cite{ShKiKw:18} \\
  \cline{4-7}
  & & & ECDH & $\blacksquare$ & Cache \textsc{Flush-Reload} & \cite{GeVaYa:17} \\
  \cline{4-7}
  & & & ECDH & $\blacksquare$ & Remote timing & \cite{KaPeVa:16} \\
  \cline{2-7}

  & \multirow{3}{*}{Modular Inverse} & \multirow{3}{*}{\makecell[l]{Binary Extended Euclidean \\Algorithm}} & RSA & $\blacksquare$ & Branch & \cite{AcGuSe:07} \\
  \cline{4-7}
  & & & RSA & $\blacksquare$ & Page Fault & \cite{WeSpBo:18} \\
  \cline{4-7}
  & & & RSA & $\blacksquare$ & Cache \textsc{Flush-Reload} & \cite{AlGaTa:18} \\
  \hline

  \multirow{11}{*}{\makecell[l]{Symmetric \\Cryptography}} & \multirow{7}{*}{\makecell[l]{Substitution-\\Permutation}}& \multirow{7}{*}{Table lookup} & MISTY1 & $\square$ & Remote timing & \cite{tsunoo2002crypt}\\
  \cline{4-7}
  & &  & DES & $\square$ & Remote timing & \cite{tsunoo2003cryptanalysis}\\
  \cline{4-7}
  & &  & AES & $\square$ & Remote timing & \cite{BoMi:06,aciiccmez2007cache,Be:05} \\
  \cline{4-7}
  & &  & CLEFIA & $\square$ & Remote timing & \cite{rebeiro2009cache}\\
  \cline{4-7}
  & &  & AES & $\square$ & Cache \textsc{Prime-Probe} & \cite{OsShTr:06,neve2006advances} \\
  \cline{4-7}
  & &  & AES & $\square$ & Cache \textsc{Evict-Time} & \cite{OsShTr:06} \\
  \cline{4-7}
  & &  & AES & $\square$ & Cache \textsc{Flush-Reload} & \cite{GuBaKr:11, IrInEi:14} \\
  \cline{2-7}
  & \multirow{4}{*}{Shift register} & \multirow{4}{*}{Table lookup} & eSTREAM & $\square$ & Remote timing & \cite{gierlichs2008susceptibility} \\
  \cline{4-7}  
  & &  & HC-256 & $\square$ & Remote timing & \cite{zenner2008cache} \\
  \cline{4-7}
  & &  & LFSR & $\square$ & Remote timing & \cite{leander2009cache} \\
  \cline{4-7}
  & &  & SNOW 3G & $\square$ & Remote timing & \cite{brumley2010consecutive} \\
  \hline

  \multirow{7}{*}{\makecell[l]{Post Quantum \\Cryptography}}  & \multirow{3}{*}{\makecell[l]{Distribution\\Sampling}} & \multirow{1}{*}{CDT sampling} & BLISS & $\square$ & Cache \textsc{Flush-Reload} & \cite{BrHuLa:16,PeBrYa:17} \\
  \cline{3-7}
  & & \multirow{2}{*}{Rejection sampling} & BLISS & $\square$ & Cache \textsc{Flush-Reload} & \cite{BrHuLa:16,PeBrYa:17} \\
  \cline{4-7}
  & & & BLISS & $\blacksquare$ & Branch & \cite{EsFoGr:17, TiWa:19} \\
  \cline{2-7}

  & \multirow{2}{*}{\makecell[l]{Failure Rate \\Reduction}} & \multirow{2}{*}{Error Correcting Code} & \multirow{2}{*}{Ring-LWE} & \multirow{2}{*}{$\blacksquare$} & \multirow{2}{*}{Remote timing} & \multirow{2}{*}{\cite{DaTiVe:19}} \\
  & & & & & & \\
  \cline{2-7}
  & \multirow{2}{*}{\makecell[l]{Message \\Randomization}} & \multirow{2}{*}{Padding-Hash} & \multirow{2}{*}{NTRU} & \multirow{2}{*}{$\blacksquare$} & \multirow{2}{*}{Remote timing} & \multirow{2}{*}{\cite{SiWh:07}} \\
  & & & & & & \\
  \hline

  \multirow{6}{*}{\makecell[l]{Cryptographic \\Protocol}} & \multirow{3}{*}{RSA-PAD} & \multirow{3}{*}{Uniform response message} & TLS & $\blacksquare$ & Page, Cache, Branch & \cite{XiLiCh:17} \\
  \cline{4-7}
  & & & TLS & $\blacksquare$ & Cache, Branch & \cite{RoGiGe:19} \\
  \cline{4-7}
  & & & XML Encryption & $\blacksquare$ & Cache \textsc{Flush-Reload} & \cite{ZhJuRe:14} \\
  \cline{2-7}

  & \multirow{3}{*}{CBC-MAC-PAD} & \multirow{3}{*}{Constant-time compression} & TLSv1.1, TLSv1.2 & $\blacksquare$ & Cache \textsc{Flush-Reload} & \cite{IrLnEi:15} \\
  \cline{4-7}
  & &  & TLS & $\blacksquare$ & Page, Cache, Branch & \cite{XiLiCh:17} \\
  \cline{4-7}
  & &  & TLS & $\blacksquare$ & Cache \textsc{Prime-Probe} & \cite{RoPaSh:18} \\

  \hline
  \end{tabular}
  \end{threeparttable}}
\caption{Side-channel vulnerabilities. ($\blacksquare$: control flow, $\square$: data flow)}
\vspace{-25pt}
\label{table:summary_crypto}
\end{table*}

\subsection{Asymmetric Cryptography}
\label{sec:asycrypto}





\bheading{Modular multiplication.}
Given three integers $x$, $y$ and $m$, this operation is to calculate 
$x*y \mod m$.
Both OpenSSL and GnuPG implement two multiplication routines: naive multiplication and
Karatsuba multiplication \cite{KaOf:62}. The selection of the routine is based on the
operand size: the naive routine is taken for small multiplicands, while Karatsuba
routine is adopted for large ones. Such implementation introduces control-flow side channels about the operands: Karatsuba
routine is typically faster than the native routine. An adversary can measure the execution time
to infer the sizes of the operands, and then recover the secret key \cite{BrBo:05}.





\bheading{Modular exponentiation \& scalar multiplication.}
We consider the two operations together as they share similar implementations and 
vulnerabilities. Modular exponentiation is to calculate $x^y \mod m$, where $x$, $y$ and $m$ 
are three integers.
Scalar multiplication is to calculate $yx$ where $y$ is a scalar and $x$ is a point on the 
elliptic curve. The implementations of these two operations can reveal the secret key $y$ in
RSA and ElGamal, or the secret scalar $y$ in ECC via side channels.



The first implementation of modular exponentiation is \emph{square-and-multiply} \cite{Go:98}, where the calculation is converted into a sequence of \texttt{SQUARE} and \texttt{MULTIPLY} operations. The binary representation of $y$ is denoted as $y_{n-1}y_{n-2}...y_0$. Then starting from $n-1$ to 0, for each bit $y_i$, \texttt{SQUARE} is called. If $y_i$ is 1, \texttt{MULTIPLY} is also called. Similarly, scalar multiplication adopts the \emph{double-and-add} implementation \cite{HaMeVa:05}, which runs a sequence of \texttt{PointDouble} and \texttt{PointAdd} based on each bit $y_i$.
Such implementations are vulnerable to control-flow attacks: the execution of
\texttt{MULTIPLY} or \texttt{PointAdd} depends on bit $y_i$. By observing the traces of
\texttt{SQUARE} and \texttt{MULTIPLY} in modular exponentiation, or \texttt{PointDouble} and
\texttt{PointAdd} in scalar multiplication, an adversary can fully recover $y$. In earlier days, this implementation has been attacked via side channels \cite{Ko:96, dhem1998practical}. More recently, successful
attacks were demonstrated against RSA in GnuPG via cache \textsc{Prime-Probe}
\cite{ZhJuRe:12, LiYaGe:15} and \textsc{Flush-Reload} \cite{YaFa:14} attacks, and against 
EdDSA via TLB attacks \cite{GrRaBo:18}.


The second implementation of modular exponentiation is \emph{square-and-multiply-always}, which was designed to mitigate the above vulnerability. It always executes both \texttt{SQUARE} and \texttt{MULTIPLY} operations for each bit, and selects the output of \texttt{SQUARE} if $y_i$ is 0, or the output of \texttt{MULTIPLY} following \texttt{SQUARE} if $y_i$ is 1. Similarly, \emph{double-and-add-always} \cite{HaMeVa:05} was proposed for scalar multiplication in ECC.
These implementations execute a fixed number of \texttt{SQUARE} (\texttt{PointDouble}) and \texttt{MULTIPLY} (\texttt{PointAdd}) operations, defeating remote timing attacks.
However output selection still requires a secret-dependent branch, which is usually
smaller than one cache line. If it fits within the same cache line with the preceding and succeeding
code, then it is not vulnerable to microarchitectural attacks. However, Doychev and K{\"o}pf \cite{DoKo:17}
showed that for Libgcrypt, some compiler options can put this branch into separate cache lines, making
this implementation vulnerable to cache-based attacks. Gras et al. \cite{GrRaBo:18} showed that this
branch can be put into separate pages, and the implementation is subject to TLB-based attacks.



The third implementation is \emph{sliding window} \cite{BoCo:89}. For modular exponentiation, the exponent $y$ is represented as a sequence of windows $d_i$. Each window starts and ends with bit $1$, and the window length cannot exceed a fixed parameter $w$.
So the value of any window is an odd number between $1$ and $2^w-1$. This method pre-computes
$x^v \mod m$ for each odd value $v \in [1, 2^w-1]$, and stores these results in a table indexed
by $i\in[0, (v-1)/2]$. Then it scans every window, squares and multiplies the corresponding entry
in the table. Similarly, for scalar multiplication, the scalar $y$ is represented as a $w$-ary non-adjacent
form ($w$NAF), with each window value $d_i \in \{0,\pm1,\pm3,...,\pm(2^{w-1}-1)\}$.
It first pre-computes the values of $\{1, 3, ..., 2^{w-1}-1\}x$, and stores them into a table. 
Then it scans each window, doubles and adds $d_ix$ (in case $d_i<0$, adding $d_ix$ becomes subtracting $(-d_i)x$).



Two types of vulnerabilities exist in such implementations. The first one is 
a secret-dependent control flow: different routines will be called depending on
whether a window is zero. By monitoring the execution trace of those
branches, the adversary learns if each window is zero, and further recovers the secret.
Such attacks have been realized against RSA \cite{Pe:05, BeBrGe:17} and ECDSA
\cite{BeVaSm:14, VaSmYa:15, FaWaCh:16, AlBrFa:16, AlBrHa:19}.
The second one is a secret-dependent data flow: the access location in the pre-computed table
is determined by each window value. By observing the access pattern, the adversary is able to
recover each window value. Attacks exploiting this vulnerability have been mounted on
RSA \cite{InGuIr:16}, ElGamal \cite{LiYaGe:15} and ECDSA \cite{BrHa:09}.





The fourth implementation is \emph{fixed window} \cite{KaMeVa:96}, designed to approach true constant-time implementation. Similar to sliding 
window, it also divides the secret $y$ into a set of windows, pre-computes the 
exponentiation or multiplication of each window value, and stores the results in a table. The 
differences are that the window size is fixed as $w$, and the table stores both odd and even
(including zero) values. It removes the critical control flow branch at the cost of more 
memory and slower run time. To remove the critical data flow, this approach can be combined with scatter-gather 
memory layout technique \cite{BrGrSe:06}, which stores the pre-computed values in
different cache lines instead of consecutive memory locations. Specifically,
each window value is stored across multiple cache lines, and each cache line stores
parts of multiple window values. When \texttt{MULTIPLY} or \texttt{PointAdd} is executed,
multiple cache lines are fetched to reconstruct the window value, hiding the
access pattern from the adversary. Recently, Moghimi et al. \cite{moghimi2020tpm} performed a black-box timing analysis to steal private keys from the fixed window scalar multiplication inside the Intel Trusted Platform Module (TPM). 



This implementation is still vulnerable to attacks \cite{YaGeHe:17} using the cache bank,
the minimal data access unit in caches. As previously mentioned, the timing difference between hitting the same bank and hitting different banks in the same cache line
enables the adversary to infer the window values accessed during the
gathering phase, and then recover the secret bits. Garcia et al. \cite{GaBrYa:16} discovered a software bug in the DSA implementation in OpenSSL 1.0.2h: the flag to enable the fixed-window exponentiation is not correctly passed to the call site and thus the modular exponentiation still takes the insecure sliding window code path. 


%

The fifth implementation is \emph{masked window}, derived from the fixed window implementation to further hide the cache bank access patterns. The idea is to access all window values instead of just the one needed,
and then use a mask to filter out unused data. It performs a constant sequence of memory
accesses, and has been proven secure against different types of cache-based attacks \cite{DoKo:17}.

The sixth implementation, \emph{Montgomery ladder} \cite{Mo:87,JoYe:02}, is a variation of double-and-add-always for scalar multiplication. It also represents $y$ in the
binary form and executes both \texttt{PointAdd} and \texttt{PointDouble} functions for each bit,
irrespective of the bit value. The outputs of the functions are assigned to the 
intermediate variables determined by the bit value. A difference from double-and-add-always
is that in Montgomery ladder, the parameter of \texttt{PointDouble} is also determined by the 
bit value. Thus, the implementation contains even more secret-dependent branches. Yarom and Benger
\cite{YaBe:14} adopted cache \textsc{Flush-Reload} technique to identify the branch patterns in 
an attack to ECDSA in OpenSSL. 
Brumley and Tuveri \cite{BrTu:11} discovered that a loop in OpenSSL 0.9.8 begins
from the most significant non-zero bit in $y$ to 0. So the number of iterations is proportional
to $log(y)$. This presents a vulnerability for remote timing attacks. 

The last approach is \emph{Branchless Montgomery ladder} \cite{LaHaTu:16}, which replaces branches in Montgomery ladder with a function that uses bitwise logic to swap two intermediate values only if the bit is 1, and thus removes the timing channel. However, the implementations of \texttt{PointAdd} and \texttt{PointDouble} can still bring side channels.
First, OpenSSL adopts a lookup table to accelerate the square operation in these two functions.
The access pattern to the table can leak information about the secret in ECDH \cite{ShKiKw:18}.
Second, the modulo operation $x \mod m$ in these two functions adopted the early exit implementation: $x$ is
directly returned if its value is smaller than $m$. This branch can be exploited by the adversary
to check whether $x$ is smaller than $m$, and then deduce secrets in ECDH \cite{GeVaYa:17}. Third,
Kaufmann et al. \cite{KaPeVa:16} discovered that in Microsoft Windows, the multiplication function
of two 64-bit integers has an operand-dependent branch: if both operands 
have their 32 least significant bits equal to 0, then the multiplication is skipped and the 
result will be 0. This early exit branch was exploited to attack ECDH. 

\bheading{Modular Inverse.}
This operation is to calculate the integer $x^{-1}$ given $x$ and $m$ such that $x^{-1}x \equiv 1 \mod m$.
It can also be used to check if two integers are co-prime.
One possible implementation is \emph{Binary Extended Euclidean Algorithm} (BEEA) \cite{KaMeVa:96}, which uses arithmetic shift, comparison and subtraction to replace division. It is
particularly efficient for big integers, but suffers from control flow vulnerabilities
due to the introduction of operand-dependent branches.
Branch prediction \cite{AcGuSe:07} attacks were demonstrated to recover the value of $m$ in ECDSA and RSA. Page fault \cite{WeSpBo:18}
and cache \textsc{Flush-Reload} \cite{AlGaTa:18} techniques were adopted to attack the $gcd$ operation
in RSA key generation.


An alternative approach is \emph{Extended Euclidean Algorithm} (EEA). It calculates quotients and remainders in each step without introducing 
secret-dependent branches. It is less efficient but secure against control flow side-channel attacks.
Currently there are no side-channel vulnerabilities discovered in this implementation. However, some cryptographic libraries have software bugs that may disable this implementation accidentally. Garc{\'\i}a and Brumley \cite{GaPeBr:16} discovered that in OpenSSL 1.0.1u, the constant-time flag in ECDSA is not set for the secret nonces. Thus, the modular inverse computation still calls the vulnerable BEEA function instead of the secure EEA one.

\subsection{Symmetric Cryptography}
\label{sec:aes}
Symmetric ciphers, including block ciphers and stream ciphers, are also vulnerable to microarchitectural side-channel attacks. Different from asymmetric ciphers which usually contain lengthy and complex mathematical operations, symmetric ciphers typically leverage lookup tables instead of branch instructions or data-dependent rotations in computation.
As a result, symmetric ciphers are more susceptible to data-flow side-channel vulnerabilities than control-flow ones.

\bheading{Substitution-Permutation.}
This is a series of linked mathematical operations used in block ciphers.
It takes a block of the plaintext and the key as inputs, and applies several alternating 
rounds of substitution boxes and permutation boxes to produce the 
ciphertext block. 

The most common implementation of Substitution-Permutation is table lookup, which converts the algebraic operations in each round into memory accesses. Since the accessed entries of the lookup tables are determined by the secret keys and the plaintexts, an adversary can capture such access patterns to infer secrets. There are generally three types of attacks.

The first one is to steal the keys based on the entire execution time. Tsunoo et al. \cite{tsunoo2002crypt} discovered that the numbers of cache hits and misses when accessing the lookup table can affect the encryption time. Based on this observation, a cache timing attack was mounted on the block cipher MISTY1 \cite{tsunoo2002crypt} and further other ciphers, including DES \cite{tsunoo2003cryptanalysis}, AES \cite{Be:05} and Camellia \cite{zhao2009cache}. After that, an improved attack technique \cite{BoMi:06} was designed that leverages the cache access collisions to attack AES in OpenSSL. Aciicmez et al. \cite{aciiccmez2007cache} introduced a realistic remote cache timing attack to steal the AES key by analyzing only the first two encryption rounds. 

The second type of attacks is to build ``templates'' to infer the access pattern of the lookup table during encryption. Accesses to different entries in the lookup table can cause changes in encryption time, and such timing difference is only related to the lookup indexes when the host system configuration is deterministic. Hence, the adversary can construct a template trace of execution time on a system with the same configuration as the victim one, and then perform correlation analysis on the trace. Bernstein \cite{Be:05} used a large quantity of plaintexts to construct the timing template, and achieved remote cache timing attack on the first round of AES in OpenSSL. A few follow-up studies \cite{o2005investigation,neve2006cache,canteaut2006understanding} reproduced the attack and analyzed why it can reveal the key remotely. This technique was further applied to other block ciphers, e.g., CLEFIA \cite{rebeiro2009cache}.

The third type checks the state of the cache during or after the encryption to infer the internal state of the target cipher. Osvik et al. \cite{OsShTr:06} analyzed the cache state after encryption to deduce lookup operations in the first two rounds of AES. They carried out the attacks in both synchronous and asynchronous modes. Neve et al. \cite{neve2006advances} followed this method to target the last round of AES in OpenSSL, and successfully recovered the key with only 14 plaintexts.
Recent work tried to sniff the cache state of each table lookup operation to obtain finer-grained side-channel information. Gullasch et al. \cite{GuBaKr:11} manipulated the OS scheduler to craft a spy process that steals the cache set address of each table lookup very precisely. Irazoqui et al. \cite{IrInEi:14} adopted the cache \textsc{Flush-Reload} technique to obtain detailed access pattern of the victim cipher.

\bheading{Shift register.}
This critical component in stream ciphers is designed to generate pseudorandom key streams. In some stream ciphers, the core logic is implemented as lookup table operation to achieve efficiency. This gives rise to data flow side-channel vulnerabilities.

Gierlichs et al.
\cite{gierlichs2008susceptibility} performed a theoretical analysis on the susceptibility of eSTREAM candidates against side-channel attacks, and discovered that some of them are vulnerable from cache timing attacks due to the usage of lookup tables. 
Zenner \cite{zenner2008cache} described a cache timing attack on the HC-256 stream cipher and offered multiple suggestions for hardening the cipher. 
He further analyzed eSTREAM finalists, and pointed out that most stream ciphers are surprisingly resistant to cache timing attacks, as long as the lookup table is not adopted \cite{zenner2009cache}. Leander et al. \cite{leander2009cache} applied the cache timing analysis on LFSR-based stream ciphers, and proposed a general framework showing that the internal state of these ciphers can be recovered with very little computational effort. On this basis, Brumley et al. \cite{brumley2010consecutive} presented a cache timing attack on the SNOW 3G stream cipher, recovering the full cipher state in a short time.

\subsection{Post-Quantum Cryptography}
\label{sec:quantum}
Although post-quantum cryptography is secure against quantum computer based attacks, the
implementations of those algorithms may contain side-channel vulnerabilities that are subject to
attacks even by a classical computer.

\bheading{Distribution sampling.}
This operation is to sample an integer from a distribution. It is essential for BLISS
\cite{DuDuLe:13} to make the signature statistically independent of the secrets. However, an 
adversary can adopt side-channel attacks to recover the sampled data, and hence the secrets.


One popular and efficient sampling method is Cumulative Distribution Table (CDT) sampling \cite{Pe:10}, which pre-computes a table $\text{T}[i]=\mathbb{P}[x\leq i | x \sim D_{\sigma}]$. At the sampling phase, a random number $r$ is uniformly
chosen from $[0, 1)$, 
and the target $i$ is identified from T that satisfies $r \in [T[i-1], T[i])$. Some 
implementations adopt a guide table I to restrict the search space and accelerate the search 
process. BLISS adopts this approach to sample blinding values from a discrete Gaussian distribution, and add them to the signature.
However, the access pattern to the two tables reveals information about the sampled values. An adversary can adopt the cache \textsc{Flush-Reload} technique to recover the blinding values,
and further the secret key in BLISS \cite{BrHuLa:16, PeBrYa:17}.

Rejection Sampling \cite{GePeVa:08}, alternatively, samples a bit from a Bernoulli distribution $B(\text{exp}(-x/2\sigma^2))$. The implementation can bring side-channel opportunities to steal the secret $x$:
(1) a lookup table $\text{ET}[i] = \text{exp}(-2^i/(2\sigma^2))$ is pre-computed to 
accelerate the bit sampling, causing a data flow vulnerability; (2) the sampling
process needs to iterate over each secret bit and different branches will be executed for 
different bit values, resulting in a control flow vulnerability.

Practical attacks that exploit those vulnerabilities exist. First, rejection
sampling can replace CDT sampling for blinding value generation. An
adversary could utilize cache \cite{BrHuLa:16,PeBrYa:17} or branch \cite{EsFoGr:17} based attacks
to recover the sampled values in BLISS. Second, this approach can also be used to sample
random bits to probabilistically determine whether the blinding value is positive or negative,
and whether the signature should be accepted or rejected.
An adversary can infer the secret from this process via cache or branch
traces \cite{EsFoGr:17, TiWa:19}.

\bheading{Failure rate reduction.}
Post-quantum schemes may have certain failure rate during encryption or decryption
due to its statistic nature. Thus it is necessary to devise mechanisms to reduce the
possibility of failure.
Error Correcting Code (ECC) is adopted to significantly reduce the failure rate, but 
its implementation can reveal whether the ciphertext contains an error via timing channels: 
a ciphertext without an error is much faster to decode than one with errors. An adversary 
can exploit such information to recover the key \cite{DaTiVe:19}.

\bheading{Message randomization.}
Some post-quantum schemes require the message to be randomized during encryption and decryption.
This process can also create side-channel vulnerabilities.
For instance, encryption and decryption in NTRU use hash functions to randomize the messages.
However, the number of hash function invocations highly depends on the input message. As a result, the 
total execution time of encryption or decryption will vary on different inputs. By 
measuring such time information, an adversary is able to recover the secret input
\cite{SiWh:07}. 

\subsection{Cryptographic Protocol}
\label{sec:protocolpad}
In addition to cryptographic algorithms, side-channel attacks were proposed to target cryptographic protocols (specifically, their padding mechanisms).

\bheading{RSA-PAD.}
SSL/TLS commonly adopts RSA to exchange the symmetric key $K$, following Public Key Cryptography Standards (PKCS). The client pads the key $K$, encrypts it using RSA and sends the ciphertext to the server. When the server receives the ciphertext, she accepts the decrypted message only if the first two bytes are $0x00||0x02$. Otherwise, she sends an error message back to the sender and aborts the connection. This message serves as a side channel to recover the plaintext \cite{Bl:98}: when the client sends out a ciphertext, the adversary can intercept the message and send a modified one to the server. From the server's response, the adversary can learn if the first two bytes of the plaintext are $0x00||0x02$ (PKCS conforming) or not. This can reduce the scope of the plaintext. The attacker can repeat this process until the scope is narrowed down to one single value. 

A common defense is to unify the responses for valid and invalid paddings: if the decrypted message structure is not PKCS conforming, the receiver generates a random string as the plaintext, and performs all subsequent handshake computations on it. Thus, the adversary cannot distinguish valid ciphertexts from invalid ones based on the responses. However, the adversary can still adopt microarchitectural attack techniques to identify such information. Xiao et al. \cite{XiLiCh:17} adopted cache and control inference attacks to identify several control-flow vulnerabilities, which are mainly for improper error logging and reporting mechanisms. Ronen et al. \cite{RoGiGe:19} evaluated TLS implementations in several applications, and found seven of them were vulnerable to cache \textsc{Flush-Reload} and branch prediction attacks, due to secret-dependent control flows in data conversion, padding verification and padding oracle mitigation. Zhang et al. \cite{ZhJuRe:14} adopted the cache \textsc{Flush-Reload} technique to capture the access traces as the padding oracle of XML encryption.

\bheading{CBC-MAC-PAD.}
Standard network protocols adopt CBC-MAC to encrypt messages. The plaintext $P$ is composed of the message, its Message Authentication Code (MAC) and a padding string \texttt{pad}, and is encrypted in CBC mode. The receiver decrypts the ciphertext and validates the padding format and the MAC. If both are correct, she accepts the original intact message $m$. If the format is invalid, she rejects the message with a \emph{decryption\_failed} error response. If the format is correct but the MAC is incorrect, she rejects the message with a \emph{bad\_record\_mac} error response. These three conditions with three different responses create a side channel \cite{Va:02}: an adversary can modify the ciphertext and send it to the receiver for decryption. Based on the response, he can learn whether the chosen ciphertext is decrypted into an incorrect padding. This oracle enables the adversary to learn each byte of an arbitrary plaintext block.

Different countermeasures were designed to mitigate such side channels: responses for both invalid padding format error and invalid MAC error are unified to be indistinguishable to the adversary \cite{Mo:12}. Dummy MAC validation, padding data and compression operations were added to make the validation constant-time. However, 
these implementations still contain secret-dependent control flows, rendering them
vulnerable to microarchitectural attacks. An adversary can obtain the padding validation results
via cache \textsc{Flush-Reload} \cite{IrLnEi:15}, \textsc{Prime-Probe} \cite{RoPaSh:18}
or control-flow inference \cite{XiLiCh:17} techniques.

\section{Summary of Side-channel Countermeasures}
\label{sec:insights}

In this section, we summarize the potential defenses against microarchitectural attacks, and categorize them into three application-level, four system-level, and three hardware-level strategies. 

\subsection{Application-level Strategies}

\bheading{Runtime behavior unification.}
Control flow vulnerabilities exist when different secret values lead to different code paths that are distinguishable by the adversary using certain side-channel techniques. Two strategies can be used to remove such control flow.


The first strategy is \emph{always-execute-and-select-by-condition}. 
All possible code paths are executed regardless of the branching condition. 
Based on the secret value, the correct result is assigned to the return variable. This technique is adopted in modular exponentiation (square-and-multiply-always), scalar multiplication (double-and-add-always) and CBC-MAC-PAD (constant-time compression). 
This strategy is effective against remote timing attacks. However, the control flow in result selection can still be observed by a local adversary via microarchitectural attacks. Additionally, if the values for all code paths are pre-computed and stored in memory, the adversary can also infer the secret via data flow, exemplified by sliding window implementations in modular exponentiation and scalar multiplication.
The second strategy is \emph{always-execute-and-select-by-bit}. The difference between this strategy and the previous one is that the selection phase employs bitwise operations of the secret, 
and thus avoids introducing  branches or access patterns. The branchless Montgomery ladder algorithm adopts this solution for constant-time conditional swap in scalar multiplication.

Data flow vulnerabilities exist when different values of the secret lead to different
memory accesses that can be observed by the adversary.
Two strategies can remove such data flow.
The first strategy is \emph{always-access-and-select-by-bit}. It accesses all critical locations, and selects the correct value based on the bitwise operation. 
It is adopted in masked window modular exponentiation and scalar multiplication. 
The second strategy is \emph{calculate-on-the-fly}. We can calculate the value every time it is used instead of pre-computing all values and storing them into a table, particularly when the calculation is inexpensive and does not introduce secret-dependent control flows. Branchless Montgomery ladder adopts this method in the square operation of scalar multiplication.


\bheading{Runtime behavior randomization.}
For asymmetric ciphers, one possible approach is cryptographic blinding. There are generally two
types of blinding techniques.


We can adopt the \emph{key blinding} technique: a random factor is blended into the secret key, but the original key and the randomized key generate the same cryptographic result. The adversary can only obtain the randomized key via side-channel attacks, which is useless without knowing the blended random factor.
For ECDSA and ECDH, the randomized key is $k+sr$ where $r$ is a random number and $s$ is
the group order. The scalar multiplication generates $(k+sr)G$, the same as $kG$ \cite{Co:99}.
For RSA and ElGamal, the randomized key is $d+r\phi(n)$ where $r$ is a random number and
$\phi$ is the Euler's totient function. The decryption yields $c^{d+r\phi(n)} \mod n$, which is the same
as $c^d \mod n$. In both cases, the true value of $k$ is hidden from side-channel adversaries.



We can also utilize \emph{plaintext/ciphertext blinding}, which randomizes the plaintexts or ciphertexts that are adaptively chosen by the adversary.
The randomized texts cause the adversary to recover a wrong key via
side-channel analysis. This solution works only if correct ciphertexts can be produced from
randomized plaintexts and vice versa. 
For ECDSA and ECDH, we can choose a random point $R$ and use $G'=G+R$ in computation. The adversary
cannot recover $k$ from the side-channel observation without the knowledge of $R$ \cite{Co:99}, but
we can easily reproduce the correct result $kG$ by subtracting $kR$ from $kG'$. For RSA and
ElGamal, we can generate a random value $r$, and replace $c$ with $c*r^e$. Now the decryption
process is randomized to be $(c*r^e)^d \mod n = c^d * r^{ed} \mod n$.
To get $c^d \mod n$ we can simply multiply the result by $r^{-1}$, as $r^{ed} * r^{-1} \equiv 1 \mod n$.

For symmetric ciphers, the main target of side-channel attacks is the lookup table. One simple mitigation idea is to periodically randomize the entries in the table at runtime. Brickell et al. \cite{brickell2006software} designed compact and frequently randomized S-box for AES. Although the implementation is nearly 2$\times$ slower, it can indeed defeat the cache timing attack proposed in \cite{Be:05}.




\bheading{Software vulnerability identification.}
Some static approaches were designed to identify or verify potential side-channel vulnerabilities in commodity software. Abstract interpretation is a common approach to analyze the
source code and measure the information leakage (bounds). CacheAudit
\cite{DoKoMa:15, DoKo:17} modeled the relationship between the adversary's
observation and program's execution traces as a Markov chain, and quantified
the upper bound of the adversary's probability of success and the information leakage.
Molnar et al. \cite{MoPiSc:05} modeled control-flow side channels with a
program counter transcript, in which the value of the program counter at
each step is leaked to an adversary. FlowTracker \cite{RoQuAr:16} adopted
information flow tracking to analyze the assembly instructions and identify
the implicit flow edges in constant-time implementation of Elliptic Curve
Cryptography. Irazoqui et al. \cite{irazoqui2016mascat} designed a static code analysis tool to look up implicit features of microarchitectural attacks. 
SC-Eliminator \cite{wu2018eliminating} eliminated side-channel leakage using program repair, which conducts code transformations on unbalanced conditional jumps and cross cache line memory accesses to equalize the execution time. 
Wang et al. \cite{WaBaLi:19} proposed Secret-Augmented Symbolic
domain to track program secrets and their dependencies for precision and
coarse-grained public information for scalability. Various approaches were proposed \cite{BaBeGu:14, AlBaBa:16, BaPiTr:17,
DeFrHo:17, ReBaVe:17, BoHaKa:17, kopf2012automatic} to verify constant-time behavior of a program,
and check if it has secret-dependent conditional jumps or memory accesses.

There are also some dynamic analysis approaches, which focus on concrete program executions and identify vulnerabilities
from runtime execution traces. Zankl et al. \cite{ZaHeSi:16} profiled the number
of executions in a modular exponentiation operation, and calculated the Pearson correlation
coefficient between this number and the Hamming weights of the exponent to identify
information leakage during modular exponentiation. Wang et al. \cite{WaWaLi:17} proposed
CacheD to identify the vulnerable instructions by feeding different secrets into
the program and checking if each instruction has memory accesses to different
cache locations. Shin et al. \cite{ShKiKw:18} used the \textsc{Flush-Reload}
technique to collect two cache activity traces with two different secret inputs, 
and applied K-means clustering algorithm to check the dependency between cache activities
and secret inputs. Mutual information was adopted \cite{IrCoGu:17, WiMoEi:18} to measure
the side-channel information leakage and the relationship between secret inputs and memory
activities. Weiser et al. \cite{WeZaSp:18} exploited Intel Pin tool to collect
execution addresses, and applied statistic Kuiper's test and Randomized Dependence Coefficient to
discover vulnerabilities at the granularity of byte addresses. Xiao et al. \cite{XiLiCh:17}
proposed a framework to identify padding oracle attacks in SSL/TLS protocols in SGX secure
enclaves.

\subsection{System-level Strategies}

\bheading{Process execution partitioning.}
Since one enabling factor of microarchitectural attacks is the shared hardware components, we can enforce resource isolation for each process. The strategy is spatial partitioning, i.e., assigning different parts of the hardware units to processes. For instance, in the cloud scenario, hypervisor-based solutions were designed to defeat LLC attacks by partitioning the LLC, via page coloring \cite{ShSoCh:11}, page locking \cite{KiPeMa:12}, and Intel Cache Allocation Technology \cite{ShSoCh:11}.
Zhou et al. \cite{zhou2016software} prevented \textsc{Flush-Reload} attacks by managing dynamic page mapping to avoid cache line sharing, and thwarted \textsc{Prime-Probe} attacks through reduction of cross-domain cache line eviction. Hardware Transactional Memory was leveraged to eliminate the cache interference and prevent the adversary from evicting the victim's memory lines out of the cache \cite{gruss2017strong,chen2018leveraging}. To defeat side-channel attacks in web browsers, Schwarz et al. \cite{schwarz2018javascript} designed a fine-grained permission system to restrict the behaviors of JavaScript interface and functions.

\noindent\textbf{Process scheduling.}
Since a lot of microarchitectural side-channel attacks require the attacker and victim programs to run concurrently on the same machine, one possible strategy is to carefully schedule different programs to achieve temporal partitioning. Zhang and Reiter \cite{ZhRe:13} introduced an OS-based solution, which frequently flushes the local microarchitectural states (BTB, TLB, caches) to reduce side-channel leakage during context switches. Similar ideas were proposed in \cite{godfrey2013server, varadarajan2014scheduler}, where CPU caches are flushed during VM switches to defeat cache side-channel attacks in the cloud. To reduce the overhead of state cleansing operations, Sprabery et al. \cite{sprabery2018scheduling} implemented the scheduling as an extension to the Completely-Fair-Scheduler in Linux. 

\bheading{Measurement randomization.}
This idea is to add randomization to the adversary's measurements, making it difficult or infeasible to capture accurate information based on the observations. This was first proposed in \cite{hu1992reducing} to fuzz the timing information to reduce timing channels. Vattikonda et al. \cite{VaDaSh:11} modified the \emph{rdtsc} instruction from the hypervisor to randomize the emulated timer. Martin et al. \cite{martin2012timewarp} optimized this approach by adding random noise in each predefined epoch. Li et al. \cite{LiGaRe:14} introduced Stopwatch, which disables precise timing measurement in the cloud server to mitigate timing-channel attacks. 

An alternative way is to add randomization inside the application during compiling. Crane et al. \cite{crane2015thwarting} designed an approach to dynamically randomize the control flow in the application to defeat cache side-channel attacks. Braun et al. \cite{braun2015robust} inserted random temporal paddings into the source application to obfuscate the adversary's observations. 

\bheading{Attack Detection.}
In addition to prevent side-channel attacks, another direction is to detect the occurrence of side-channel attacks at runtime. The key insight is that the victim and attacking programs involved in a microarchitectural side-channel attack exhibit unusual behaviors compared to normal applications, which can be observed by the privileged software. For instance, signature-based detection systems \cite{demme2013feasibility,chiappetta2016real,payer2016hexpads} were proposed to detect side-channel attacks using hardware performance counters. NIGHTs-WATCH \cite{mushtaq2018nights} leveraged machine learning techniques to detect cache attacks based on the performance counter values. Zhang et al. \cite{ZhZhLe:16} combined both signature-based and anomaly-based detection methods to identify cache \textsc{Prime-Probe} and \textsc{Flush-Reload} attacks with high fidelity. Hunger et al. \cite{hunger2015understanding} proposed mimicking the behaviors of the victim to attract and identify the adversary by monitoring its common characteristics.
To detect page-level side-channel attacks in Intel SGX, multiple techniques \cite{chen2017detecting, strackx2017heisenberg, shih2017t} took advantage of the Intel TSX feature.


\subsection{Architecture-level Strategies}

\bheading{Hardware resource partitioning.}
Computer architects and researchers have designed security-aware architectures to protect critical applications from side-channel attacks. One straightforward way is to partition the shared resources to prevent processes from interfering with one another.


As the CPU cache is the most popular target for microarchitectural attacks, a lot of efforts have been spent to enhance the security of cache architectures. The simplest way is to divide the cache into multiple partitions by ways, and allocate them to different processes exclusively, such as
Statically Partitioned cache \cite{he2017secure} and SecVerilog cache \cite{zhang2012language,zhang2015hardware}. However, statically partitioned caches can significantly reduce the effective cache capacity for each process, causing huge performance degradation and unfairness.  


A more promising direction is dynamic partitioning. Partition-Locked cache \cite{WaLe:07} assigns a protection bit to each memory line to denote whether it needs to be locked in the cache. Green et al. \cite{green2017autolock} identified an undocumented feature \emph{AutoLock} on ARM processors to prevent evictions of lines in core-private caches. 
Vantage \cite{sanchez2011vantage} enforces fine-grained partitioning during replacement instead of physically restricting the placement of cache lines.
NoMo Cache \cite{DoJaLo:12} reserves certain blocks in every cache set for each thread, which cannot be evicted by other threads. The number of those blocks is dynamically adjusted based on the activity of the thread. 
To maintain high associativity while partitioning the cache, Futility Scaling \cite{wang2014futility} keeps evicting cache lines with the largest scaled futility to retain a number of useful lines. 
SecDCP Cache \cite{wang2016secdcp} improved over SecVerilog cache by dynamically partitioning the cache for different processes based on the cache miss rate of instructions at runtime. The work was further enhanced in FairSDP \cite{sari2019fairsdp}, which improves the fairness among competing threads. 
SHARP Cache \cite{yan2017secure} implements Core Valid Bits to cache lines, enabling the OS to prioritize the cache lines for eviction when cache conflict occurs. This can reduce the interference among different processes. DAWG \cite{kiriansky2018dawg} introduces minimal modifications on the hardware to fully isolate cache his/misses and metadata updates across protection domains in the cache set. ZBM \cite{saileshwar2019lookout} modifies the replacement policy to equalize the latencies of cache hits and misses on certain lines invalidated due to flush-caused invalidation.


This partitioning strategy is also adopted by other platforms and scenarios. To mitigate side-channel attacks in Trusted Execution Environment, Sanctum Cache \cite{costan2016sanctum} assigns different memory regions to different enclaves or OSes to disable cache sharing, and flushes the caches during context switching from the enclave mode to the non-enclave mode. 
Following this work, HybCache \cite{dessouky2020hybcache} selectively applies partitioning only for isolated execution domains, making the sharing of cache resources more flexible and efficient. 

For other microarchitectural components, Static-Partition TLB \cite{deng2019secure} was introduced, which follows the idea of static-partitioning cache and Sanctum to isolate the TLB accesses between the victim and the attacker. Wang et al. \cite{wang2012efficient} designed approaches to prevent information leakage via on-chip networks by restricting low-security traffic. 
To protect the memory controllers, Wang et al. \cite{wang2014timing} adopted temporal partitioning to group memory access requests in queues according to their security domains, and separate the requests serving different domains in different time slots. This design was further improved by the lattice priority scheduling \cite{ferraiuolo2016lattice} and quantitative security guarantee \cite{wang2017secure} to reduce performance overhead and increased scalability.


\bheading{Resource usage randomization.}
This strategy is to randomize the resource usage of the processes to obfuscate the side-channel observation. 
Random Permutation cache \cite{WaLe:07} maintains a dynamic and random memory-to-cache mapping table for each process to ensure the accessed set is unpredictable. Newcache \cite{liu2016newcache, wang2008novel} introduced a virtual Logical Direct-Mapped (LDM) Cache with two mappings: the one from memory addresses to the LDM cache is direct-mapped for high performance, while the other from the LDM cache to the physical cache is fully-associative for strong security.
Non Deterministic Cache \cite{keramidas2008non} leverages cache access delay to obscure the relationship between cache accesses and the observed timing information.
Random Fill Cache \cite{LiLe:14} randomizes cache prefetching by bringing the accessed memory line along with random neighbors to the cache, so the memory access pattern is convoluted. TSCache \cite{trilla2018cache} makes the cache access pattern and timing leakage unpredictable in the embedded devices with injection of randomized cache timing behaviors. 
CEASER Cache \cite{qureshi2018ceaser} encrypts and remaps the addresses of memory lines, which can efficiently decrease the probability of conflict caused by cache misses. An improved version, CEASER-S \cite{qureshi2019new} was then designed to divide the cache into multi-way partitions and adopt random
placement among these partitions, further increasing the uncertainty
in memory-to-cache mapping. SCATTER Cache \cite{werner2019scattercache} translates the memory address and process information to a random cache set index, and guarantees that each cache way has its own specific index.  
Phantom Cache \cite{tanphantomcache} proposed a novel localized randomization method, which randomly places a loaded memory block at a location in its fixed mapping range. 
Purnal et al. \cite{purnal2020systematic} introduced a generic model for randomization-based caches with systematic analysis, which can also serve as a baseline for future cache design. 

Randomization can be applied to other microarchitectural components as well. Random-Fill TLB \cite{deng2019secure} decorrelates the memory access from the actual TLB entry by randomizing the address translation for TLB misses, which can result in non-deterministic observations.
Camouflage \cite{zhou2017camouflage} hides the timing information by shaping the time of memory operations into a predetermined distribution and adding fake memory traffic as needed.

\noindent\textbf{Hardware vulnerability identification.}
Multiple approaches were designed to model security-aware hardware architectures and measure their vulnerabilities under different attacks. Demme et al. \cite{demme2012side} proposed the Side-channel Vulnerability Factor (SVF), a metric to quantify the system's vulnerabilities by measuring the correlation between the attacker's observations and the victim's actual execution. \cite{zhang2013side, bhattacharya2012hardware} introduced improved metrics over SVF to assess other hardware components and attacks. Zhang et al. \cite{zhang2014new} modeled the cache architectures as finite-state machines and then quantitatively revealed potential side channel leakage. He et al. \cite{he2017secure} used probabilistic information flow graph to model the interaction of the attacker and the victim in the cache architecture and measure the cache's resilience against side channels. More analyses of secure cache architectures have been done using computation tree logic \cite{deng2018cache}, three-step model \cite{deng2019analysis}, attack graphs \cite{wang2020analyzing}, and neural networks \cite{zhang2018analyzing}.

\subsection{Discussions of These Defenses}
Among aforementioned three categories of defense strategies, the application-level ones are the most widely adopted in the cryptographic community due to two reasons: (1) they are easy to implement and patch the vulnerabilities immediately; (2) these approaches bring little computational overhead to the applications. However, these solutions are not very general, and the designs require manual analysis by experts. Hence, it is difficult to guarantee modern applications are free of side-channel vulnerabilities inside the tremendous amount of code. We will perform a comprehensive study about two popular cryptographic applications in the next section.

There are relatively fewer system-level or architecture-level strategies applied to real-world platforms or products, although a lot of general solutions have been well developed in academia. These solutions may have performance issues, and are much harder to implement in the existing machines, especially for the new hardware designs. So currently the most practical defense solutions for side-channel attacks are still based on modifying cryptographic algorithms and implementations.


\section{Evaluation of Cryptographic Libraries}
\label{sec:history}

From a practical perspective, we review, analyze and evaluate the development of side-channel
attacks and defenses in two commonly used cryptographic libraries: OpenSSL and GNU Crypto (GnuPG,
Libgcrypt and GnuTLS). We collected the history of side-channel vulnerabilities and countermeasures (1999 -- 2019) from Common
Vulnerabilities and Exposures (CVE), the version control history and the source code. Tables
\ref{table:openssl-history} and \ref{table:gnupg-history_crypto} show the evolution of the libraries.


\begin{table*}[ht]
\centering
\resizebox{\linewidth}{!}{
\begin{threeparttable}
\begin{tabular}{lllllllllll}
  \hline
  \multirow{2}{*}{\textbf{ID}} & \multirow{2}{*}{\textbf{Patch Date}} & \multirow{2}{*}{\textbf{Version}} & \multirow{2}{*}{\textbf{Operations}} & \multirow{2}{*}{\textbf{Implementation}} & \multicolumn{2}{c}{\textbf{CVE}} & \multicolumn{3}{c}{\textbf{CVSS}} & \multirow{2}{*}{\textbf{Countermeasures}} \\
  \cline{6-10}
  & & & & & Date & ID & B & E & I & \\
  \hline
  \hline
  1 & 2001/07/09 & 0.9.6b & RSA-PAD & Uniform error message & & &   & & & Fix bugs\\
  \hline
  2 & 2003/02/19 & 0.9.6i, 0.9.7a & CBC-MAC-PAD & Uniform error message  & 2003/03/03  &  2003-0078 & 5 & 10 & 2.9 & Dummy checking for TLS\\
  \hline
  \multirow{2}{*}{3} & \multirow{2}{*}{2003/04/10} & \multirow{2}{*}{0.9.6j, 0.9.7b} & Modular multiplication & \makecell[l]{Basic and Karatsuba \\multiplication} & 2003/03/31 & 2003-0147 &  5 & 10 & 2.9 & RSA blinding \\
  \cline{4-11}
  & & & RSA-PAD & Uniform error message &  2003/03/24 & 2003-0131 &  5 & 10 & 2.9 & Uniform version error message \\
  \hline
  4 & 2005/07/05 & 0.9.8 & \multirow{2}{*}{Modular exponentiation} & \multirow{2}{*}{Sliding window} & & &  & & & \multirow{2}{*}{Fixed window} \\
  \cline{1-3}
  5 & 2005/10/11 & 0.9.7h &  &  &  &  &  & & &  \\
  \hline
  6 & 2007/10/11 & 0.9.8f & Modular inversion & \makecell[l]{Binary Extended Euclidean \\Algorithm}  &  &  &  & & & Euclidean Extended Algorithm \\
  \hline
  7 & 2011/09/06 & 1.0.0e & \multirow{2}{*}{Scalar multiplication} & \multirow{2}{*}{Montgomery ladder} &  \multirow{2}{*}{2011/05/31} &  \multirow{2}{*}{ 2011-1945} & \multirow{2}{*}{2.6} & \multirow{2}{*}{4.9} & \multirow{2}{*}{2.9} & \multirow{2}{*}{\makecell[l]{Make the bit length of scalar \\constant}} \\
  \cline{1-3}
  \multirow{3}{*}{8} & \multirow{3}{*}{2012/01/04} & 0.9.8s & & & &  &  & & & \\
  \cline{3-11}
  & & \multirow{2}{*}{0.9.8s, 1.0.0f} & \multirow{2}{*}{CBC-MAC-PAD} & Uniform error message & 2012/01/05 &  2011-4108 & 4.3 & 8.6 & 2.9 & Dummy checking for DTLS \\
  \cline{5-11}
  & & & & Padding data initialization & 2012/01/05 &  2011-4576 & 5 & 10 & 2.9 & Fix bugs \\
  \hline
  9 & 2012/03/12 & 0.9.8u, 1.0.0h & RSA-PAD  (PKCS\#7,CMS) & Error message & 2012/03/12 &  2012-0884 &  5 & 10 & 2.9 & \makecell[l]{Uniform error message and \\dummy checking} \\
  \hline
  \multirow{2}{*}{10} & \multirow{2}{*}{2012/03/14} & \multirow{2}{*}{1.0.1} & Scalar multiplication & Sliding window &  &  &  & & & Masked window \\
  \cline{4-11}
  & & & Substitution-Permutation & T-box lookup & &  &  & & & AES-NI support \\
  \hline
  11 & 2013/02/05 & \makecell[l]{0.9.8y, 1.0.0k,\\1.0.1d} & CBC-MAC-PAD & Dummy MAC checking & 2013/02/08 &  2013-0169 & 2.6 & 4.9 &2.9 & Dummy data padding \\
  \hline
  12 & 2014/04/07 & 1.0.1g & \multirow{2}{*}{Scalar multiplication} & \multirow{2}{*}{Montgomery ladder} & \multirow{2}{*}{2014/03/25} &  \multirow{2}{*}{ 2014-0076} &  \multirow{2}{*}{1.9} & \multirow{2}{*}{3.4} & \multirow{2}{*}{2.9} & \multirow{2}{*}{Branchless Montgomery ladder} \\
  \cline{1-3}
  13 & 2014/06/05 & 0.9.8za, 1.0.0m & & & &  &  & & & \\
  \hline
  14 & 2014/10/15 & \makecell[l]{0.9.8zc, 1.0.0o,\\1.0.1j} & CBC-MAC-PAD & Error message & 2014/10/14 &  2014-3566 & 4.3 &8.6 &2.9 & Disable fallback of SSLv3.0 \\
  \hline
  \multirow{4}{*}{15} & \multirow{4}{*}{2016/01/28} & \multirow{4}{*}{1.0.1r, 1.0.2f} & \multirow{4}{*}{RSA-PAD} & \multirow{4}{*}{Uniform error message} & 2016/02/14 &  2015-3197 &4.3  &8.6 &2.9 & \multirow{4}{*}{Disable SSLv2 ciphers}\\
  & & & & & 2016/03/02 &  2016-0703 & 4.3  &8.6 &2.9 &  \\
  & & & & & 2016/03/02 &  2016-0704 &  4.3  &8.6 &2.9 &  \\
  & & & & & 2016/03/01 &  2016-0800 &  4.3  &8.6 &2.9 &  \\
  \hline
  \multirow{5}{*}{16} & \multirow{5}{*}{2016/03/01} & \multirow{5}{*}{1.0.1s, 1.0.2g} & \multirow{4}{*}{RSA-PAD} & \multirow{4}{*}{Uniform error message} & 2016/02/14 &  2015-3197 &  4.3  &8.6 &2.9 & \multirow{4}{*}{Disable SSLv2 protocols} \\
  & & & & &  2016/03/02&  2016-0703 &  4.3  &8.6 &2.9 &  \\
  & & & & & 2016/03/02 &  2016-0704 &  4.3  &8.6 &2.9 &  \\
  & & & & & 2016/03/01 &  2016-0800 & 4.3  &8.6 &2.9 &  \\
  \cline{4-11}
  & & & Modular exponentiation & Fixed window& 2016/03/03 &  2016-0702 & 1.9 &3.4 &2.9 & Masked window \\
  \hline
  17 & 2016/05/03 & 1.0.1t, 1.0.2h & CBC-MAC-PAD (AES-NI) & Dummy data padding & 2016/05/04 &  2016-2107 & 2.6 &4.9 &2.9 & Fix bugs \\
  \hline
  18 & 2016/09/22 & 1.0.1u, 1.0.2i & Modular exponentiation & Fixed window & 2016/06/19 &  2016-2178 & 2.1 &3.9 &2.9 & Fix bugs \\
  \hline
  \multirow{4}{*}{19} & \multirow{4}{*}{2018/08/14} & \multirow{3}{*}{1.0.2p, 1.1.0i} & Scalar multiplication & \makecell[l]{Branchless Montgomery \\ladder} & &  &  & & & \makecell[l]{On-the-fly calculation to replace \\lookup table} \\
  \cline{4-11}
  & & & Modular inversion & \makecell[l]{Binary Greatest Common \\Divisor} & 2018/04/16 &  2018-0737 & 4.3 &8.6 &2.9 & Extended Euclidean Algorithm \\
  \cline{4-11}
  & & & Modulo & Early exit &  &  &  & & & ECDSA and DSA blinding \\
  \cline{3-11}
  & & 1.1.0i & Scalar multiplication & Sliding window & &  &  & & & Branchless Montgomery ladder \\
  \hline  
  \multirow{6}{*}{20} & \multirow{6}{*}{2018/09/11} & \multirow{6}{*}{1.1.1} & \multirow{4}{*}{Scalar multiplication} & \multirow{2}{*}{\makecell[l]{Branchless Montgomery \\ladder}} & &  &  & & & Differential addition-and-doubling \\
  \cline{11-11}
  & & & & & &  &  & & & Coordinate blinding \\
  \cline{5-11}  
  & & & & Masked window & &  &  & & & Branchless Montgomery ladder \\
  \cline{5-11}  
  & & & & Sliding window & &  &  & & & Branchless Montgomery ladder \\
  \cline{4-11}
  & & & \multirow{2}{*}{Modular inversion} & \multirow{2}{*}{\makecell[l]{Extended Euclidean \\Algorithm}} & &  &  & & & New constant-time function for EC \\
  \cline{11-11}
  & & & & & &  &  & & & Input blinding \\
  \hline
  \multirow{3}{*}{21} & \multirow{3}{*}{2018/11/20} & 1.0.2q& Scalar multiplication & Sliding window & 2018/11/15 &  2018-5407 & 1.9  &3.4 &2.9 & Branchless Montgomery ladder \\
  \cline{3-11}
  & & \multirow{2}{*}{\makecell[l]{1.0.2q, 1.1.0j,\\1.1.1a}} & DSA sign setup & Space preallocation & 2018/10/30 &  2018-0734 & 4.3 &8.6 &2.9 & Fix bugs \\
  \cline{4-11}
  & & & Scalar multiplication & Space preallocation & 2018/10/29 &  2018-0735 & 4.3 &8.6 &2.9 & Fix bugs \\
  \hline
  \multirow{2}{*}{22} & \multirow{2}{*}{2019/02/26} & 1.0.2r & CBC-MAC-PAD & Protocol error handling & 2019/02/27 &  2019-1559 & 4.3 &8.6 &2.9 & Fix bugs \\
  \cline{3-11}
  & & 1.1.1b & Modular inversion (EC) & \makecell[l]{Binary Extended Euclidean \\Algorithm} &  &  &  & & & \makecell[l]{EC-specific inversion function with \\input blinding} \\
  \hline
 \multirow{2}{*}{23} & \multirow{2}{*}{2019/09/10} & \multirow{2}{*}{\makecell[l]{1.1.1d, 1.1.0l,\\1.0.2t}} & EC Group & Cofactor & 2019/09/10 &  2019-1547 &1.9  &3.4 & 2.9& Fix bugs \\
  \cline{4-11}  
 & & & RSA-PAD & CMS and PKCS7 decrypt & 2019/09/10 &  2019-1563 & 4.3 &8.6 &2.9 & Fix bugs \\
  \hline
\end{tabular}
\end{threeparttable}}
\caption{Vulnerabilities in OpenSSL (For CVSS column, B: Base; E: Exploitability; I: Impact)}
\vspace{-25pt}
\label{table:openssl-history}
\end{table*}

\begin{table*}[ht]
\centering
\resizebox{\linewidth}{!}{
\begin{threeparttable}
\begin{tabular}{lllllllllll}
  \hline
  \multirow{2}{*}{\textbf{ID}} & \multirow{2}{*}{\textbf{Patch Date}} & \multirow{2}{*}{\textbf{Version}} & \multirow{2}{*}{\textbf{Operations}} & \multirow{2}{*}{\textbf{Implementation}} & \multicolumn{2}{c}{\textbf{CVE}} & \multicolumn{3}{c}{\textbf{CVSS}} & \multirow{2}{*}{\textbf{Countermeasures}} \\
  \cline{6-10}
  & & & & & Date & ID & B & E & I & \\
  \hline
  \hline
  1 & 2006/09/08 & T1.4.3 & \multirow{2}{*}{RSA-PAD} & \multirow{2}{*}{Error Message} & &  &  & & & \multirow{2}{*}{Uniform error message} \\
  \cline{1-3}
  2 & 2006/09/21 & T1.5.1 & & & &  &  & & & \\
  \hline
  3 & 2011/06/29 & L1.5.0 & Substitution-permutation & T-box lookup & & &  &  & & AES-NI support \\
  \hline
  4 & 2012/01/06 & T3.0.11 & CBC-MAC-PAD & Uniform error message & 2012/01/05 &  2012-0390 & 4.3 &4.9 &2.9 & Dummy checking for DTLS \\
  \hline
  5 & 2013/02/04 & \makecell[l]{T2.12.23, T3.0.28, \\T3.1.7} & CBC-MAC-PAD & Dummy MAC checking & 2013/05/19 &  2013-1619 & 4.3 &4.9 &2.9 & Dummy data padding \\
  \hline
  6 & 2013/07/25 & P1.4.14,  L1.5.3 & Modular exponentiation & Square-and-Multiply & 2013/08/19 &  2013-4242 &1.9  &3.4 &2.9 & Square-and-Multiply-always \\
  \hline
  7 & 2013/12/16 & L1.6.0 & Modular exponentiation & Square-and-Multiply &  2013/08/19 & 2013-4242 & 1.9  &3.4 &2.9 & Square-and-Multiply-always \\
  \hline
  8 & 2013/12/18 & P1.4.16 & Modular multiplication & \makecell[l]{Basic and Karatsuba \\multiplication} & 2013/12/20 &  2013-4576 & 2.1 &3.9 &2.9 & Exponentiation blinding \\
  \hline
  9 & 2014/08/07 & L1.5.4 & Modular multiplication & \makecell[l]{Basic and Karatsuba \\multiplication} & 2014/10/19 &  2014-5270 &  2.1 &3.9 &2.9 & Exponentiation blinding \\
  \hline
  \multirow{2}{*}{10} & \multirow{2}{*}{2015/02/27} & \multirow{2}{*}{P1.4.19, L1.6.3} & Modular multiplication & \makecell[l]{Basic and Karatsuba \\multiplication} & 2015/02/27 &  2014-3591 & 1.9 & 3.4&2.9 & ElGamal Blinding \\
  \cline{4-11}
  & & & Modular exponentiation & Sliding window & 2015/02/27 &  2015-0837 & 4.3 &8.6 &2.9 & Remove control flow \\
  \hline
  11 & 2016/02/09 & L1.6.5 & \multirow{2}{*}{Scalar multiplication} & \multirow{2}{*}{Sliding window} & \multirow{2}{*}{2016/04/19} &  \multirow{2}{*}{ 2015-7511} & \multirow{2}{*}{1.9} &\multirow{2}{*}{3.4} &\multirow{2}{*}{2.9} & \multirow{2}{*}{Double-and-Add-always} \\
  \cline{1-3}
  \multirow{2}{*}{12} & \multirow{2}{*}{2016/02/18} & \multirow{2}{*}{L1.5.5} & & & &  &  & & & \\
  \cline{4-11}
  & & & \multirow{2}{*}{Modular multiplication} & \multirow{2}{*}{\makecell[l]{Basic and Karatsuba \\multiplication}} & \multirow{2}{*}{2015/02/27} &  \multirow{2}{*}{ 2014-3591} &  \multirow{2}{*}{1.9} &\multirow{2}{*}{3.4} &\multirow{2}{*}{2.9} & \multirow{2}{*}{ElGamal Blinding} \\
  \cline{1-3}
  \multirow{3}{*}{13} & \multirow{3}{*}{2016/04/15} & \multirow{3}{*}{L1.7.0} & & & &  & & & \\
  \cline{4-11}
  & & & Modular exponentiation & Sliding window & 2015/02/27 &  2015-0837 & 4.3 & 8.6& 2.9& Remove control flow \\
  \cline{4-11}
  & & & Scalar multiplication & Sliding window & 2016/04/19 &  2015-7511 & 1.9 &3.4 & 2.9& Double-and-Add-always \\
  \hline
  14 & 2017/06/29 & L1.7.8 & \multirow{3}{*}{Modular exponentiation} & \multirow{3}{*}{Sliding window} & \multirow{3}{*}{2018/07/26} & \multirow{3}{*}{ 2017-7526} &  \multirow{3}{*}{4.3} &\multirow{3}{*}{8.6} &\multirow{3}{*}{2.9} & \multirow{3}{*}{RSA blinding} \\
  \cline{1-3}
  15 & 2017/07/18 & L1.8.0 & & & &  &  & & & \\
  \cline{1-3}
  16 & 2017/07/19 & P1.4.22 & & & &  &  & & & \\
  \hline
  17 & 2017/08/27 & L1.7.9, L1.8.1 & Scalar multiplication & \makecell[l]{Branchless montgomery \\ladder} & 2017/08/29 &  2017-0379 & 5 &10 & 2.9& Input validation \\
  \hline
  18 & 2018/06/13 & L1.7.10, L1.8.3 & Modulo & Early exit & 2018/06/13 &  2018-0495 & 1.9 &3.4 &2.9 & ECDSA blinding \\
  \hline
  \multirow{3}{*}{19} & \multirow{3}{*}{2018/07/16} & \multirow{3}{*}{\makecell[l]{T3.3.30, T3.5.19,\\T3.6.3}} & \multirow{3}{*}{CBC-MAC-PAD} & \multirow{3}{*}{Pseudo constant time} &  2018/08/22 &  2018-10844 &  1.9 &3.4 &2.9 & \multirow{3}{*}{\makecell[l]{New variant of pseudo \\constant time (Not fully \\mitigated)}} \\
  \cline{6-10}
  & & & & & 2018/08/22 &  2018-10845 &  1.9 &3.4 &2.9 & \\
  \cline{6-10}
  & & & & & 2018/08/22 &  2018-10846 &  1.9 &3.4 &2.9 & \\
  \hline
  20 & 2018/12/01 & T3.6.5 & RSA-PAD & Pseudo constant time & 2018/12/03 &  2018-16868 &  1.9 &3.4 &2.9 & Hide access pattern \& timing \\
  \hline
\end{tabular}
\end{threeparttable}}
\caption{Vulnerabilities in GNU crypto (For the version column, P: GnuPG; L: Libgcrypt; T: GnuTLS. For CVSS column, B: Base; E: Exploitability; I: Impact)}
\vspace{-15pt}
\label{table:gnupg-history_crypto}
\end{table*}

\subsection{Vulnerability Severity}
\label{sec:vul-severe}


We examine the severity and practicality of side-channel attacks as well as the attention
developers paid to them. We establish the measurements for these threats and compare them
with other vulnerability categories. We adopt the Common Vulnerability Scoring System (CVSS)\footnote{The latest CVSS version 
is v3.0. We adopt CVSS v2.0, as old vulnerabilities were not 
assigned CVSS v3.0 scores.} to assess each CVE. 
CVSS is a widely-accepted industry standard to identify and assess vulnerabilities across diverse platforms. It contains three metric groups: \emph{Base}, \emph{Temporal}, and \emph{Environmental}, each consisting of a set of metrics. 
We consider the \emph{Base} score that well represents the inherent quality of a 
vulnerability. It comprises two sub-scores, \emph{Exploitability} that defines 
the difficulty to attack the software and \emph{Impact} that defines the 
level of damage to certain properties of the software under a successful
attack. The score ranges from 0 (least severe) to 10 (most severe). Detailed computation of those scores can be found in the National Vulnerability Database (NVD) website\footnote{\url{https://nvd.nist.gov/vuln-metrics/cvss/v2-calculator}}. The score of each side-channel vulnerability is collected from the NVD \footnote{\url{https://nvd.nist.gov/}} and CVE\footnote{\url{https://cve.mitre.org/}} websites. It is worth noting that CVSS is a general metric for characterizing software vulnerabilities. It does not contain evaluation criteria for microarchitectural threats, and may not comprehensively reveal the inherent features of microarchitectural attacks. However, it can reflect the attitude of the cryptographic community towards side-channel attacks in a practical way.

For OpenSSL and GNU Crypto, the top vulnerabilities are denial-of-service, arbitrary 
code execution, buffer overflow, and memory corruption.
Table \ref{table:compare-vul} compares the average scores and quantities of these vulnerability
categories\footnote{There are some mistakes in CVEs: (1) all 
side-channel vulnerabilities should only have partial confidentiality impact, while 
CVE-2003-0131, CVE-2013-1619 and CVE-2018-16868 were also assigned partial integrity or 
availability impact. (2) CVE-2018-10844, CVE-2018-10845 and CVE-2018-10846 should have local 
access vector, but they were assigned network access vector. We corrected them 
in our analysis.}. We observe that \emph{side-channel vulnerabilities are regarded less severe 
than other types} due to lower \emph{Exploitability} and \emph{Impact} sub-scores.
Side-channel attacks usually require stronger adversarial capabilities, in-depth knowledge about the
underlying platforms, and a large amount of attack sessions, but only cause partial
confidentiality breach as they leak (part of) keys or plaintexts. In contrast, other vulnerabilities
may enable less experienced attackers to exploit them for executing arbitrary code or disabling the services entirely.

\begin{table*}[h]
\begin{subtable}{0.47\linewidth}
\centering
\resizebox{\linewidth}{!}{
\begin{tabular}{l|ccc|c}
\hline
                    & \multicolumn{3}{c|}{CVSS}    & \multirow{2}{*}{Count} \\
                    \cline{2-4}
                    & Base & Exploit & Impact &                         \\
                    \hline
Denial of Service   & 5.74 & 9.46        & 4.33   & 157                     \\
\hline
Buffer Overflow     & 6.76 & 9.43        & 5.77   & 43                      \\
\hline
\textbf{Side channel} & \textbf{3.32} & \textbf{6.27}        & \textbf{2.90}    & \textbf{37}                      \\
\hline
Code Execution      & 7.98 & 9.08        & 7.74   & 35                      \\
\hline
Mem Corruption   & 6.56 & 9.72        & 5.29   & 20                      \\
\hline
\end{tabular}
}
\caption{Comparisons with different vulnerabilities}
\label{table:compare-vul}
\end{subtable}%
\hspace*{10pt}
\begin{subtable}{0.49\linewidth}
\centering
\resizebox{\linewidth}{!}{
\begin{tabular}{l|c|c|c|c}
\hline
\multirow{2}{*}{}           & \multicolumn{3}{c|}{CVSS}   & \multirow{2}{*}{Count} \\ \cline{2-4}
                            & Base & Exploit & Impact &                        \\ \hline
Asymmetric Crypto    & 2.86 & 5.48         & 2.90    & 18                     \\ \hline
Crypto Protocol      & 3.76 & 7.02        & 2.90    & 19                     \\ \hline\hline
\textbf{Microarchitecture}  & \textbf{3.04} & \textbf{5.87}        & \textbf{2.90}    & \textbf{16}                     \\ \hline
Physical & 1.97 & 3.57        & 2.90    & 3                      \\ \hline
Network            & 3.87 & 7.22        & 2.90    & 18                     \\ \hline
\end{tabular}
}
  \caption{Comparisons of side-channel vulnerabilities}
  \label{table:compare-sc}
  \end{subtable}
\caption{Severity and number of Software Vulnerabilities}
\vspace{-20pt}
\label{table:crypto-history}
\end{table*}


Next we break down and compare different types of side-channel vulnerabilities, as shown in Table \ref{table:compare-sc}. We first consider the two categories of operations: asymmetric ciphers
and protocol padding. We did not find any side-channel CVEs related to post-quantum cryptography as its development is still at an early stage. We also skip the vulnerabilities in symmetric ciphers, as there are only two reported CVEs (row 10 in Table \ref{table:openssl-history} and row 3 in Table \ref{table:gnupg-history_crypto}). This is because symmetric ciphers are simpler and thus less vulnerable than asymmetric ones, and the widely-adopted AES-NI instruction set extension can effectively mitigate existing vulnerabilities. 
We observe that \emph{vulnerabilities in protocol
padding are generally more severe than those in asymmetric ciphers due to higher Exploitability}.
The underlying reason is that \emph{Exploitability} is determined by the access vector: network vector and local vector are neck and neck for vulnerabilities in asymmetric 
ciphers, but the former dominates access vectors of padding oracle attacks, rendering 
them more exploitable. We also compare microarchitectural attacks with network and physical attacks. From Table \ref{table:compare-sc} we can observe that network attacks are most severe due to high \emph{Exploitability}. Scores of microarchitectural attacks are also higher than that of physical attacks, as some microarchitectural vulnerabilities can be exploited via remote timing measurement, while all physical attacks must be conducted on site.




\subsection{Vulnerability Response}
\label{sec:vul-resp}


We evaluate the responses to discovered side-channel vulnerabilities from application developers.

\bheading{Response speed.}
For each vulnerability, we measure the \emph{vulnerability window}, defined as the duration from the vulnerability publication date\footnote{A side-channel vulnerability may be published in different ways, including online archives, formal academic publications and the CVE system. We use the earliest of all such dates.} to the patch release date. If the patch release date is earlier than the vulnerability publication date, the vulnerability window is negative. Obviously a narrower vulnerability window (in case of positive) leads to fewer chances of exploit and less damage.

\begin{figure}[h]
  \centering
  \begin{subfigure}{0.32\linewidth}
    \centering
    \includegraphics[width=\linewidth]{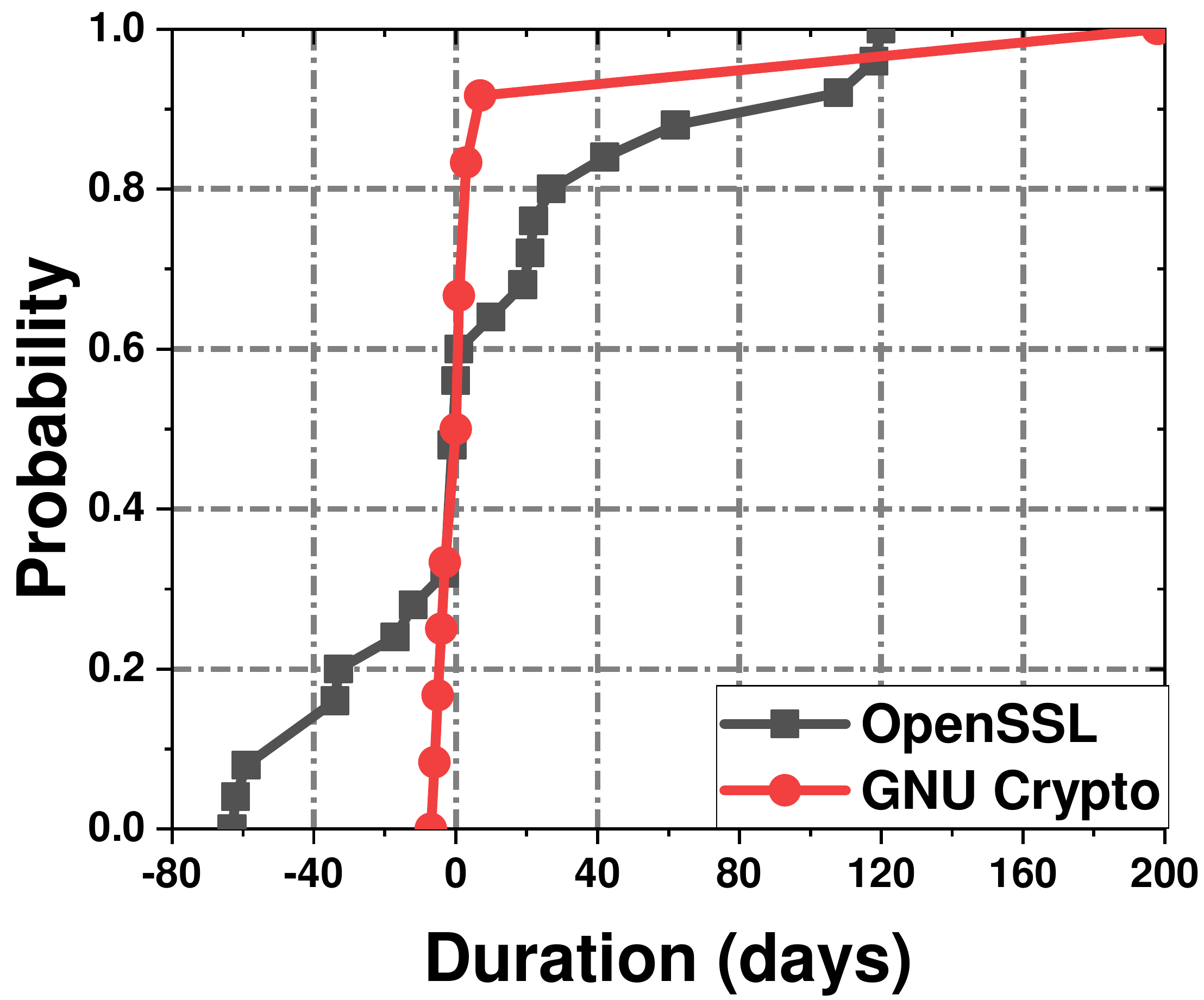}
    \caption{Different libraries}
    \label{fig:cve-3}
  \end{subfigure}%
  \begin{subfigure}{0.32\linewidth}
    \centering
    \includegraphics[width=\linewidth]{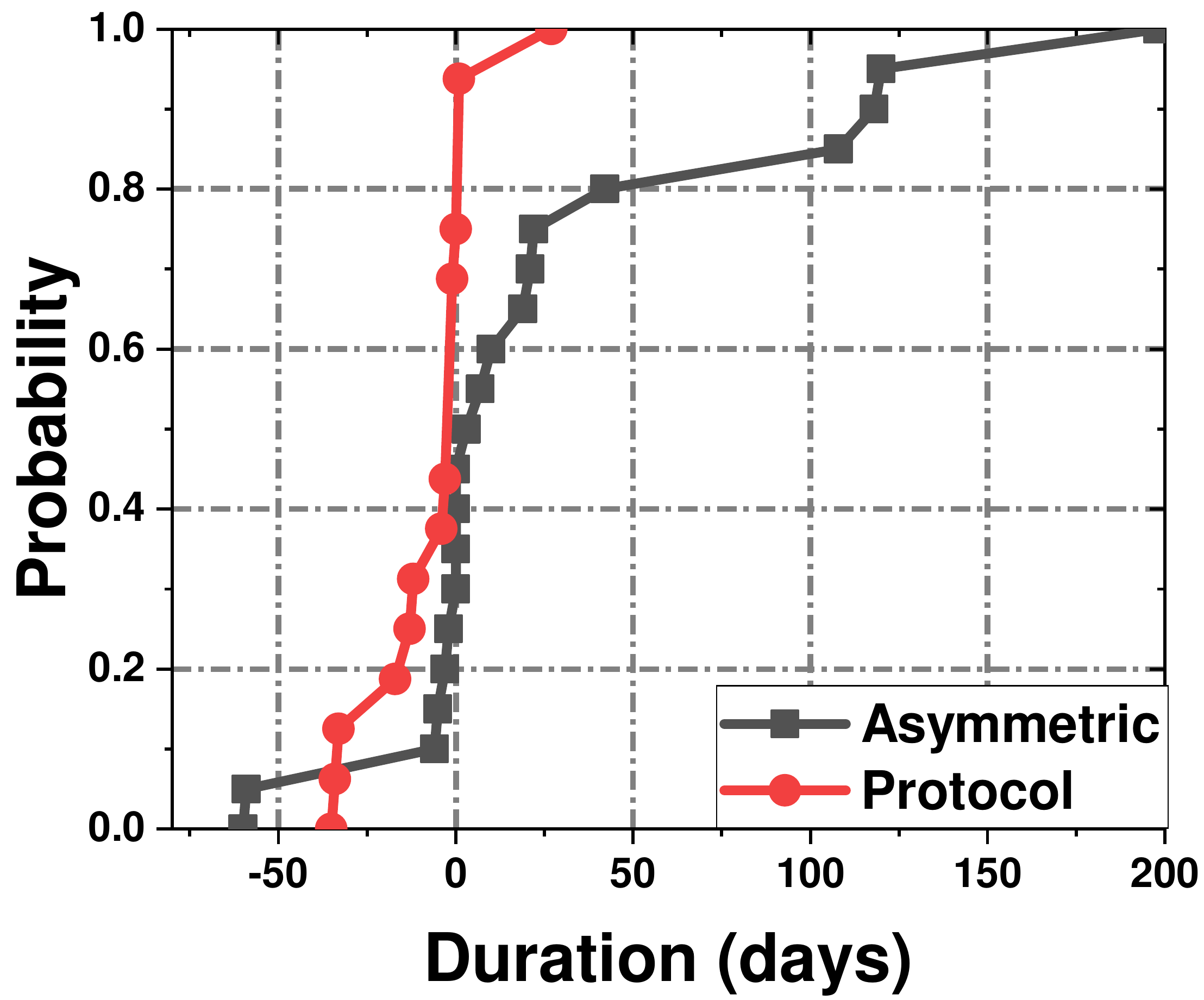}
    \caption{Different operations}
    \label{fig:cve-ops}
  \end{subfigure}
  \begin{subfigure}{0.32\linewidth}
    \centering
    \includegraphics[width=\linewidth]{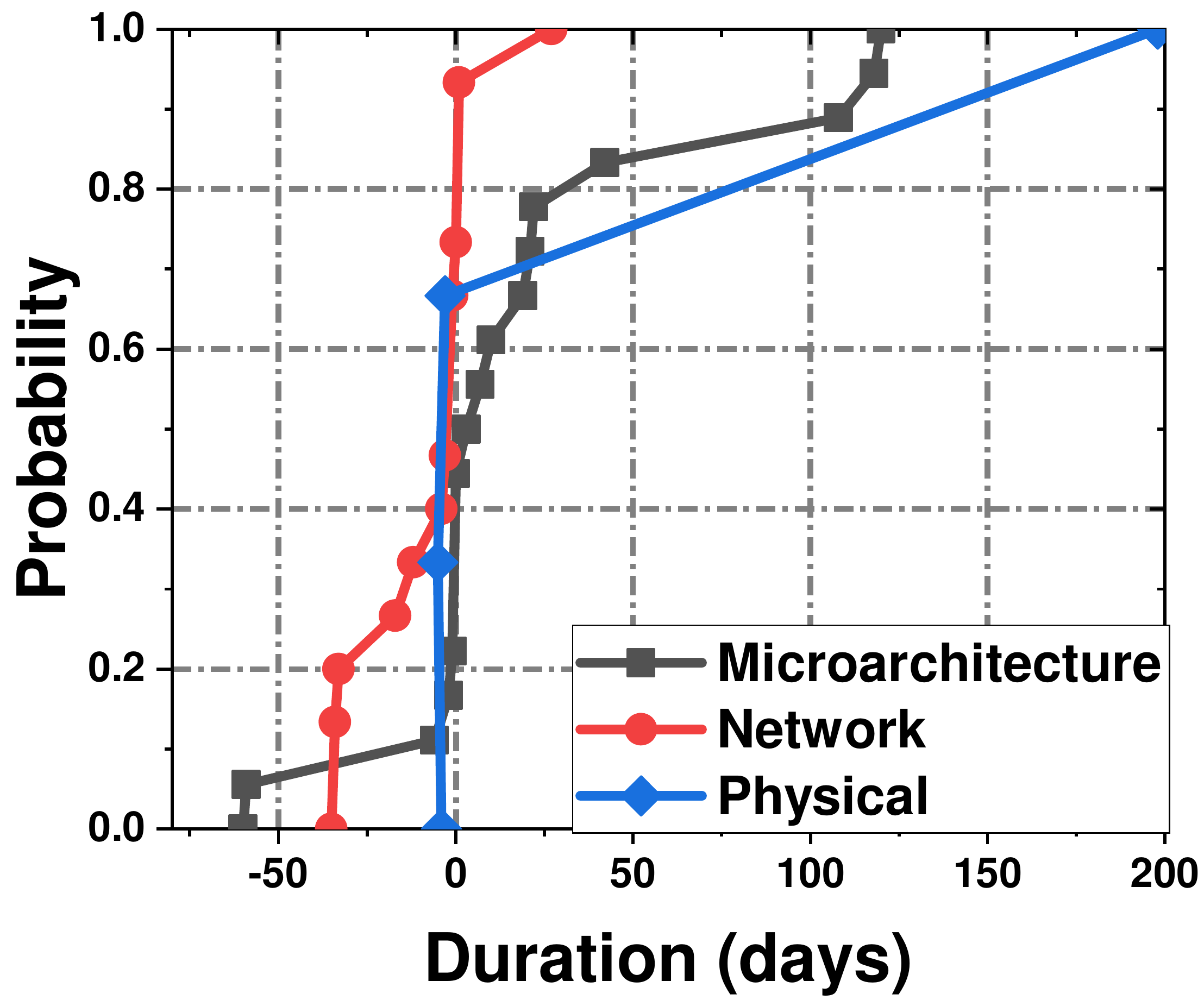}
    \caption{Different access vectors}
    \label{fig:cve-6}
  \end{subfigure}
  \vspace{-10pt}
  \caption{Cumulative distributions of vulnerability windows}
  \vspace{-10pt}
  \label{fig:cumu-dist}
\end{figure}

Figure \ref{fig:cve-3} shows the cumulative distribution of vulnerability windows for OpenSSL
and GNU Crypto. We can see that \emph{both libraries responded to side-channel vulnerabilities 
very actively}: 56\% and 50\% of the vulnerabilities were fixed by the two libraries respectively before
publication; more than 80\% of the vulnerabilities were fixed within one month of their
disclosure; each library has only one case that spanned more than 4 months, the longest 
being 198 days in GnuPG.
Figures \ref{fig:cve-ops} and \ref{fig:cve-6} compare the vulnerability windows of different operations and attack types, respectively. \emph{Although network attacks are more severe than local attacks, they were fixed at similar
speeds.}

\bheading{Response coverage.}
We found that \emph{the majority of discovered vulnerabilities were addressed in OpenSSL and GNU Crypto, except that
microarchitectural padding oracle vulnerabilities \cite{XiLiCh:17, RoPaSh:18} still exist in both
libraries at the time of writing}. One possible reason is that such host attacks require stronger adversarial
capabilities and can only work in limited contexts, and thus are less severe. 


\subsection{Cross-branch Patch Consistency}
\label{sec:cross-branch}


An application usually maintains different development branches concurrently. An adversary can still attack unpatched branches if the fix is not applied to all live
branches at the same time. For instance, OpenSSL replaced the vulnerable sliding window scalar multiplication with
branchless Montgomery ladder in version 1.1.0i on August 14, 2018, but not in the 1.0.2 branch.
This left a chance for port-based attacks \cite{AlBrHa:19} to work on the sliding window
implementation in OpenSSL 1.0.2, which urged the developers to apply the patch to 1.0.2q on
November 20, 2018.

For each vulnerability, we measure the \emph{cross-branch vulnerability window}, defined as
the duration from the first patch release date to the date when all live branches are patched.
Table \ref{table:cross-library} shows the number of patches in different vulnerability windows
for both libraries. \emph{In most cases, a patch was applied to all live branches at the same time (0 days).
Some patches are however still missing in certain branches at the time of writing (never).}
For example, OpenSSL 1.0.1 introduced masked-window multiplication
and AES-NI support that were never ported to 0.9.8 and 1.0.0 branches before
their end of life. OpenSSL 1.0.2r includes a bug fix for protocol error handling, but it
is not applied to 1.1.0 and 1.1.1. Some new side-channel bug fixes,
not critical though, in OpenSSL 1.1.1 and 1.1.1b were not included in 1.0.2 and 1.1.0. For GNU Crypto,
CVE-2015-0837 was fixed in GnuPG 1.4.19 and Libgcrypt 1.6.3, but not in Libgcrypt 1.5.x. Fortunately
this branch has reached its end of life on December 31, 2016.

\begin{table*}[h]
\begin{subtable}{0.47\linewidth}
\centering
\resizebox{\linewidth}{!}{
\begin{tabular}{l|cccccc}
\hline
Duration (days) & 0  & 59 & 98 & 120 & 196 & never \\ \hline
Counts          & 15 & 2  & 1  & 1   & 1   & 6     \\ \hline
\end{tabular}
}
\caption{OpenSSL}
\label{table:cross-openssl}
\end{subtable}%
\hspace*{10pt}
\begin{subtable}{0.49\linewidth}
\centering
\resizebox{\linewidth}{!}{
\begin{tabular}{l|ccccccc}
\hline
Duration (days) & 0 & 9 & 13 & 20 & 232 & 356 & never \\ \hline
Counts          & 8 & 1 & 1  & 1  & 1   & 1   & 7     \\ \hline
\end{tabular}
}
  \caption{GNU Crypto}
  \label{table:cross-gnu}
  \end{subtable}
\caption{Number of patches for cross-branch windows}
\vspace{-20pt}
\label{table:cross-library}
\end{table*}

\subsection{Countermeasure Type}
\label{sec:count-type}

\begin{figure}[h]
  \centering
  \begin{subfigure}{0.4\linewidth}
    \centering
    \includegraphics[width=\linewidth]{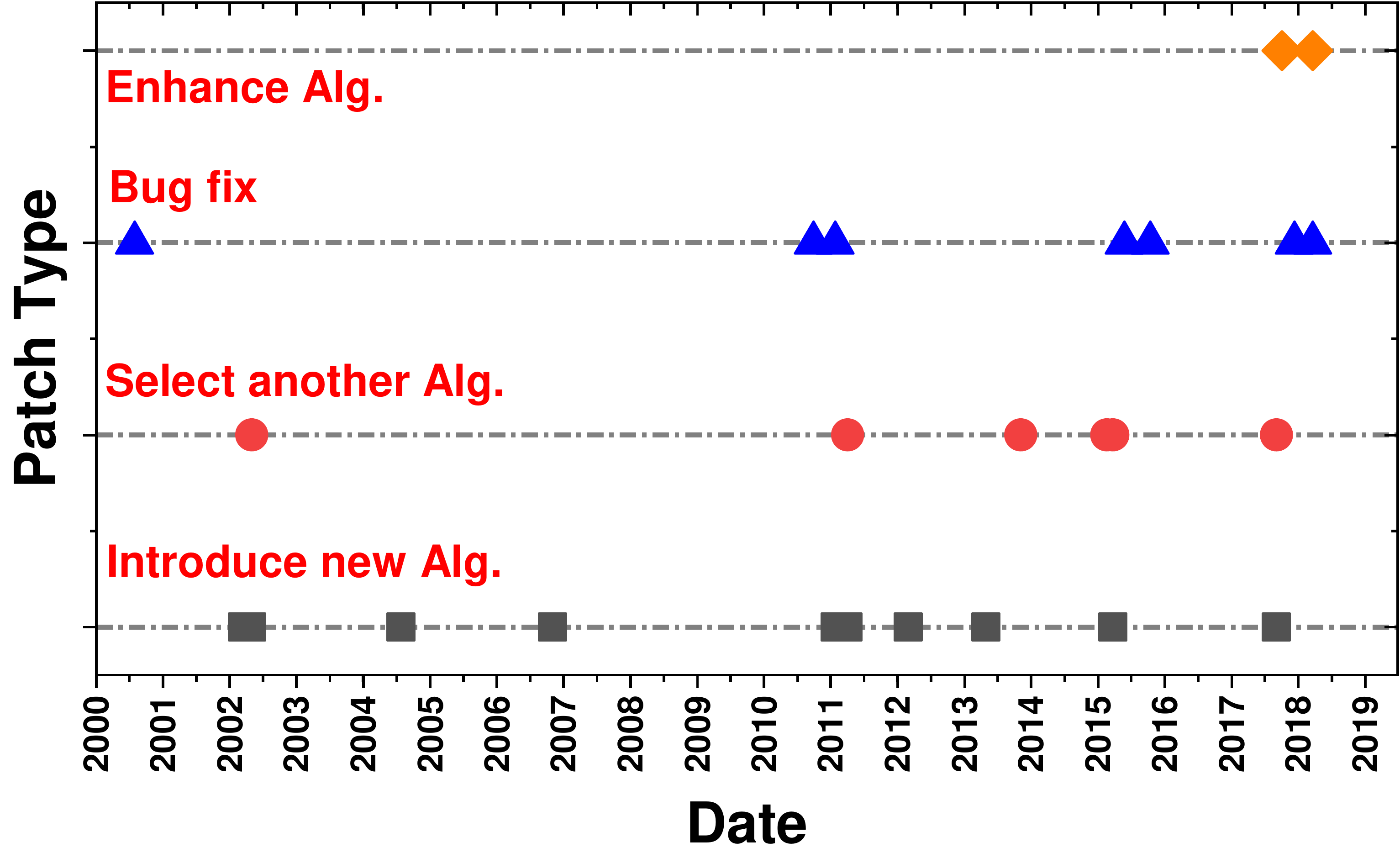}
    \caption{OpenSSL}
    \label{fig:cve-9}
  \end{subfigure}%
  \hspace*{10pt}
  \begin{subfigure}{0.4\linewidth}
    \centering
    \includegraphics[width=\linewidth]{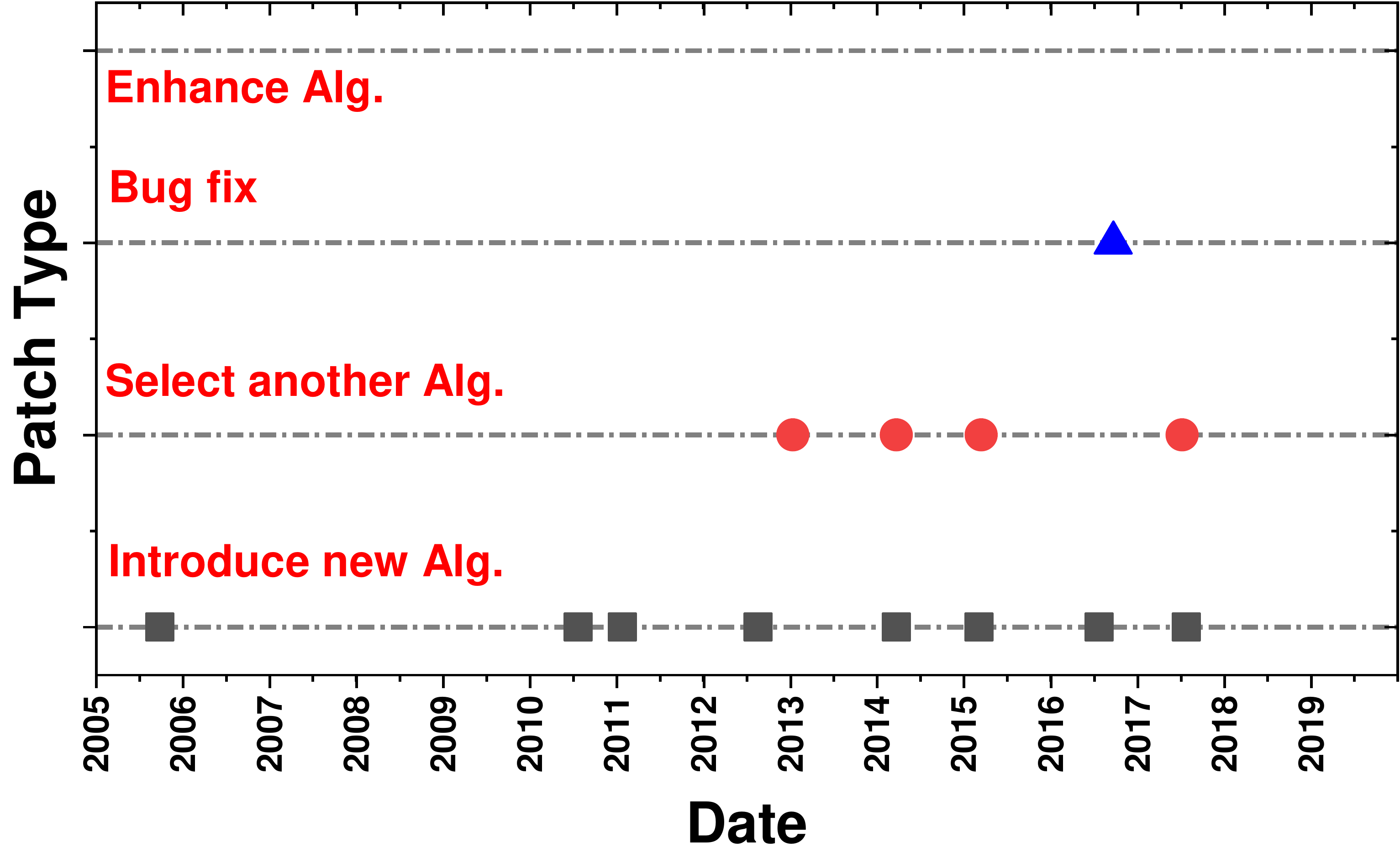}
    \caption{GNU crypto}
    \label{fig:cve-10}
  \end{subfigure}
  \vspace{-10pt}
  \caption{Countermeasure types of the two libraries}
  \vspace{-10pt}
  \label{fig:fix-type}
\end{figure}

We study four categories of countermeasures commonly adopted by cryptographic libraries to fix 
side-channel vulnerabilities: (1) introducing
brand new implementations; (2) selecting existing secure implementations; (3) fixing software bugs;
(4) enhancing robustness of existing implementations. Countermeasure classification for
OpenSSL and GNU Crypto is shown in Figure \ref{fig:fix-type}.

In the earlier days, the primary fix for side-channel vulnerabilities in OpenSSL was to introduce new
implementations. \emph{After many years' evolution, every cryptographic operation now has secure 
implementations, and brand new solutions become unnecessary.} Recent patches
were often minor bug fixes. Besides, previously developers only patched the 
code upon revelation of new issues. Now they proactively fortify the
implementation without the evidence of potential vulnerabilities. This definitely improves the
security of the library against side-channel attacks. 

GNU Crypto has fewer vulnerabilities and patches compared to OpenSSL, and prefers to use
traditional solutions for some common issues. For instance, to 
mitigate the vulnerability in sliding window scalar multiplication, OpenSSL
adopted a new solution, masked-window multiplication, while Libgcrypt regressed to
less efficient double-and-add-always. Besides, development of GNU Crypto is generally
several years behind that of OpenSSL. 

\subsection{Comparisons with Other Libraries}
\label{sec:cross-library-comp}

Finally we summarize side-channel CVEs in other cryptographic applications (Table \ref{table:cve-other}).

\begin{table*}[t]
\begin{subtable}{0.5\linewidth}
\centering
\resizebox{\linewidth}{!}{
\begin{tabular}{lllllll}
  \hline
\textbf{ID}  & \textbf{CVE Date}  & \textbf{C.} & \textbf{Application} & \textbf{Operations} & \textbf{CVE} & \textbf{Patch date} \\ \hline
1& 2001/06/27	&	$\blacksquare$	&	OpenSSH,	AppGate,	ssh-1	&	RSA-PAD	&	 2001-0361	&	2001/01/29	\\	\hline				
2& 2004/12/31	&	$\blacksquare$	&	MatrixSSL	&	Modular	Multiplication	&	 2004-2682	&	2004/06/01	\\	\hline					
3& 2009/08/31	&	$\blacksquare$	&	XySSL	&	RSA-PAD	&	 2008-7128	&	$\CIRCLE$	\\	\hline						
4& 2010/09/22	&	$\square$	&	Microsoft	IIS	&	Padding	oracle	attack	&	 2010-3332	&	2010/09/27	\\	\hline			
5& 2010/10/20	&	$\blacksquare$	&	Apache	MyFaces	&	Padding	oracle	attack	&	 2010-2057	&	2010/06/10	\\	\hline			
6& 2010/10/20	&	$\blacksquare$	&	Oracle	Mojarra	&	Padding	oracle	attack	&	 2010-4007	&	2010/06/10	\\	\hline			
7& 2013/02/08	&	$\blacksquare$	&	Rack	&	HMAC	comparison	&	 2013-0263	&	2013/02/07	\\	\hline					
8& 2013/02/08	&	$\blacksquare$	&	Mozilla	NSS	&	MAC-CBC-PAD	&	 2013-1620	&	2013/02/14	\\	\hline					
9& 2013/02/08	&	$\blacksquare$	&	wolfSSL	CyaSSL	&	MAC-CBC-PAD	&	 2013-1623	&	2013/02/05	\\	\hline					
10& 2013/02/08	&	$\blacksquare$	&	Bouncy	Castle	&	MAC-CBC-PAD	&	 2013-1624	&	2013/02/10	\\	\hline					
11& 2013/02/08	&	$\square$	&	Opera	&	MAC-CBC-PAD	&	 2013-1618	&	2013/02/13	\\	\hline						
12& 2013/06/21	&	$\square$	&	IBM	WebSphere	Commerce	&	Padding	oracle	attack	&	 2013-0523	&	$\Circle$	\\	\hline		
13& 2013/10/04	&	$\blacksquare$	&	PolarSSL	&	RSA-CRT	&	 2013-5915	&	2013/10/01	\\	\hline						
14& 2013/11/17	&	$\blacksquare$	&	OpenVPN	&	Padding	oracle	attack	&	 2013-2061	&	2013/03/19	\\	\hline				
15& 2014/08/16	&	$\square$	&	IBM	WebSphere	DataPower	&	-	&	 2014-0852	&	$\Circle$	\\	\hline				
16& 2014/12/09	&	$\square$	&	F5	BIG-IP	&	MAC-CBC-PAD	&	 2014-8730	&	$\Circle$	\\	\hline					
17& 2015/07/01	&	$\blacksquare$	&	Libcrypt++	&	Rabin-Williams DSA	&	 2015-2141	&	2015/11/20	\\	\hline						
18& 2015/08/02	&	$\square$	&	Siemens	RuggedCom	ROS	&	MAC-CBC-PAD	&	 2015-5537	&	$\Circle$	\\	\hline				
19& 2015/11/08	&	$\square$	&	IBM	DataPower	Gateways 	&	Padding	oracle	attack	&	 2015-7412	&	$\Circle$	\\	\hline		
20& 2016/04/07	&	$\blacksquare$	&	Erlang/OTP	&	MAC-CBC-PAD	&	 2015-2774	&	2015/03/26	\\	\hline						
21& 2016/04/12	&	$\square$	&	EMC	RSA	BSAFE	&	RSA-CRT	&	 2016-0887	&	$\Circle$	\\	\hline				
22& 2016/04/21	&	$\square$	&	CloudForms	Mgmt. Engine	&	Padding	oracle	attack	&	 2016-3702	&	$\CIRCLE$	\\	\hline		
23& 2016/05/13	&	$\blacksquare$	&	Botan	&	MAC-CBC-PAD	&	 2015-7827	&	2015/10/26	\\	\hline						
24& 2016/05/13	&	$\blacksquare$	&	Botan	&	Modular	inversion	&	 2016-2849	&	2016/04/28	\\	\hline					
25& 2016/07/26	&	$\blacksquare$	&	Cavium	SDK	&	RSA-CRT	&	 2015-5738	&	$\Circle$	\\	\hline					
26& 2016/09/03	&	$\blacksquare$	&	jose-php	&	HMAC	comparison	&	 2016-5429	&	2016/08/30	\\	\hline					
27& 2016/09/08	&	$\square$	&	HPE	Integrated	Lights-Out	3	&	Padding	oracle	attack	&	 2016-4379	&	2016/08/30	\\	\hline	
28& 2016/10/10	&	$\blacksquare$	&	Intel	IPP	&	RSA	&	 2016-8100	&	$\Circle$	\\	\hline					
29& 2016/10/28	&	$\blacksquare$	&	Botan	&	RSA-PAD	&	 2016-8871	&	2016/10/26	\\	\hline						
30& 2016/12/13	&	$\blacksquare$	&	wolfSSL	&	AES	T-table	lookup	&	 2016-7440	&	2016/09/26	\\	\hline				
31& 2016/12/15	&	$\square$	&	Open-Xchange	OX	Guard	&	Padding	oracle	attack	&	 2016-4028	&	2016/04/21	\\	\hline		
32& 2017/01/23	&	$\blacksquare$	&	Malcolm	Fell	jwt	&	Hash	comparison	&	 2016-7037	&	2016/09/05	\\	\hline			
33& 2017/02/03	&	$\square$	&	EMC	RSA	BSAFE	&	Padding	oracle	attack	&	 2016-8217	&	2017/01/20	\\	\hline		
34& 2017/02/13	&	$\blacksquare$	&	Crypto++	&	&	 2016-3995	&	2016/09/11	\\	\hline							
35& 2017/03/03	&	$\blacksquare$	&	MatrixSSL	&	RSA-CRT	&	 2016-6882	&	2016/11/25	\\	\hline						
36& 2017/03/03	&	$\blacksquare$	&	MatrixSSL	&	RSA-PAD	&	 2016-6883	&	2016/04/18	\\	\hline						
37& 2017/03/07	&	$\blacksquare$	&	Intel	QAT	&	RSA-CRT	&	 2017-5681	&	$\Circle$	\\	\hline					
38& 2017/03/23	&	$\square$	&	Cloudera	Navigator	&	MAC-CBC-PAD	&	 2015-4078	&	$\Circle$	\\	\hline					
39& 2017/04/10	&	$\blacksquare$	&	Botan	&	MAC-CBC-PAD	&	 2015-7824	&	2015/10/26	\\	\hline						
40& 2017/04/14	&	$\blacksquare$	&	Nettle	&	Modular	exponentiation	&	 2016-6489	&	2016/08/04	\\	\hline					
41& 2017/06/30	&	$\square$	&	OSCI-Transport	&	Padding	oracle	attack	&	 2017-10668	&	$\CIRCLE$	\\	\hline				
42& 2017/07/27	&	$\blacksquare$	&	Apache	HTTP	&	Padding	oracle	attack	&	 2016-0736	&	2016/10/20	\\	\hline			
43& 2017/08/02	&	$\square$	&	Citrix	NetScaler	&	MAC-CBC-PAD	&	 2015-3642	&	$\Circle$	\\	\hline					
44& 2017/08/10	&	$\blacksquare$	&	Apache	CXF	&	MAC	comparison	&	 2017-3156	&	$\Circle$	\\	\hline				
45& 2017/08/20	&	$\blacksquare$	&	Nimbus	JOSE+JWT	&	Padding	oracle	attack	&	 2017-12973	&	2017/06/02	\\	\hline			
46& 2017/09/25	&	$\blacksquare$	&	Botan	&	Modular	exponentiation	&	 2017-15533	&	2017/10/02	\\	\hline					
47& 2017/11/17	&	$\square$	&	F5	BIG-IP	&	RSA-PAD	&	 2017-6168	&	$\Circle$	\\	\hline					
48& 2017/12/12	&	$\blacksquare$	&	Erlang/OTP	&	RSA-PAD	&	 2017-1000385	&	2017/11/23	\\	\hline						
49& 2017/12/12	&	$\blacksquare$	&	Bouncy	Castle	&	RSA-PAD	&	 2017-13098	&	2017/12/28	\\	\hline					
50& 2017/12/12	&	$\blacksquare$	&	wolfSSL	&	RSA-PAD	&	 2017-13099	&	2017/10/31	\\	\hline						
51& 2017/12/13	&	$\square$	&	Citrix	NetScaler	&	RSA-PAD	&	 2017-17382	&	$\Circle$	\\	\hline					
52& 2017/12/13	&	$\square$	&	Radware	Alteon	&	RSA-PAD	&	 2017-17427	&	$\Circle$	\\	\hline					
53& 2017/12/15	&	$\square$	&	Cisco	ASA	&	RSA-PAD	&	 2017-12373	&	2018/01/05	\\	\hline					
54& 2017/12/28	&	$\blacksquare$	&	Intel	IPP	&	&	 2018-3691	&	2018/05/22	\\	\hline						
55& 2018/01/02	&	$\blacksquare$	&	Linaro	OP-TEE	&	RSA	Montgomery	&	 2017-1000413	&	2017/07/07	\\	\hline				

  \end{tabular}
}
\end{subtable}%
\hspace*{3pt}
\begin{subtable}{0.5\linewidth}
\centering
\resizebox{\linewidth}{!}{
\begin{tabular}{lllllll}
  \hline
\textbf{ID}  & \textbf{CVE Date}  & \textbf{C.} & \textbf{Application} & \textbf{Operations} & \textbf{CVE} & \textbf{Patch date} \\ \hline
56& 2018/01/10	&	$\square$	&	Palo	Alto	Networks	PAN-OS	&	RSA-PAD	&	 2017-17841	&	$\Circle$	\\	\hline			
57& 2018/02/05	&	$\square$	&	Cavium	Nitrox	and	TurboSSL	&	RSA-PAD	&	 2017-17428	&	$\Circle$	\\	\hline			
58& 2018/02/07	&	$\square$	&	IBM	GSKit	&	Padding	oracle	attack	&	 2018-1388	&	$\Circle$	\\	\hline			
59& 2018/02/26	&	$\square$	&	Unisys	ClearPath	MCP	&	RSA-PAD	&	 2018-5762	&	$\Circle$	\\	\hline				
60& 2018/05/17	&	$\square$	&	Symantec	SSL	Visibility	&	RSA-PAD	&	 2017-15533	&	2018/01/12	\\	\hline				
61& 2018/05/17	&	$\square$	&	Symantec	IntelligenceCenter	&	RSA-PAD	&	 2017-18268	&	$\Circle$	\\	\hline					
62& 2018/06/04	&	$\blacksquare$	&	Bouncy	Castle	&	DSA	&	 2016-1000341	&	2016/12/23	\\	\hline					
63& 2018/06/04	&	$\blacksquare$	&	Bouncy	Castle	&	Padding	oracle	attack	&	 2016-1000345	&	2016/12/23	\\	\hline			
64& 2018/06/14	&	$\blacksquare$	&	Mozilla	NSS	&	Padding	oracle	attack	&	 2018-12404	&	2018/12/07	\\	\hline			
65& 2018/06/14	&	$\blacksquare$	&	LibreSSL	&	Modulo	primitive	&	 2018-12434	&	2018/06/13	\\	\hline					
66& 2018/06/14	&	$\blacksquare$	&	Botan	&	Modulo	primitive	&	 2018-12435	&	2018/07/02	\\	\hline					
67& 2018/06/14	&	$\blacksquare$	&	wolfssl	&	Modulo	primitive	&	 2018-12436	&	2018/05/27	\\	\hline					
68& 2018/06/14	&	$\blacksquare$	&	LibTomCrypt	&	Modulo	primitive	&	 2018-12437	&	$\CIRCLE$	\\	\hline					
69& 2018/06/14	&	$\blacksquare$	&	LibSunEC	&	Modulo	primitive	&	 2018-12438	&	$\CIRCLE$	\\	\hline					
70& 2018/06/14	&	$\blacksquare$	&	MatrixSSL	&	Modulo	primitive	&	 2018-12439	&	2018/09/13	\\	\hline					
71& 2018/06/14	&	$\blacksquare$	&	BoringSSL	&	Modulo	primitive	&	 2018-12440	&	$\CIRCLE$	\\	\hline					
72& 2018/07/28	&	$\blacksquare$	&	ARM	mbed	TLS	&	MAC-CBC-PAD	&	 2018-0497	&	2018/07/24	\\	\hline				
73& 2018/07/28	&	$\blacksquare$	&	ARM	mbed	TLS	&	MAC-CBC-PAD	&	 2018-0498	&	2018/07/24	\\	\hline				
74& 2018/07/31	&	$\square$	&	Huawei	products	&	RSA-PAD	&	 2017-17174	&	$\Circle$	\\	\hline					
75& 2018/08/15	&	$\square$	&	Clavister	cOS	Core	&	RSA-PAD	&	 2018-8753	&	$\Circle$	\\	\hline				
76& 2018/08/15	&	$\square$	&	ZyXEL	ZyWALL/USG	&	RSA-PAD	&	 2018-9129	&	$\Circle$	\\	\hline					
77& 2018/08/21	&	$\square$	&	Huawei	products	&	RSA-PAD	&	 2017-17305	&	$\Circle$	\\	\hline					
78& 2018/08/23	&	$\blacksquare$	&	Cloud	Foundry	Bits	Service	&	&	 2018-15800	&	2018/12/05	\\	\hline				
79& 2018/08/31	&	$\square$	&	RSA	BSAFE	Edition	Suite	&	RSA-PAD	&	 2018-11057	&	$\Circle$	\\	\hline		
80& 2018/09/11	&	$\square$	&	RSA	BSAFE	SSL-J	&	RSA-PAD	&	 2018-11069	&	$\Circle$	\\	\hline				
81& 2018/09/11	&	$\square$	&	RSA	BSAFE	Crypto-J	&	RSA-PAD	&	 2018-11070	&	$\Circle$	\\	\hline				
82& 2018/09/12	&	$\square$	&	Intel	AMT	&	RSA-PAD	&	 2018-3616	&	$\Circle$	\\	\hline					
83& 2018/09/21	&	$\blacksquare$	&	Apache	Mesos	&	HMAC	comparison	&	 2018-8023	&	2018/07/25	\\	\hline				
84& 2018/12/03	&	$\blacksquare$	&	nettle	&	RSA-PAD	&	 2018-16869	&	$\CIRCLE$	\\	\hline						
85& 2018/12/03	&	$\blacksquare$	&	wolfSSL	&	RSA-PAD	&	 2018-16870	&	2018/12/27	\\	\hline						
86& 2019/01/03	&	$\square$	&	RSA	BSAFE	Crypto-C	&	&	 2019-3731	&	2019/09/11	\\	\hline					
87& 2019/01/03	&	$\square$	&	RSA	BSAFE	Crypto-C	&	&	 2019-3732	&	2018/08/28	\\	\hline					
88& 2019/01/03	&	$\square$	&	RSA	BSAFE	Crypto-J	&	&	 2019-3739	&	2019/08/11	\\	\hline					
89& 2019/01/03	&	$\square$	&	RSA	BSAFE	Crypto-J	&	&	 2019-3740	&	2019/08/11	\\	\hline					
90& 2019/02/22	&	$\square$	&	Citrix	NetScaler	Gateway	&	Padding	oracle	attack	&	 2019-6485	&	$\Circle$	\\	\hline		
91& 2019/03/01	&	$\blacksquare$	&	hostapd,	wpa\_supplicant	&	&	 2019-9494	&	2019/04/21	\\	\hline						
92& 2019/03/01	&	$\blacksquare$	&	hostapd,	wpa\_supplicant	&	&	 2019-9495	&	2019/04/21	\\	\hline						
93& 2019/03/08	&	$\blacksquare$	&	Botan	&	Scalar	multiplication	&	 2018-20187	&	2018/10/01	\\	\hline					
94& 2019/03/26	&	$\blacksquare$	&	Apache	Tapestry	&	HMAC	comparison	&	 2019-10071	&	2019/09/07	\\	\hline				
95& 2019/04/03	&	$\blacksquare$	&	elliptic	&	Scalar	multiplication	&	 2019-10764	&	$\CIRCLE$ \\	\hline						
96& 2019/04/11	&	$\square$	&	Intel	PTT,	TXE,	SPS	&	&	 2019-11090	&	$\CIRCLE$ \\	\hline			
97& 2019/07/07	&	$\blacksquare$	&	hostapd,	wpa\_supplicant	&	&	 2019-13377	&	2019/08/07	\\	\hline						
98& 2019/07/17	&	$\blacksquare$	&	wolfSSL,	wolfCrypt	&	Scalar	multiplication	&	 2019-13628	&	2019/07/22	\\	\hline				
99& 2019/07/17	&	$\blacksquare$	&	MatrixSSL	&	Scalar	multiplication	&	 2019-13629	& $\CIRCLE$	\\	\hline						
100& 2019/07/27	&	$\blacksquare$	&	Crypto++	&	Scalar	multiplication	&	 2019-14318	&	2019/07/29	\\	\hline					
101& 2019/08/27	&	$\square$	&	Fortinet	FortiOS	&	&	 2019-15703	&	2019/12/19	\\	\hline						
102& 2019/09/26	&	$\blacksquare$	&	ARM	mbed	TLS	\&	Crypto	&	&	 2019-16910	&	2019/09/06	\\	\hline			
103& 2019/10/21	&	$\blacksquare$	&	ARM	mbed	TLS	\&	Crypto	&	&	 2019-18222	&	2020/02/21	\\	\hline			
104& 2019/12/05	&	$\blacksquare$	&	Jenkins	&	TCP	secret	comparison	&	 2020-2101	&	2020/01/29	\\	\hline				
105& 2019/12/05	&	$\blacksquare$	&	Jenkins	&	HMAC	comparison	&	 2020-2102	&	2020/01/29	\\	\hline					
106& 2019/12/24	&	$\blacksquare$	&	wolfSSL	&	Modulo	multiplication	&	 2019-19960	&	2019/12/20	\\	\hline					
107& 2019/12/24	&	$\blacksquare$	&	wolfSSL	&	Modulo	inversion	&	 2019-19963	&	2019/12/20	\\	\hline					
108& 2020/01/22	&	$\blacksquare$	&	Parity	libsecp256k1-rs	&	Scalar	overflow	check	&	 2019-20399	&	2019/10/02	\\	\hline			
109& 2020/03/24	&	$\blacksquare$	&	ARM	mbed	TLS	&	Modular	inversion	&	 2020-10932	&	2020/4/14	\\	\hline			
110& 2020/04/12	&	$\blacksquare$	&	wolfSSL	&	Modulo	multiplication	&	 2020-11713	&	2020/04/22	\\	\hline					

  \end{tabular}
}
  \end{subtable}
\caption{Side-channel vulnerabilities in other applications. ($\blacksquare$: open-source, $\square$: closed-source; $\CIRCLE$: Whether this vulnerability is addressed is not revealed. $\Circle$: This vulnerability is addressed, but the date is not revealed.)}
\vspace{-20pt}
\label{table:cve-other}
\end{table*}

\bheading{Vulnerable Categories.}
Table \ref{table:cve-11} shows the breakdown of vulnerabilities in different categories.
We observe that \emph{vulnerabilities exist widely in many applications, besides OpenSSL and GNU Crypto}. We believe a lot of 
unrevealed vulnerabilities still exist in various applications, for two reasons.

First, researchers tend to study common cryptographic libraries, encouraging their developers
to continuously improve the code. Other less evaluated 
applications may still contain out-of-date vulnerabilities, but their developers are unaware or
ignorant of them. For instance, RSA padding oracle attack was proposed 20 
years ago and has been mitigated in common libraries like OpenSSL and GnuTLS, but it
still exists in about one third of top 100 Internet domains including
Facebook and PayPal, as well as widely used products from IBM, Cisco and so on \cite{HaJuCr:18}. 

Second, microarchitectural attacks usually require the source code to be available, prohibiting
researchers from discovering vulnerabilities in closed-source applications. 
For instance, Table \ref{table:cve-11} shows that the majority of vulnerabilities found in 
closed-source applications are padding oracles via remote timing or message side channels,
likely because no source code is needed to experiment with these attacks.
We do not know if they also suffer from padding oracle attacks via microarchitectural side
channels, as current studies \cite{IrLnEi:15, XiLiCh:17, RoPaSh:18, RoGiGe:19} evaluated
them only on open-source libraries. It is also unclear if they possess vulnerabilities
related to asymmetric ciphers for the similar reason.


\bheading{Response speed and coverage.}
Figure \ref{fig:cve-12} compares the response speeds of different applications.
Interestingly, \emph{they all responded to the vulnerabilities very fast.} Most vulnerabilities
were published only after the release of corresponding patches, leaving 
no vulnerability windows to exploit.

Regarding the coverage, \emph{most discovered vulnerabilities were addressed, with a few exceptions}
(annotated with $\CIRCLE$ in Table \ref{table:cve-other})
where too little public information is available. For these cases, we are unable to ascertain
whether these issues were solved or not. 

\begin{center}
\begin{tabularx}{\textwidth}{*{2}{>{\centering\arraybackslash}X}}
   \centering
    \resizebox{1.1\linewidth}{!}{
    \begin{tabular}{ll|ccc|c}
\hline
\multicolumn{2}{l|}{\multirow{2}{*}{}}     & \multicolumn{3}{c|}{CVSS}               & \multirow{2}{*}{Count} \\ \cline{3-5}
\multicolumn{2}{l|}{}                      & Base        & Exploit     & Impact      &                        \\ \hline\hline
\multirow{2}{*}{OpenSSL}       & Asymmetric & 3.14        & 6.09        & 2.90         & 10                     \\ \cline{2-6} 
                               & Protocol   & 4.25        & 8.46        & 2.90         & 13                     \\ \hline\hline
\multirow{2}{*}{GNU Crypto}    & Asymmetric & 2.58        & 4.88        & 2.90         & 9                      \\ \cline{2-6} 
                               & Protocol   & 2.70         & 3.90         & 2.90         & 6                      \\ \hline\hline
\multirow{3}{*}{Open-source}   & Asymmetric & 3.89        & 7.10       & 3.27         & 31                     \\ \cline{2-6} 
                               & Protocol   & 4.25 & 7.86 & 3.30 & 24                     \\ \cline{2-6} 
                               & Other      & 3.69         & 6.89           & 3.13         & 15                      \\ \hline\hline
\multirow{3}{*}{Closed-source} & Asymmetric & 2.60         & 4.90         & 2.90         & 1                      \\ \cline{2-6} 
                               & Protocol   & 4.36      & 8.43     & 3.09      & 32                     \\ \cline{2-6} 
                               & Other      & 4.28         & 8.47         & 2.90         & 7                      \\ \hline
\end{tabular}}
    \captionof{table}{Severity comparisons}
    \label{table:cve-11}
& 
    \includegraphics[width=0.8\linewidth,valign=m]{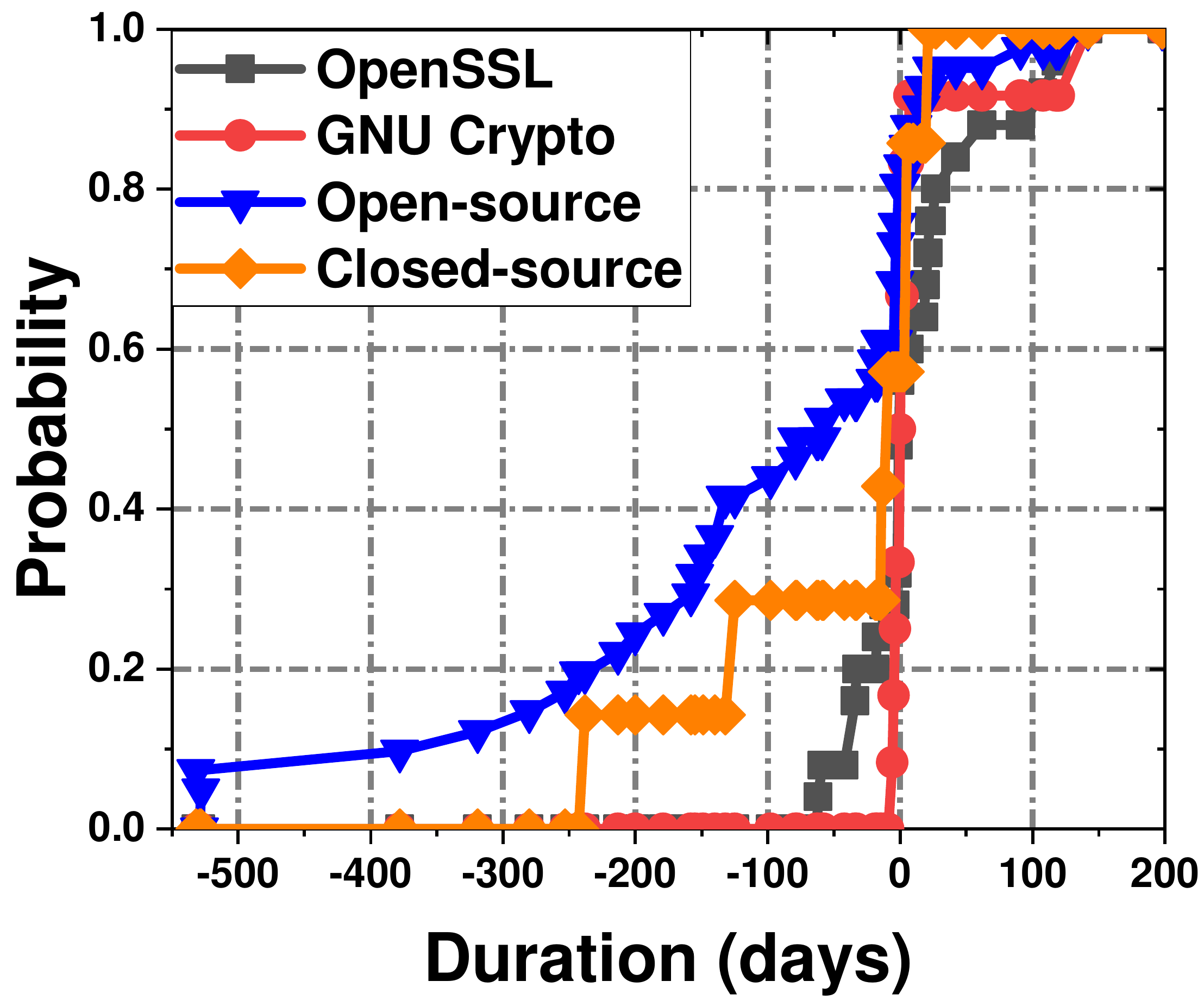} 
    \vspace{-10pt}
    \captionof{figure}{Response speed}
    \label{fig:cve-12}
\end{tabularx}
 \end{center}

\section{Conclusion}
\label{sec:conclu}
Microarchitectural side-channel attacks against cryptographic implementations have been an enduring
topic over the past 20 years. Many vulnerabilities have been discovered from previous cryptographic
implementations, but unknown ones likely still exist in today's implementations.
The good news is that the community resolved these vulnerabilities very actively,
and hence large-scale side-channel attacks causing severe real-world damages
have not happened so far. Besides, years of efforts have fortified common cryptographic
libraries and applications against side-channel attacks, and recently discovered
vulnerabilities were less significant or surprising.

Looking ahead, we expect continuous arms race between side-channel attacks and defenses.
We encourage researchers to discover new vulnerabilities and attacks,
evaluate them on a wider range of applications, and develop novel countermeasures for them.

\bibliographystyle{ACM-Reference-Format}
\bibliography{ref}


\begin{thebibliography}{243}


\ifx \showCODEN    \undefined \def \showCODEN     #1{\unskip}     \fi
\ifx \showDOI      \undefined \def \showDOI       #1{#1}\fi
\ifx \showISBNx    \undefined \def \showISBNx     #1{\unskip}     \fi
\ifx \showISBNxiii \undefined \def \showISBNxiii  #1{\unskip}     \fi
\ifx \showISSN     \undefined \def \showISSN      #1{\unskip}     \fi
\ifx \showLCCN     \undefined \def \showLCCN      #1{\unskip}     \fi
\ifx \shownote     \undefined \def \shownote      #1{#1}          \fi
\ifx \showarticletitle \undefined \def \showarticletitle #1{#1}   \fi
\ifx \showURL      \undefined \def \showURL       {\relax}        \fi
\providecommand\bibfield[2]{#2}
\providecommand\bibinfo[2]{#2}
\providecommand\natexlab[1]{#1}
\providecommand\showeprint[2][]{arXiv:#2}

\bibitem[\protect\citeauthoryear{Acii{\c{c}}mez}{Acii{\c{c}}mez}{2007}]%
        {aciiccmez2007yet}
\bibfield{author}{\bibinfo{person}{Onur Acii{\c{c}}mez}.}
  \bibinfo{year}{2007}\natexlab{}.
\newblock \showarticletitle{Yet another microarchitectural attack: exploiting
  I-cache}. In \bibinfo{booktitle}{\emph{ACM workshop on Computer security
  architecture}}.
\newblock


\bibitem[\protect\citeauthoryear{Ac{\i}i{\c{c}}mez, Brumley, and
  Grabher}{Ac{\i}i{\c{c}}mez et~al\mbox{.}}{2010}]%
        {aciiccmez2010new}
\bibfield{author}{\bibinfo{person}{Onur Ac{\i}i{\c{c}}mez},
  \bibinfo{person}{Billy~Bob Brumley}, {and} \bibinfo{person}{Philipp
  Grabher}.} \bibinfo{year}{2010}\natexlab{}.
\newblock \showarticletitle{New results on instruction cache attacks}. In
  \bibinfo{booktitle}{\emph{International Workshop on Cryptographic Hardware
  and Embedded Systems}}.
\newblock


\bibitem[\protect\citeauthoryear{Ac{\i}i{\c{c}}mez, Gueron, and
  Seifert}{Ac{\i}i{\c{c}}mez et~al\mbox{.}}{2007a}]%
        {AcGuSe:07}
\bibfield{author}{\bibinfo{person}{Onur Ac{\i}i{\c{c}}mez},
  \bibinfo{person}{Shay Gueron}, {and} \bibinfo{person}{Jean-Pierre Seifert}.}
  \bibinfo{year}{2007}\natexlab{a}.
\newblock \showarticletitle{New branch prediction vulnerabilities in OpenSSL
  and necessary software countermeasures}. In \bibinfo{booktitle}{\emph{IMA
  International Conference on Cryptography and Coding}}.
\newblock


\bibitem[\protect\citeauthoryear{Ac{\i}i{\c{c}}mez, Ko{\c{c}}, and
  Seifert}{Ac{\i}i{\c{c}}mez et~al\mbox{.}}{2007b}]%
        {aciiccmez2007predicting}
\bibfield{author}{\bibinfo{person}{Onur Ac{\i}i{\c{c}}mez},
  \bibinfo{person}{{\c{C}}etin~Kaya Ko{\c{c}}}, {and}
  \bibinfo{person}{Jean-Pierre Seifert}.} \bibinfo{year}{2007}\natexlab{b}.
\newblock \showarticletitle{Predicting secret keys via branch prediction}. In
  \bibinfo{booktitle}{\emph{Cryptographers’ Track at the RSA Conference}}.
\newblock


\bibitem[\protect\citeauthoryear{Ac{\i}i{\c{c}}mez, Schindler, and
  Ko{\c{c}}}{Ac{\i}i{\c{c}}mez et~al\mbox{.}}{2007c}]%
        {aciiccmez2007cache}
\bibfield{author}{\bibinfo{person}{Onur Ac{\i}i{\c{c}}mez},
  \bibinfo{person}{Werner Schindler}, {and} \bibinfo{person}{{\c{C}}etin~K
  Ko{\c{c}}}.} \bibinfo{year}{2007}\natexlab{c}.
\newblock \showarticletitle{Cache based remote timing attack on the AES}. In
  \bibinfo{booktitle}{\emph{Cryptographers’ track at the RSA conference}}.
\newblock


\bibitem[\protect\citeauthoryear{Aciicmez and Seifert}{Aciicmez and
  Seifert}{2007}]%
        {aciicmez2007cheap}
\bibfield{author}{\bibinfo{person}{Onur Aciicmez} {and}
  \bibinfo{person}{Jean-Pierre Seifert}.} \bibinfo{year}{2007}\natexlab{}.
\newblock \showarticletitle{Cheap hardware parallelism implies cheap security}.
  In \bibinfo{booktitle}{\emph{Workshop on Fault Diagnosis and Tolerance in
  Cryptography}}.
\newblock


\bibitem[\protect\citeauthoryear{Aldaya, Brumley, ul~Hassan, Garc{\'\i}a, and
  Tuveri}{Aldaya et~al\mbox{.}}{2019a}]%
        {AlBrHa:19}
\bibfield{author}{\bibinfo{person}{Alejandro~Cabrera Aldaya},
  \bibinfo{person}{Billy~Bob Brumley}, \bibinfo{person}{Sohaib ul Hassan},
  \bibinfo{person}{Cesar~Pereida Garc{\'\i}a}, {and} \bibinfo{person}{Nicola
  Tuveri}.} \bibinfo{year}{2019}\natexlab{a}.
\newblock \showarticletitle{Port contention for fun and profit}. In
  \bibinfo{booktitle}{\emph{IEEE Symposium on Security and Privacy}}.
\newblock


\bibitem[\protect\citeauthoryear{Aldaya, Garc{\'\i}a, Tapia, and
  Brumley}{Aldaya et~al\mbox{.}}{2019b}]%
        {AlGaTa:18}
\bibfield{author}{\bibinfo{person}{Alejandro~Cabrera Aldaya},
  \bibinfo{person}{Cesar~Pereida Garc{\'\i}a}, \bibinfo{person}{Luis
  Manuel~Alvarez Tapia}, {and} \bibinfo{person}{Billy~Bob Brumley}.}
  \bibinfo{year}{2019}\natexlab{b}.
\newblock \showarticletitle{Cache-timing attacks on RSA key generation}.
\newblock \bibinfo{journal}{\emph{IACR Transactions on Cryptographic Hardware
  and Embedded Systems}} (\bibinfo{year}{2019}).
\newblock


\bibitem[\protect\citeauthoryear{Allan, Brumley, Falkner, van~de Pol, and
  Yarom}{Allan et~al\mbox{.}}{2016}]%
        {AlBrFa:16}
\bibfield{author}{\bibinfo{person}{Thomas Allan}, \bibinfo{person}{Billy~Bob
  Brumley}, \bibinfo{person}{Katrina Falkner}, \bibinfo{person}{Joop van~de
  Pol}, {and} \bibinfo{person}{Yuval Yarom}.} \bibinfo{year}{2016}\natexlab{}.
\newblock \showarticletitle{Amplifying Side Channels Through Performance
  Degradation}. In \bibinfo{booktitle}{\emph{Annual Conference on Computer
  Security Applications}}.
\newblock


\bibitem[\protect\citeauthoryear{Almeida, Barbosa, Barthe, Dupressoir, and
  Emmi}{Almeida et~al\mbox{.}}{2016}]%
        {AlBaBa:16}
\bibfield{author}{\bibinfo{person}{Jos{\'e}~Bacelar Almeida},
  \bibinfo{person}{Manuel Barbosa}, \bibinfo{person}{Gilles Barthe},
  \bibinfo{person}{Fran{\c{c}}ois Dupressoir}, {and} \bibinfo{person}{Michael
  Emmi}.} \bibinfo{year}{2016}\natexlab{}.
\newblock \showarticletitle{Verifying Constant-Time Implementations}. In
  \bibinfo{booktitle}{\emph{USENIX Security Symposium}}.
\newblock


\bibitem[\protect\citeauthoryear{Alves}{Alves}{2004}]%
        {alves2004trustzone}
\bibfield{author}{\bibinfo{person}{Tiago Alves}.}
  \bibinfo{year}{2004}\natexlab{}.
\newblock \showarticletitle{Trustzone: Integrated hardware and software
  security}.
\newblock \bibinfo{journal}{\emph{White paper}} (\bibinfo{year}{2004}).
\newblock


\bibitem[\protect\citeauthoryear{Andrysco, Kohlbrenner, Mowery, Jhala, Lerner,
  and Shacham}{Andrysco et~al\mbox{.}}{2015}]%
        {andrysco2015subnormal}
\bibfield{author}{\bibinfo{person}{Marc Andrysco}, \bibinfo{person}{David
  Kohlbrenner}, \bibinfo{person}{Keaton Mowery}, \bibinfo{person}{Ranjit
  Jhala}, \bibinfo{person}{Sorin Lerner}, {and} \bibinfo{person}{Hovav
  Shacham}.} \bibinfo{year}{2015}\natexlab{}.
\newblock \showarticletitle{On subnormal floating point and abnormal timing}.
  In \bibinfo{booktitle}{\emph{IEEE Symposium on Security and Privacy}}.
\newblock


\bibitem[\protect\citeauthoryear{Avanzi}{Avanzi}{2005}]%
        {Av:05}
\bibfield{author}{\bibinfo{person}{Roberto~Maria Avanzi}.}
  \bibinfo{year}{2005}\natexlab{}.
\newblock \showarticletitle{Side Channel Attacks on Implementations of
  Curve-Based Cryptographic Primitives.}
\newblock \bibinfo{journal}{\emph{IACR Cryptology ePrint Archive}}
  (\bibinfo{year}{2005}).
\newblock


\bibitem[\protect\citeauthoryear{Barthe, Betarte, Campo, Luna, and
  Pichardie}{Barthe et~al\mbox{.}}{2014}]%
        {BaBeGu:14}
\bibfield{author}{\bibinfo{person}{Gilles Barthe}, \bibinfo{person}{Gustavo
  Betarte}, \bibinfo{person}{Juan Campo}, \bibinfo{person}{Carlos Luna}, {and}
  \bibinfo{person}{David Pichardie}.} \bibinfo{year}{2014}\natexlab{}.
\newblock \showarticletitle{System-level non-interference for constant-time
  cryptography}. In \bibinfo{booktitle}{\emph{ACM Conference on Computer and
  Communications Security}}.
\newblock


\bibitem[\protect\citeauthoryear{Benger, Van~de Pol, Smart, and Yarom}{Benger
  et~al\mbox{.}}{2014}]%
        {BeVaSm:14}
\bibfield{author}{\bibinfo{person}{Naomi Benger}, \bibinfo{person}{Joop Van~de
  Pol}, \bibinfo{person}{Nigel~P Smart}, {and} \bibinfo{person}{Yuval Yarom}.}
  \bibinfo{year}{2014}\natexlab{}.
\newblock \showarticletitle{"Ooh Aah... Just a Little Bit": A small amount of
  side channel can go a long way}. In \bibinfo{booktitle}{\emph{International
  Workshop on Cryptographic Hardware and Embedded Systems}}.
\newblock


\bibitem[\protect\citeauthoryear{Bernstein}{Bernstein}{2005}]%
        {Be:05}
\bibfield{author}{\bibinfo{person}{Daniel~J Bernstein}.}
  \bibinfo{year}{2005}\natexlab{}.
\newblock \showarticletitle{Cache-timing attacks on AES}.
\newblock \bibinfo{journal}{\emph{Technical Report}} (\bibinfo{year}{2005}).
\newblock


\bibitem[\protect\citeauthoryear{Bernstein, Breitner, Genkin, Bruinderink,
  Heninger, Lange, van Vredendaal, and Yarom}{Bernstein et~al\mbox{.}}{2017}]%
        {BeBrGe:17}
\bibfield{author}{\bibinfo{person}{Daniel~J Bernstein},
  \bibinfo{person}{Joachim Breitner}, \bibinfo{person}{Daniel Genkin},
  \bibinfo{person}{Leon~Groot Bruinderink}, \bibinfo{person}{Nadia Heninger},
  \bibinfo{person}{Tanja Lange}, \bibinfo{person}{Christine van Vredendaal},
  {and} \bibinfo{person}{Yuval Yarom}.} \bibinfo{year}{2017}\natexlab{}.
\newblock \showarticletitle{Sliding right into disaster: Left-to-right sliding
  windows leak}. In \bibinfo{booktitle}{\emph{International Conference on
  Cryptographic Hardware and Embedded Systems}}.
\newblock


\bibitem[\protect\citeauthoryear{Betz, Westhoff, and M{\"u}ller}{Betz
  et~al\mbox{.}}{2017}]%
        {BeWeMu:17}
\bibfield{author}{\bibinfo{person}{Johann Betz}, \bibinfo{person}{Dirk
  Westhoff}, {and} \bibinfo{person}{G{\"u}nter M{\"u}ller}.}
  \bibinfo{year}{2017}\natexlab{}.
\newblock \showarticletitle{Survey on covert channels in virtual machines and
  cloud computing}.
\newblock \bibinfo{journal}{\emph{Transactions on Emerging Telecommunications
  Technologies}} (\bibinfo{year}{2017}).
\newblock


\bibitem[\protect\citeauthoryear{Bhattacharya and Mukhopadhyay}{Bhattacharya
  and Mukhopadhyay}{2016}]%
        {bhattacharya2016curious}
\bibfield{author}{\bibinfo{person}{Sarani Bhattacharya} {and}
  \bibinfo{person}{Debdeep Mukhopadhyay}.} \bibinfo{year}{2016}\natexlab{}.
\newblock \showarticletitle{Curious case of rowhammer: flipping secret exponent
  bits using timing analysis}. In \bibinfo{booktitle}{\emph{International
  Conference on Cryptographic Hardware and Embedded Systems}}.
\newblock


\bibitem[\protect\citeauthoryear{Bhattacharya, Rebeiro, and
  Mukhopadhyay}{Bhattacharya et~al\mbox{.}}{2012}]%
        {bhattacharya2012hardware}
\bibfield{author}{\bibinfo{person}{Sarani Bhattacharya},
  \bibinfo{person}{Chester Rebeiro}, {and} \bibinfo{person}{Debdeep
  Mukhopadhyay}.} \bibinfo{year}{2012}\natexlab{}.
\newblock \showarticletitle{Hardware prefetchers leak: A revisit of SVF for
  cache-timing attacks}. In \bibinfo{booktitle}{\emph{Annual IEEE/ACM
  International Symposium on Microarchitecture Workshops}}.
\newblock


\bibitem[\protect\citeauthoryear{Biswas, Ghosal, and Nagaraja}{Biswas
  et~al\mbox{.}}{2017}]%
        {BiGhNa:17}
\bibfield{author}{\bibinfo{person}{Arnab~Kumar Biswas}, \bibinfo{person}{Dipak
  Ghosal}, {and} \bibinfo{person}{Shishir Nagaraja}.}
  \bibinfo{year}{2017}\natexlab{}.
\newblock \showarticletitle{A survey of timing channels and countermeasures}.
\newblock \bibinfo{journal}{\emph{ACM Computing Surveys (CSUR)}}
  (\bibinfo{year}{2017}).
\newblock


\bibitem[\protect\citeauthoryear{Blazy, Pichardie, and Trieu}{Blazy
  et~al\mbox{.}}{2017}]%
        {BaPiTr:17}
\bibfield{author}{\bibinfo{person}{Sandrine Blazy}, \bibinfo{person}{David
  Pichardie}, {and} \bibinfo{person}{Alix Trieu}.}
  \bibinfo{year}{2017}\natexlab{}.
\newblock \showarticletitle{Verifying constant-time implementations by abstract
  interpretation}. In \bibinfo{booktitle}{\emph{European Symposium on Research
  in Computer Security}}.
\newblock


\bibitem[\protect\citeauthoryear{Bleichenbacher}{Bleichenbacher}{1998}]%
        {Bl:98}
\bibfield{author}{\bibinfo{person}{Daniel Bleichenbacher}.}
  \bibinfo{year}{1998}\natexlab{}.
\newblock \showarticletitle{Chosen ciphertext attacks against protocols based
  on the RSA encryption standard PKCS\# 1}. In \bibinfo{booktitle}{\emph{Annual
  International Cryptology Conference}}.
\newblock


\bibitem[\protect\citeauthoryear{B{\"o}ck, Somorovsky, and Young}{B{\"o}ck
  et~al\mbox{.}}{2018}]%
        {HaJuCr:18}
\bibfield{author}{\bibinfo{person}{Hanno B{\"o}ck}, \bibinfo{person}{Juraj
  Somorovsky}, {and} \bibinfo{person}{Craig Young}.}
  \bibinfo{year}{2018}\natexlab{}.
\newblock \showarticletitle{Return Of Bleichenbacher’s Oracle Threat
  (ROBOT)}. In \bibinfo{booktitle}{\emph{Usenix Security Symposium}}.
\newblock


\bibitem[\protect\citeauthoryear{Bond, Hawblitzel, Kapritsos, Leino, Lorch,
  Parno, Rane, Setty, and Thompson}{Bond et~al\mbox{.}}{2017}]%
        {BoHaKa:17}
\bibfield{author}{\bibinfo{person}{Barry Bond}, \bibinfo{person}{Chris
  Hawblitzel}, \bibinfo{person}{Manos Kapritsos}, \bibinfo{person}{K~Rustan~M
  Leino}, \bibinfo{person}{Jacob~R Lorch}, \bibinfo{person}{Bryan Parno},
  \bibinfo{person}{Ashay Rane}, \bibinfo{person}{Srinath Setty}, {and}
  \bibinfo{person}{Laure Thompson}.} \bibinfo{year}{2017}\natexlab{}.
\newblock \showarticletitle{Vale: Verifying high-performance cryptographic
  assembly code}. In \bibinfo{booktitle}{\emph{USENIX Security Symposium}}.
\newblock


\bibitem[\protect\citeauthoryear{Bonneau and Mironov}{Bonneau and
  Mironov}{2006}]%
        {BoMi:06}
\bibfield{author}{\bibinfo{person}{Joseph Bonneau} {and} \bibinfo{person}{Ilya
  Mironov}.} \bibinfo{year}{2006}\natexlab{}.
\newblock \showarticletitle{Cache-collision timing attacks against AES}. In
  \bibinfo{booktitle}{\emph{International Workshop on Cryptographic Hardware
  and Embedded Systems}}.
\newblock


\bibitem[\protect\citeauthoryear{Bos and Coster}{Bos and Coster}{1989}]%
        {BoCo:89}
\bibfield{author}{\bibinfo{person}{Jurjen Bos} {and} \bibinfo{person}{Matthijs
  Coster}.} \bibinfo{year}{1989}\natexlab{}.
\newblock \showarticletitle{Addition chain heuristics}. In
  \bibinfo{booktitle}{\emph{Conference on the Theory and Application of
  Cryptology}}.
\newblock


\bibitem[\protect\citeauthoryear{Brasser, M{\"u}ller, Dmitrienko, Kostiainen,
  Capkun, and Sadeghi}{Brasser et~al\mbox{.}}{2017}]%
        {brasser2017software}
\bibfield{author}{\bibinfo{person}{Ferdinand Brasser}, \bibinfo{person}{Urs
  M{\"u}ller}, \bibinfo{person}{Alexandra Dmitrienko}, \bibinfo{person}{Kari
  Kostiainen}, \bibinfo{person}{Srdjan Capkun}, {and}
  \bibinfo{person}{Ahmad-Reza Sadeghi}.} \bibinfo{year}{2017}\natexlab{}.
\newblock \showarticletitle{Software grand exposure: SGX cache attacks are
  practical}. In \bibinfo{booktitle}{\emph{USENIX Workshop on Offensive
  Technologies}}.
\newblock


\bibitem[\protect\citeauthoryear{Braun, Jana, and Boneh}{Braun
  et~al\mbox{.}}{2015}]%
        {braun2015robust}
\bibfield{author}{\bibinfo{person}{Benjamin~A Braun}, \bibinfo{person}{Suman
  Jana}, {and} \bibinfo{person}{Dan Boneh}.} \bibinfo{year}{2015}\natexlab{}.
\newblock \showarticletitle{Robust and efficient elimination of cache and
  timing side channels}.
\newblock \bibinfo{journal}{\emph{arXiv preprint arXiv:1506.00189}}
  (\bibinfo{year}{2015}).
\newblock


\bibitem[\protect\citeauthoryear{Brickell, Graunke, Neve, and Seifert}{Brickell
  et~al\mbox{.}}{2006b}]%
        {brickell2006software}
\bibfield{author}{\bibinfo{person}{Ernie Brickell}, \bibinfo{person}{Gary
  Graunke}, \bibinfo{person}{Michael Neve}, {and} \bibinfo{person}{Jean-Pierre
  Seifert}.} \bibinfo{year}{2006}\natexlab{b}.
\newblock \showarticletitle{Software mitigations to hedge AES against
  cache-based software side channel vulnerabilities.}
\newblock \bibinfo{journal}{\emph{IACR Cryptol. ePrint Arch.}}
  (\bibinfo{year}{2006}).
\newblock


\bibitem[\protect\citeauthoryear{Brickell, Graunke, and Seifert}{Brickell
  et~al\mbox{.}}{2006a}]%
        {BrGrSe:06}
\bibfield{author}{\bibinfo{person}{Ernie Brickell}, \bibinfo{person}{Gary
  Graunke}, {and} \bibinfo{person}{Jean-Pierre Seifert}.}
  \bibinfo{year}{2006}\natexlab{a}.
\newblock \showarticletitle{Mitigating cache/timing based side-channels in AES
  and RSA software implementations}. In \bibinfo{booktitle}{\emph{RSA
  Conference 2006 session DEV-203}}.
\newblock


\bibitem[\protect\citeauthoryear{Briongos, Malag{\'o}n, de~Goyeneche, and
  Moya}{Briongos et~al\mbox{.}}{2019}]%
        {briongos2019cache}
\bibfield{author}{\bibinfo{person}{Samira Briongos}, \bibinfo{person}{Pedro
  Malag{\'o}n}, \bibinfo{person}{Juan-Mariano de Goyeneche}, {and}
  \bibinfo{person}{Jose~M Moya}.} \bibinfo{year}{2019}\natexlab{}.
\newblock \showarticletitle{Cache misses and the recovery of the full AES 256
  key}.
\newblock \bibinfo{journal}{\emph{Applied Sciences}} (\bibinfo{year}{2019}).
\newblock


\bibitem[\protect\citeauthoryear{Briongos, Malag{\'o}n, Moya, and
  Eisenbarth}{Briongos et~al\mbox{.}}{2020}]%
        {briongos2020reload+}
\bibfield{author}{\bibinfo{person}{Samira Briongos}, \bibinfo{person}{Pedro
  Malag{\'o}n}, \bibinfo{person}{Jos{\'e}~M Moya}, {and}
  \bibinfo{person}{Thomas Eisenbarth}.} \bibinfo{year}{2020}\natexlab{}.
\newblock \showarticletitle{RELOAD+ REFRESH: Abusing Cache Replacement Policies
  to Perform Stealthy Cache Attacks}. In \bibinfo{booktitle}{\emph{USENIX
  Security Symposium}}.
\newblock


\bibitem[\protect\citeauthoryear{Bruinderink, H{\"u}lsing, Lange, and
  Yarom}{Bruinderink et~al\mbox{.}}{2016}]%
        {BrHuLa:16}
\bibfield{author}{\bibinfo{person}{Leon~Groot Bruinderink},
  \bibinfo{person}{Andreas H{\"u}lsing}, \bibinfo{person}{Tanja Lange}, {and}
  \bibinfo{person}{Yuval Yarom}.} \bibinfo{year}{2016}\natexlab{}.
\newblock \showarticletitle{Flush, Gauss, and Reload--a cache attack on the
  BLISS lattice-based signature scheme}. In
  \bibinfo{booktitle}{\emph{International Conference on Cryptographic Hardware
  and Embedded Systems}}.
\newblock


\bibitem[\protect\citeauthoryear{Brumley and Hakala}{Brumley and
  Hakala}{2009}]%
        {BrHa:09}
\bibfield{author}{\bibinfo{person}{Billy~Bob Brumley} {and}
  \bibinfo{person}{Risto~M Hakala}.} \bibinfo{year}{2009}\natexlab{}.
\newblock \showarticletitle{Cache-timing template attacks}. In
  \bibinfo{booktitle}{\emph{International Conference on the Theory and
  Application of Cryptology and Information Security}}.
\newblock


\bibitem[\protect\citeauthoryear{Brumley, Hakala, Nyberg, and Sovio}{Brumley
  et~al\mbox{.}}{2010}]%
        {brumley2010consecutive}
\bibfield{author}{\bibinfo{person}{Billy~Bob Brumley}, \bibinfo{person}{Risto~M
  Hakala}, \bibinfo{person}{Kaisa Nyberg}, {and} \bibinfo{person}{Sampo
  Sovio}.} \bibinfo{year}{2010}\natexlab{}.
\newblock \showarticletitle{Consecutive S-box lookups: A Timing Attack on SNOW
  3G}. In \bibinfo{booktitle}{\emph{International Conference on Information and
  Communications Security}}.
\newblock


\bibitem[\protect\citeauthoryear{Brumley and Tuveri}{Brumley and
  Tuveri}{2011}]%
        {BrTu:11}
\bibfield{author}{\bibinfo{person}{Billy~Bob Brumley} {and}
  \bibinfo{person}{Nicola Tuveri}.} \bibinfo{year}{2011}\natexlab{}.
\newblock \showarticletitle{Remote timing attacks are still practical}. In
  \bibinfo{booktitle}{\emph{European Symposium on Research in Computer
  Security}}.
\newblock


\bibitem[\protect\citeauthoryear{Brumley and Boneh}{Brumley and Boneh}{2005}]%
        {BrBo:05}
\bibfield{author}{\bibinfo{person}{David Brumley} {and} \bibinfo{person}{Dan
  Boneh}.} \bibinfo{year}{2005}\natexlab{}.
\newblock \showarticletitle{Remote timing attacks are practical}.
\newblock \bibinfo{journal}{\emph{Computer Networks}} (\bibinfo{year}{2005}).
\newblock


\bibitem[\protect\citeauthoryear{Bulck, Minkin, Weisse, Genkin, Kasikci,
  Piessens, Silberstein, Wenisch, Yarom, and Strackx}{Bulck
  et~al\mbox{.}}{2018}]%
        {JoMaOf:18}
\bibfield{author}{\bibinfo{person}{Jo~Van Bulck}, \bibinfo{person}{Marina
  Minkin}, \bibinfo{person}{Ofir Weisse}, \bibinfo{person}{Daniel Genkin},
  \bibinfo{person}{Baris Kasikci}, \bibinfo{person}{Frank Piessens},
  \bibinfo{person}{Mark Silberstein}, \bibinfo{person}{Thomas~F. Wenisch},
  \bibinfo{person}{Yuval Yarom}, {and} \bibinfo{person}{Raoul Strackx}.}
  \bibinfo{year}{2018}\natexlab{}.
\newblock \showarticletitle{Foreshadow: Extracting the Keys to the Intel {SGX}
  Kingdom with Transient Out-of-Order Execution}. In
  \bibinfo{booktitle}{\emph{USENIX Security Symposium}}.
\newblock


\bibitem[\protect\citeauthoryear{Canella, Genkin, Giner, Gruss, Lipp, Minkin,
  Moghimi, Piessens, Schwarz, Sunar, et~al\mbox{.}}{Canella
  et~al\mbox{.}}{2019a}]%
        {canella2019fallout}
\bibfield{author}{\bibinfo{person}{Claudio Canella}, \bibinfo{person}{Daniel
  Genkin}, \bibinfo{person}{Lukas Giner}, \bibinfo{person}{Daniel Gruss},
  \bibinfo{person}{Moritz Lipp}, \bibinfo{person}{Marina Minkin},
  \bibinfo{person}{Daniel Moghimi}, \bibinfo{person}{Frank Piessens},
  \bibinfo{person}{Michael Schwarz}, \bibinfo{person}{Berk Sunar},
  {et~al\mbox{.}}} \bibinfo{year}{2019}\natexlab{a}.
\newblock \showarticletitle{Fallout: Leaking data on meltdown-resistant cpus}.
  In \bibinfo{booktitle}{\emph{ACM SIGSAC Conference on Computer and
  Communications Security}}.
\newblock


\bibitem[\protect\citeauthoryear{Canella, Van~Bulck, Schwarz, Lipp, Von~Berg,
  Ortner, Piessens, Evtyushkin, and Gruss}{Canella et~al\mbox{.}}{2019b}]%
        {ClJoMi:18}
\bibfield{author}{\bibinfo{person}{Claudio Canella}, \bibinfo{person}{Jo
  Van~Bulck}, \bibinfo{person}{Michael Schwarz}, \bibinfo{person}{Moritz Lipp},
  \bibinfo{person}{Benjamin Von~Berg}, \bibinfo{person}{Philipp Ortner},
  \bibinfo{person}{Frank Piessens}, \bibinfo{person}{Dmitry Evtyushkin}, {and}
  \bibinfo{person}{Daniel Gruss}.} \bibinfo{year}{2019}\natexlab{b}.
\newblock \showarticletitle{A systematic evaluation of transient execution
  attacks and defenses}. In \bibinfo{booktitle}{\emph{USENIX Security
  Symposium}}.
\newblock


\bibitem[\protect\citeauthoryear{Canteaut, Lauradoux, and Seznec}{Canteaut
  et~al\mbox{.}}{2006}]%
        {canteaut2006understanding}
\bibfield{author}{\bibinfo{person}{Anne Canteaut}, \bibinfo{person}{Cedric
  Lauradoux}, {and} \bibinfo{person}{Andre Seznec}.}
  \bibinfo{year}{2006}\natexlab{}.
\newblock \showarticletitle{Understanding cache attacks}.
\newblock  (\bibinfo{year}{2006}).
\newblock


\bibitem[\protect\citeauthoryear{Chen, Chen, Xiao, Zhang, Lin, and Lai}{Chen
  et~al\mbox{.}}{2019}]%
        {ChChXi:19}
\bibfield{author}{\bibinfo{person}{Guoxing Chen}, \bibinfo{person}{Sanchuan
  Chen}, \bibinfo{person}{Yuan Xiao}, \bibinfo{person}{Yinqian Zhang},
  \bibinfo{person}{Zhiqiang Lin}, {and} \bibinfo{person}{Ten~H. Lai}.}
  \bibinfo{year}{2019}\natexlab{}.
\newblock \showarticletitle{SGXPECTRE: Stealing Intel Secrets from SGX Enclaves
  via Speculative Execution}. In \bibinfo{booktitle}{\emph{IEEE European
  Symposium on Security and Privacy}}.
\newblock


\bibitem[\protect\citeauthoryear{Chen, Liu, Mi, Zhang, Lee, Chen, and
  Wang}{Chen et~al\mbox{.}}{2018}]%
        {chen2018leveraging}
\bibfield{author}{\bibinfo{person}{Sanchuan Chen}, \bibinfo{person}{Fangfei
  Liu}, \bibinfo{person}{Zeyu Mi}, \bibinfo{person}{Yinqian Zhang},
  \bibinfo{person}{Ruby~B Lee}, \bibinfo{person}{Haibo Chen}, {and}
  \bibinfo{person}{XiaoFeng Wang}.} \bibinfo{year}{2018}\natexlab{}.
\newblock \showarticletitle{Leveraging hardware transactional memory for cache
  side-channel defenses}. In \bibinfo{booktitle}{\emph{ACM Asia Conference on
  Computer and Communications Security}}.
\newblock


\bibitem[\protect\citeauthoryear{Chen, Zhang, Reiter, and Zhang}{Chen
  et~al\mbox{.}}{2017}]%
        {chen2017detecting}
\bibfield{author}{\bibinfo{person}{Sanchuan Chen}, \bibinfo{person}{Xiaokuan
  Zhang}, \bibinfo{person}{Michael~K Reiter}, {and} \bibinfo{person}{Yinqian
  Zhang}.} \bibinfo{year}{2017}\natexlab{}.
\newblock \showarticletitle{Detecting privileged side-channel attacks in
  shielded execution with D{\'e}j{\'a} Vu}. In \bibinfo{booktitle}{\emph{ACM
  Asia Conference on Computer and Communications Security}}.
\newblock


\bibitem[\protect\citeauthoryear{Chiappetta, Savas, and Yilmaz}{Chiappetta
  et~al\mbox{.}}{2016}]%
        {chiappetta2016real}
\bibfield{author}{\bibinfo{person}{Marco Chiappetta}, \bibinfo{person}{Erkay
  Savas}, {and} \bibinfo{person}{Cemal Yilmaz}.}
  \bibinfo{year}{2016}\natexlab{}.
\newblock \showarticletitle{Real time detection of cache-based side-channel
  attacks using hardware performance counters}.
\newblock \bibinfo{journal}{\emph{Applied Soft Computing}}
  (\bibinfo{year}{2016}).
\newblock


\bibitem[\protect\citeauthoryear{Coron}{Coron}{1999}]%
        {Co:99}
\bibfield{author}{\bibinfo{person}{Jean-S{\'e}bastien Coron}.}
  \bibinfo{year}{1999}\natexlab{}.
\newblock \showarticletitle{Resistance against differential power analysis for
  elliptic curve cryptosystems}. In \bibinfo{booktitle}{\emph{International
  Workshop on Cryptographic Hardware and Embedded Systems}}.
\newblock


\bibitem[\protect\citeauthoryear{Costan and Devadas}{Costan and
  Devadas}{2016}]%
        {costan2016intel}
\bibfield{author}{\bibinfo{person}{Victor Costan} {and}
  \bibinfo{person}{Srinivas Devadas}.} \bibinfo{year}{2016}\natexlab{}.
\newblock \showarticletitle{Intel SGX Explained.}
\newblock \bibinfo{journal}{\emph{IACR Cryptol. ePrint Arch.}}
  (\bibinfo{year}{2016}).
\newblock


\bibitem[\protect\citeauthoryear{Costan, Lebedev, and Devadas}{Costan
  et~al\mbox{.}}{2016}]%
        {costan2016sanctum}
\bibfield{author}{\bibinfo{person}{Victor Costan}, \bibinfo{person}{Ilia
  Lebedev}, {and} \bibinfo{person}{Srinivas Devadas}.}
  \bibinfo{year}{2016}\natexlab{}.
\newblock \showarticletitle{Sanctum: Minimal hardware extensions for strong
  software isolation}. In \bibinfo{booktitle}{\emph{USENIX Security
  Symposium}}.
\newblock


\bibitem[\protect\citeauthoryear{Crane, Homescu, Brunthaler, Larsen, and
  Franz}{Crane et~al\mbox{.}}{2015}]%
        {crane2015thwarting}
\bibfield{author}{\bibinfo{person}{Stephen Crane}, \bibinfo{person}{Andrei
  Homescu}, \bibinfo{person}{Stefan Brunthaler}, \bibinfo{person}{Per Larsen},
  {and} \bibinfo{person}{Michael Franz}.} \bibinfo{year}{2015}\natexlab{}.
\newblock \showarticletitle{Thwarting Cache Side-Channel Attacks Through
  Dynamic Software Diversity}. In \bibinfo{booktitle}{\emph{Network and
  Distributed System Security Symposium}}.
\newblock


\bibitem[\protect\citeauthoryear{Dall, De~Micheli, Eisenbarth, Genkin,
  Heninger, Moghimi, and Yarom}{Dall et~al\mbox{.}}{2018}]%
        {dall2018cachequote}
\bibfield{author}{\bibinfo{person}{Fergus Dall}, \bibinfo{person}{Gabrielle
  De~Micheli}, \bibinfo{person}{Thomas Eisenbarth}, \bibinfo{person}{Daniel
  Genkin}, \bibinfo{person}{Nadia Heninger}, \bibinfo{person}{Ahmad Moghimi},
  {and} \bibinfo{person}{Yuval Yarom}.} \bibinfo{year}{2018}\natexlab{}.
\newblock \showarticletitle{Cachequote: Efficiently recovering long-term
  secrets of SGX EPID via cache attacks}.
\newblock  (\bibinfo{year}{2018}).
\newblock


\bibitem[\protect\citeauthoryear{D'Anvers, Tiepelt, Vercauteren, and
  Verbauwhede}{D'Anvers et~al\mbox{.}}{2019}]%
        {DaTiVe:19}
\bibfield{author}{\bibinfo{person}{Jan-Pieter D'Anvers},
  \bibinfo{person}{Marcel Tiepelt}, \bibinfo{person}{Frederik Vercauteren},
  {and} \bibinfo{person}{Ingrid Verbauwhede}.} \bibinfo{year}{2019}\natexlab{}.
\newblock \showarticletitle{Timing Attacks on Error Correcting Codes in
  Post-Quantum Schemes}. In \bibinfo{booktitle}{\emph{ACM Workshop on Theory of
  Implementation Security Workshop}}.
\newblock


\bibitem[\protect\citeauthoryear{Dehesa-Azuara, Fredrikson, Hoffmann,
  et~al\mbox{.}}{Dehesa-Azuara et~al\mbox{.}}{2017}]%
        {DeFrHo:17}
\bibfield{author}{\bibinfo{person}{Mario Dehesa-Azuara},
  \bibinfo{person}{Matthew Fredrikson}, \bibinfo{person}{Jan Hoffmann},
  {et~al\mbox{.}}} \bibinfo{year}{2017}\natexlab{}.
\newblock \showarticletitle{Verifying and synthesizing constant-resource
  implementations with types}. In \bibinfo{booktitle}{\emph{IEEE Symposium on
  Security and Privacy}}.
\newblock


\bibitem[\protect\citeauthoryear{Demme, Martin, Waksman, and
  Sethumadhavan}{Demme et~al\mbox{.}}{2012}]%
        {demme2012side}
\bibfield{author}{\bibinfo{person}{John Demme}, \bibinfo{person}{Robert
  Martin}, \bibinfo{person}{Adam Waksman}, {and} \bibinfo{person}{Simha
  Sethumadhavan}.} \bibinfo{year}{2012}\natexlab{}.
\newblock \showarticletitle{Side-channel vulnerability factor: A metric for
  measuring information leakage}. In \bibinfo{booktitle}{\emph{Annual
  International Symposium on Computer Architecture}}.
\newblock


\bibitem[\protect\citeauthoryear{Demme, Maycock, Schmitz, Tang, Waksman,
  Sethumadhavan, and Stolfo}{Demme et~al\mbox{.}}{2013}]%
        {demme2013feasibility}
\bibfield{author}{\bibinfo{person}{John Demme}, \bibinfo{person}{Matthew
  Maycock}, \bibinfo{person}{Jared Schmitz}, \bibinfo{person}{Adrian Tang},
  \bibinfo{person}{Adam Waksman}, \bibinfo{person}{Simha Sethumadhavan}, {and}
  \bibinfo{person}{Salvatore Stolfo}.} \bibinfo{year}{2013}\natexlab{}.
\newblock \showarticletitle{On the feasibility of online malware detection with
  performance counters}.
\newblock \bibinfo{journal}{\emph{ACM SIGARCH Computer Architecture News}}
  (\bibinfo{year}{2013}).
\newblock


\bibitem[\protect\citeauthoryear{Deng, Xiong, and Szefer}{Deng
  et~al\mbox{.}}{2018}]%
        {deng2018cache}
\bibfield{author}{\bibinfo{person}{Shuwen Deng}, \bibinfo{person}{Wenjie
  Xiong}, {and} \bibinfo{person}{Jakub Szefer}.}
  \bibinfo{year}{2018}\natexlab{}.
\newblock \showarticletitle{Cache timing side-channel vulnerability checking
  with computation tree logic}. In \bibinfo{booktitle}{\emph{International
  Workshop on Hardware and Architectural Support for Security and Privacy}}.
\newblock


\bibitem[\protect\citeauthoryear{Deng, Xiong, and Szefer}{Deng
  et~al\mbox{.}}{2019a}]%
        {deng2019analysis}
\bibfield{author}{\bibinfo{person}{Shuwen Deng}, \bibinfo{person}{Wenjie
  Xiong}, {and} \bibinfo{person}{Jakub Szefer}.}
  \bibinfo{year}{2019}\natexlab{a}.
\newblock \showarticletitle{Analysis of secure caches using a three-step model
  for timing-based attacks}.
\newblock \bibinfo{journal}{\emph{Journal of Hardware and Systems Security}}
  (\bibinfo{year}{2019}).
\newblock


\bibitem[\protect\citeauthoryear{Deng, Xiong, and Szefer}{Deng
  et~al\mbox{.}}{2019b}]%
        {deng2019secure}
\bibfield{author}{\bibinfo{person}{Shuwen Deng}, \bibinfo{person}{Wenjie
  Xiong}, {and} \bibinfo{person}{Jakub Szefer}.}
  \bibinfo{year}{2019}\natexlab{b}.
\newblock \showarticletitle{Secure tlbs}. In
  \bibinfo{booktitle}{\emph{International Symposium on Computer Architecture}}.
\newblock


\bibitem[\protect\citeauthoryear{Dessouky, Frassetto, and Sadeghi}{Dessouky
  et~al\mbox{.}}{2020}]%
        {dessouky2020hybcache}
\bibfield{author}{\bibinfo{person}{Ghada Dessouky}, \bibinfo{person}{Tommaso
  Frassetto}, {and} \bibinfo{person}{Ahmad-Reza Sadeghi}.}
  \bibinfo{year}{2020}\natexlab{}.
\newblock \showarticletitle{HybCache: Hybrid side-channel-resilient caches for
  trusted execution environments}. In \bibinfo{booktitle}{\emph{USENIX Security
  Symposium}}.
\newblock


\bibitem[\protect\citeauthoryear{Dhem, Koeune, Leroux, Mestr{\'e}, Quisquater,
  and Willems}{Dhem et~al\mbox{.}}{1998}]%
        {dhem1998practical}
\bibfield{author}{\bibinfo{person}{Jean-Francois Dhem},
  \bibinfo{person}{Francois Koeune}, \bibinfo{person}{Philippe-Alexandre
  Leroux}, \bibinfo{person}{Patrick Mestr{\'e}}, \bibinfo{person}{Jean-Jacques
  Quisquater}, {and} \bibinfo{person}{Jean-Louis Willems}.}
  \bibinfo{year}{1998}\natexlab{}.
\newblock \showarticletitle{A practical implementation of the timing attack}.
  In \bibinfo{booktitle}{\emph{International Conference on Smart Card Research
  and Advanced Applications}}.
\newblock


\bibitem[\protect\citeauthoryear{Disselkoen, Kohlbrenner, Porter, and
  Tullsen}{Disselkoen et~al\mbox{.}}{2017}]%
        {disselkoen2017prime+}
\bibfield{author}{\bibinfo{person}{Craig Disselkoen}, \bibinfo{person}{David
  Kohlbrenner}, \bibinfo{person}{Leo Porter}, {and} \bibinfo{person}{Dean
  Tullsen}.} \bibinfo{year}{2017}\natexlab{}.
\newblock \showarticletitle{Prime+ abort: A timer-free high-precision l3 cache
  attack using intel TSX}. In \bibinfo{booktitle}{\emph{USENIX Security
  Symposium}}.
\newblock


\bibitem[\protect\citeauthoryear{Domnitser, Jaleel, Loew, Abu-Ghazaleh, and
  Ponomarev}{Domnitser et~al\mbox{.}}{2012}]%
        {DoJaLo:12}
\bibfield{author}{\bibinfo{person}{Leonid Domnitser}, \bibinfo{person}{Aamer
  Jaleel}, \bibinfo{person}{Jason Loew}, \bibinfo{person}{Nael Abu-Ghazaleh},
  {and} \bibinfo{person}{Dmitry Ponomarev}.} \bibinfo{year}{2012}\natexlab{}.
\newblock \showarticletitle{Non-monopolizable Caches: Low-complexity Mitigation
  of Cache Side Channel Attacks}.
\newblock \bibinfo{journal}{\emph{ACM Trans. Archit. Code Optim.}}
  (\bibinfo{year}{2012}).
\newblock


\bibitem[\protect\citeauthoryear{Doychev and K{\"o}pf}{Doychev and
  K{\"o}pf}{2017}]%
        {DoKo:17}
\bibfield{author}{\bibinfo{person}{Goran Doychev} {and} \bibinfo{person}{Boris
  K{\"o}pf}.} \bibinfo{year}{2017}\natexlab{}.
\newblock \showarticletitle{Rigorous analysis of software countermeasures
  against cache attacks}. In \bibinfo{booktitle}{\emph{ACM SIGPLAN Notices}}.
\newblock


\bibitem[\protect\citeauthoryear{Doychev, K{\"o}pf, Mauborgne, and
  Reineke}{Doychev et~al\mbox{.}}{2015}]%
        {DoKoMa:15}
\bibfield{author}{\bibinfo{person}{Goran Doychev}, \bibinfo{person}{Boris
  K{\"o}pf}, \bibinfo{person}{Laurent Mauborgne}, {and} \bibinfo{person}{Jan
  Reineke}.} \bibinfo{year}{2015}\natexlab{}.
\newblock \showarticletitle{Cacheaudit: A tool for the static analysis of cache
  side channels}.
\newblock \bibinfo{journal}{\emph{ACM Transactions on Information and System
  Security}} \bibinfo{volume}{18}, \bibinfo{number}{1} (\bibinfo{year}{2015}).
\newblock


\bibitem[\protect\citeauthoryear{Ducas, Durmus, Lepoint, and
  Lyubashevsky}{Ducas et~al\mbox{.}}{2013}]%
        {DuDuLe:13}
\bibfield{author}{\bibinfo{person}{L{\'e}o Ducas}, \bibinfo{person}{Alain
  Durmus}, \bibinfo{person}{Tancr{\`e}de Lepoint}, {and} \bibinfo{person}{Vadim
  Lyubashevsky}.} \bibinfo{year}{2013}\natexlab{}.
\newblock \showarticletitle{Lattice signatures and bimodal Gaussians}. In
  \bibinfo{booktitle}{\emph{Annual Cryptology Conference}}.
\newblock


\bibitem[\protect\citeauthoryear{El~Mrabet, Fournier, Goubin, and
  Lashermes}{El~Mrabet et~al\mbox{.}}{2015}]%
        {el2015survey}
\bibfield{author}{\bibinfo{person}{Nadia El~Mrabet},
  \bibinfo{person}{Jacques~JA Fournier}, \bibinfo{person}{Louis Goubin}, {and}
  \bibinfo{person}{Ronan Lashermes}.} \bibinfo{year}{2015}\natexlab{}.
\newblock \showarticletitle{A survey of fault attacks in pairing based
  cryptography}.
\newblock \bibinfo{journal}{\emph{Cryptography and Communications}}
  (\bibinfo{year}{2015}).
\newblock


\bibitem[\protect\citeauthoryear{ElGamal}{ElGamal}{1985}]%
        {El:85}
\bibfield{author}{\bibinfo{person}{Taher ElGamal}.}
  \bibinfo{year}{1985}\natexlab{}.
\newblock \showarticletitle{A public key cryptosystem and a signature scheme
  based on discrete logarithms}.
\newblock \bibinfo{journal}{\emph{IEEE transactions on information theory}}
  (\bibinfo{year}{1985}).
\newblock


\bibitem[\protect\citeauthoryear{Espitau, Fouque, G{\'e}rard, and
  Tibouchi}{Espitau et~al\mbox{.}}{2017}]%
        {EsFoGr:17}
\bibfield{author}{\bibinfo{person}{Thomas Espitau},
  \bibinfo{person}{Pierre-Alain Fouque}, \bibinfo{person}{Beno{\^\i}t
  G{\'e}rard}, {and} \bibinfo{person}{Mehdi Tibouchi}.}
  \bibinfo{year}{2017}\natexlab{}.
\newblock \showarticletitle{Side-channel attacks on BLISS lattice-based
  signatures: Exploiting branch tracing against strongswan and electromagnetic
  emanations in microcontrollers}. In \bibinfo{booktitle}{\emph{ACM Conference
  on Computer and Communications Security}}.
\newblock


\bibitem[\protect\citeauthoryear{Evtyushkin, Riley, Abu-Ghazaleh, and
  Ponomarev}{Evtyushkin et~al\mbox{.}}{2018}]%
        {EvRiAb:18}
\bibfield{author}{\bibinfo{person}{Dmitry Evtyushkin}, \bibinfo{person}{Ryan
  Riley}, \bibinfo{person}{Nael Abu-Ghazaleh}, {and} \bibinfo{person}{Dmitry
  Ponomarev}.} \bibinfo{year}{2018}\natexlab{}.
\newblock \showarticletitle{BranchScope: A New Side-Channel Attack on
  Directional Branch Predictor}. In \bibinfo{booktitle}{\emph{International
  Conference on Architectural Support for Programming Languages and Operating
  Systems}}.
\newblock


\bibitem[\protect\citeauthoryear{Fan, Guo, De~Mulder, Schaumont, Preneel, and
  Verbauwhede}{Fan et~al\mbox{.}}{2010}]%
        {FaGuDe:10}
\bibfield{author}{\bibinfo{person}{Junfeng Fan}, \bibinfo{person}{Xu Guo},
  \bibinfo{person}{Elke De~Mulder}, \bibinfo{person}{Patrick Schaumont},
  \bibinfo{person}{Bart Preneel}, {and} \bibinfo{person}{Ingrid Verbauwhede}.}
  \bibinfo{year}{2010}\natexlab{}.
\newblock \showarticletitle{State-of-the-art of secure ECC implementations: a
  survey on known side-channel attacks and countermeasures}. In
  \bibinfo{booktitle}{\emph{IEEE International Symposium on Hardware-Oriented
  Security and Trust}}.
\newblock


\bibitem[\protect\citeauthoryear{Fan and Verbauwhede}{Fan and
  Verbauwhede}{2012}]%
        {FaVe:12}
\bibfield{author}{\bibinfo{person}{Junfeng Fan} {and} \bibinfo{person}{Ingrid
  Verbauwhede}.} \bibinfo{year}{2012}\natexlab{}.
\newblock \showarticletitle{An updated survey on secure ECC implementations:
  Attacks, countermeasures and cost}.
\newblock In \bibinfo{booktitle}{\emph{Cryptography and Security: From Theory
  to Applications}}.
\newblock


\bibitem[\protect\citeauthoryear{Fan, Wang, and Cheng}{Fan
  et~al\mbox{.}}{2016}]%
        {FaWaCh:16}
\bibfield{author}{\bibinfo{person}{Shuqin Fan}, \bibinfo{person}{Wenbo Wang},
  {and} \bibinfo{person}{Qingfeng Cheng}.} \bibinfo{year}{2016}\natexlab{}.
\newblock \showarticletitle{Attacking openssl implementation of ECDSA with a
  few signatures}. In \bibinfo{booktitle}{\emph{ACM Conference on Computer and
  Communications Security}}.
\newblock


\bibitem[\protect\citeauthoryear{Ferraiuolo, Wang, Zhang, Myers, and
  Suh}{Ferraiuolo et~al\mbox{.}}{2016}]%
        {ferraiuolo2016lattice}
\bibfield{author}{\bibinfo{person}{Andrew Ferraiuolo}, \bibinfo{person}{Yao
  Wang}, \bibinfo{person}{Danfeng Zhang}, \bibinfo{person}{Andrew~C Myers},
  {and} \bibinfo{person}{G~Edward Suh}.} \bibinfo{year}{2016}\natexlab{}.
\newblock \showarticletitle{Lattice priority scheduling: Low-overhead
  timing-channel protection for a shared memory controller}. In
  \bibinfo{booktitle}{\emph{IEEE International Symposium on High Performance
  Computer Architecture (HPCA)}}.
\newblock


\bibitem[\protect\citeauthoryear{Frigo, Vannacci, Hassan, van~der Veen, Mutlu,
  Giuffrida, Bos, and Razavi}{Frigo et~al\mbox{.}}{2020}]%
        {frigo2020trrespass}
\bibfield{author}{\bibinfo{person}{Pietro Frigo}, \bibinfo{person}{Emanuele
  Vannacci}, \bibinfo{person}{Hasan Hassan}, \bibinfo{person}{Victor van~der
  Veen}, \bibinfo{person}{Onur Mutlu}, \bibinfo{person}{Cristiano Giuffrida},
  \bibinfo{person}{Herbert Bos}, {and} \bibinfo{person}{Kaveh Razavi}.}
  \bibinfo{year}{2020}\natexlab{}.
\newblock \showarticletitle{TRRespass: Exploiting the Many Sides of Target Row
  Refresh}.
\newblock \bibinfo{journal}{\emph{arXiv preprint arXiv:2004.01807}}
  (\bibinfo{year}{2020}).
\newblock


\bibitem[\protect\citeauthoryear{Garc{\'\i}a and Brumley}{Garc{\'\i}a and
  Brumley}{2017}]%
        {GaPeBr:16}
\bibfield{author}{\bibinfo{person}{Cesar~Pereida Garc{\'\i}a} {and}
  \bibinfo{person}{Billy~Bob Brumley}.} \bibinfo{year}{2017}\natexlab{}.
\newblock \showarticletitle{Constant-time callees with variable-time callers}.
  In \bibinfo{booktitle}{\emph{USENIX Security Symposium}}.
\newblock


\bibitem[\protect\citeauthoryear{Ge, Yarom, Cock, and Heiser}{Ge
  et~al\mbox{.}}{2018}]%
        {GeYaCo:18}
\bibfield{author}{\bibinfo{person}{Qian Ge}, \bibinfo{person}{Yuval Yarom},
  \bibinfo{person}{David Cock}, {and} \bibinfo{person}{Gernot Heiser}.}
  \bibinfo{year}{2018}\natexlab{}.
\newblock \showarticletitle{A survey of microarchitectural timing attacks and
  countermeasures on contemporary hardware}.
\newblock \bibinfo{journal}{\emph{Journal of Cryptographic Engineering}}
  (\bibinfo{year}{2018}).
\newblock


\bibitem[\protect\citeauthoryear{Genkin, Pachmanov, Tromer, and Yarom}{Genkin
  et~al\mbox{.}}{2018}]%
        {genkin2018drive}
\bibfield{author}{\bibinfo{person}{Daniel Genkin}, \bibinfo{person}{Lev
  Pachmanov}, \bibinfo{person}{Eran Tromer}, {and} \bibinfo{person}{Yuval
  Yarom}.} \bibinfo{year}{2018}\natexlab{}.
\newblock \showarticletitle{Drive-by key-extraction cache attacks from portable
  code}. In \bibinfo{booktitle}{\emph{International Conference on Applied
  Cryptography and Network Security}}.
\newblock


\bibitem[\protect\citeauthoryear{Genkin, Pipman, and Tromer}{Genkin
  et~al\mbox{.}}{2015}]%
        {GePiTr:15}
\bibfield{author}{\bibinfo{person}{Daniel Genkin}, \bibinfo{person}{Itamar
  Pipman}, {and} \bibinfo{person}{Eran Tromer}.}
  \bibinfo{year}{2015}\natexlab{}.
\newblock \showarticletitle{Get your hands off my laptop: Physical side-channel
  key-extraction attacks on PCs}.
\newblock \bibinfo{journal}{\emph{Journal of Cryptographic Engineering}}
  (\bibinfo{year}{2015}).
\newblock


\bibitem[\protect\citeauthoryear{Genkin, Shamir, and Tromer}{Genkin
  et~al\mbox{.}}{2014}]%
        {GeShTr:14}
\bibfield{author}{\bibinfo{person}{Daniel Genkin}, \bibinfo{person}{Adi
  Shamir}, {and} \bibinfo{person}{Eran Tromer}.}
  \bibinfo{year}{2014}\natexlab{}.
\newblock \showarticletitle{RSA key extraction via low-bandwidth acoustic
  cryptanalysis}. In \bibinfo{booktitle}{\emph{International cryptology
  conference}}.
\newblock


\bibitem[\protect\citeauthoryear{Genkin, Valenta, and Yarom}{Genkin
  et~al\mbox{.}}{2017}]%
        {GeVaYa:17}
\bibfield{author}{\bibinfo{person}{Daniel Genkin}, \bibinfo{person}{Luke
  Valenta}, {and} \bibinfo{person}{Yuval Yarom}.}
  \bibinfo{year}{2017}\natexlab{}.
\newblock \showarticletitle{May the fourth be with you: A microarchitectural
  side channel attack on several real-world applications of Curve25519}. In
  \bibinfo{booktitle}{\emph{ACM Conference on Computer and Communications
  Security}}.
\newblock


\bibitem[\protect\citeauthoryear{Gentry, Peikert, and Vaikuntanathan}{Gentry
  et~al\mbox{.}}{2008}]%
        {GePeVa:08}
\bibfield{author}{\bibinfo{person}{Craig Gentry}, \bibinfo{person}{Chris
  Peikert}, {and} \bibinfo{person}{Vinod Vaikuntanathan}.}
  \bibinfo{year}{2008}\natexlab{}.
\newblock \showarticletitle{Trapdoors for hard lattices and new cryptographic
  constructions}. In \bibinfo{booktitle}{\emph{Annual ACM symposium on Theory
  of computing}}.
\newblock


\bibitem[\protect\citeauthoryear{Gierlichs, Batina, Clavier, Eisenbarth,
  Gouget, Handschuh, Kasper, Lemke-Rust, Mangard, Moradi,
  et~al\mbox{.}}{Gierlichs et~al\mbox{.}}{2008}]%
        {gierlichs2008susceptibility}
\bibfield{author}{\bibinfo{person}{Benedikt Gierlichs}, \bibinfo{person}{Lejla
  Batina}, \bibinfo{person}{Christophe Clavier}, \bibinfo{person}{Thomas
  Eisenbarth}, \bibinfo{person}{Aline Gouget}, \bibinfo{person}{Helena
  Handschuh}, \bibinfo{person}{Timo Kasper}, \bibinfo{person}{Kerstin
  Lemke-Rust}, \bibinfo{person}{Stefan Mangard}, \bibinfo{person}{Amir Moradi},
  {et~al\mbox{.}}} \bibinfo{year}{2008}\natexlab{}.
\newblock \showarticletitle{Susceptibility of eSTREAM candidates towards side
  channel analysis}.
\newblock  (\bibinfo{year}{2008}).
\newblock


\bibitem[\protect\citeauthoryear{Godfrey and Zulkernine}{Godfrey and
  Zulkernine}{2013}]%
        {godfrey2013server}
\bibfield{author}{\bibinfo{person}{Michael Godfrey} {and}
  \bibinfo{person}{Mohammad Zulkernine}.} \bibinfo{year}{2013}\natexlab{}.
\newblock \showarticletitle{A server-side solution to cache-based side-channel
  attacks in the cloud}. In \bibinfo{booktitle}{\emph{IEEE Sixth International
  Conference on Cloud Computing}}.
\newblock


\bibitem[\protect\citeauthoryear{Gordon}{Gordon}{1998}]%
        {Go:98}
\bibfield{author}{\bibinfo{person}{Daniel~M. Gordon}.}
  \bibinfo{year}{1998}\natexlab{}.
\newblock \showarticletitle{A Survey of Fast Exponentiation Methods}.
\newblock \bibinfo{journal}{\emph{J. Algorithms}} (\bibinfo{year}{1998}).
\newblock


\bibitem[\protect\citeauthoryear{G{\"o}tzfried, Eckert, Schinzel, and
  M{\"u}ller}{G{\"o}tzfried et~al\mbox{.}}{2017}]%
        {gotzfried2017cache}
\bibfield{author}{\bibinfo{person}{Johannes G{\"o}tzfried},
  \bibinfo{person}{Moritz Eckert}, \bibinfo{person}{Sebastian Schinzel}, {and}
  \bibinfo{person}{Tilo M{\"u}ller}.} \bibinfo{year}{2017}\natexlab{}.
\newblock \showarticletitle{Cache attacks on Intel SGX}. In
  \bibinfo{booktitle}{\emph{European Workshop on Systems Security}}.
\newblock


\bibitem[\protect\citeauthoryear{Gras, Razavi, Bos, and Giuffrida}{Gras
  et~al\mbox{.}}{2018}]%
        {GrRaBo:18}
\bibfield{author}{\bibinfo{person}{Ben Gras}, \bibinfo{person}{Kaveh Razavi},
  \bibinfo{person}{Herbert Bos}, {and} \bibinfo{person}{Cristiano Giuffrida}.}
  \bibinfo{year}{2018}\natexlab{}.
\newblock \showarticletitle{Translation Leak-aside Buffer: Defeating Cache
  Side-channel Protections with TLB Attacks}. In
  \bibinfo{booktitle}{\emph{USENIX Security Symposium}}.
\newblock


\bibitem[\protect\citeauthoryear{Green, Rodrigues-Lima, Zankl, Irazoqui,
  Heyszl, and Eisenbarth}{Green et~al\mbox{.}}{2017}]%
        {green2017autolock}
\bibfield{author}{\bibinfo{person}{Marc Green}, \bibinfo{person}{Leandro
  Rodrigues-Lima}, \bibinfo{person}{Andreas Zankl}, \bibinfo{person}{Gorka
  Irazoqui}, \bibinfo{person}{Johann Heyszl}, {and} \bibinfo{person}{Thomas
  Eisenbarth}.} \bibinfo{year}{2017}\natexlab{}.
\newblock \showarticletitle{AutoLock: Why Cache Attacks on ARM Are Harder Than
  You Think}. In \bibinfo{booktitle}{\emph{USENIX Security Symposium}}.
\newblock


\bibitem[\protect\citeauthoryear{Gruss, Lettner, Schuster, Ohrimenko, Haller,
  and Costa}{Gruss et~al\mbox{.}}{2017}]%
        {gruss2017strong}
\bibfield{author}{\bibinfo{person}{Daniel Gruss}, \bibinfo{person}{Julian
  Lettner}, \bibinfo{person}{Felix Schuster}, \bibinfo{person}{Olya Ohrimenko},
  \bibinfo{person}{Istvan Haller}, {and} \bibinfo{person}{Manuel Costa}.}
  \bibinfo{year}{2017}\natexlab{}.
\newblock \showarticletitle{Strong and efficient cache side-channel protection
  using hardware transactional memory}. In \bibinfo{booktitle}{\emph{USENIX
  Security Symposium}}.
\newblock


\bibitem[\protect\citeauthoryear{Gruss, Maurice, Wagner, and Mangard}{Gruss
  et~al\mbox{.}}{2016}]%
        {gruss2016flush+}
\bibfield{author}{\bibinfo{person}{Daniel Gruss},
  \bibinfo{person}{Cl{\'e}mentine Maurice}, \bibinfo{person}{Klaus Wagner},
  {and} \bibinfo{person}{Stefan Mangard}.} \bibinfo{year}{2016}\natexlab{}.
\newblock \showarticletitle{Flush+ Flush: a fast and stealthy cache attack}. In
  \bibinfo{booktitle}{\emph{International Conference on Detection of Intrusions
  and Malware, and Vulnerability Assessment}}.
\newblock


\bibitem[\protect\citeauthoryear{Gruss, Spreitzer, and Mangard}{Gruss
  et~al\mbox{.}}{2015}]%
        {gruss2015cache}
\bibfield{author}{\bibinfo{person}{Daniel Gruss}, \bibinfo{person}{Raphael
  Spreitzer}, {and} \bibinfo{person}{Stefan Mangard}.}
  \bibinfo{year}{2015}\natexlab{}.
\newblock \showarticletitle{Cache template attacks: Automating attacks on
  inclusive last-level caches}. In \bibinfo{booktitle}{\emph{USENIX Security
  Symposium}}.
\newblock


\bibitem[\protect\citeauthoryear{Gullasch, Bangerter, and Krenn}{Gullasch
  et~al\mbox{.}}{2011}]%
        {GuBaKr:11}
\bibfield{author}{\bibinfo{person}{David Gullasch}, \bibinfo{person}{Endre
  Bangerter}, {and} \bibinfo{person}{Stephan Krenn}.}
  \bibinfo{year}{2011}\natexlab{}.
\newblock \showarticletitle{Cache games--Bringing access-based cache attacks on
  AES to practice}. In \bibinfo{booktitle}{\emph{IEEE Symposium on Security and
  Privacy}}.
\newblock


\bibitem[\protect\citeauthoryear{H{\"a}hnel, Cui, and Peinado}{H{\"a}hnel
  et~al\mbox{.}}{2017}]%
        {hahnel2017high}
\bibfield{author}{\bibinfo{person}{Marcus H{\"a}hnel}, \bibinfo{person}{Weidong
  Cui}, {and} \bibinfo{person}{Marcus Peinado}.}
  \bibinfo{year}{2017}\natexlab{}.
\newblock \showarticletitle{High-resolution side channels for untrusted
  operating systems}. In \bibinfo{booktitle}{\emph{USENIX Annual Technical
  Conference}}.
\newblock


\bibitem[\protect\citeauthoryear{Hankerson, Menezes, and Vanstone}{Hankerson
  et~al\mbox{.}}{2005}]%
        {HaMeVa:05}
\bibfield{author}{\bibinfo{person}{Darrel Hankerson}, \bibinfo{person}{Alfred~J
  Menezes}, {and} \bibinfo{person}{Scott Vanstone}.}
  \bibinfo{year}{2005}\natexlab{}.
\newblock \showarticletitle{Guide to elliptic curve cryptography}.
\newblock \bibinfo{journal}{\emph{Computing Reviews}} (\bibinfo{year}{2005}).
\newblock


\bibitem[\protect\citeauthoryear{He and Lee}{He and Lee}{2017}]%
        {he2017secure}
\bibfield{author}{\bibinfo{person}{Zecheng He} {and} \bibinfo{person}{Ruby~B
  Lee}.} \bibinfo{year}{2017}\natexlab{}.
\newblock \showarticletitle{How secure is your cache against side-channel
  attacks?}. In \bibinfo{booktitle}{\emph{Annual IEEE/ACM International
  Symposium on Microarchitecture}}.
\newblock


\bibitem[\protect\citeauthoryear{Hoffstein, Pipher, and Silverman}{Hoffstein
  et~al\mbox{.}}{1998}]%
        {HoPiSi:98}
\bibfield{author}{\bibinfo{person}{Jeffrey Hoffstein}, \bibinfo{person}{Jill
  Pipher}, {and} \bibinfo{person}{Joseph~H Silverman}.}
  \bibinfo{year}{1998}\natexlab{}.
\newblock \showarticletitle{NTRU: A ring-based public key cryptosystem}. In
  \bibinfo{booktitle}{\emph{International Algorithmic Number Theory
  Symposium}}.
\newblock


\bibitem[\protect\citeauthoryear{Hou, Breier, Zhang, and Liu}{Hou
  et~al\mbox{.}}{2019}]%
        {hou2019fully}
\bibfield{author}{\bibinfo{person}{Xiaolu Hou}, \bibinfo{person}{Jakub Breier},
  \bibinfo{person}{Fuyuan Zhang}, {and} \bibinfo{person}{Yang Liu}.}
  \bibinfo{year}{2019}\natexlab{}.
\newblock \showarticletitle{Fully automated differential fault analysis on
  software implementations of block ciphers}.
\newblock \bibinfo{journal}{\emph{IACR Transactions on Cryptographic Hardware
  and Embedded Systems}} (\bibinfo{year}{2019}).
\newblock


\bibitem[\protect\citeauthoryear{Hu}{Hu}{1992}]%
        {hu1992reducing}
\bibfield{author}{\bibinfo{person}{Wei-Ming Hu}.}
  \bibinfo{year}{1992}\natexlab{}.
\newblock \showarticletitle{Reducing timing channels with fuzzy time}.
\newblock \bibinfo{journal}{\emph{Journal of computer security}}
  (\bibinfo{year}{1992}).
\newblock


\bibitem[\protect\citeauthoryear{Hund, Willems, and Holz}{Hund
  et~al\mbox{.}}{2013}]%
        {HuWiHo:13}
\bibfield{author}{\bibinfo{person}{Ralf Hund}, \bibinfo{person}{Carsten
  Willems}, {and} \bibinfo{person}{Thorsten Holz}.}
  \bibinfo{year}{2013}\natexlab{}.
\newblock \showarticletitle{Practical timing side channel attacks against
  kernel space ASLR}. In \bibinfo{booktitle}{\emph{IEEE Symposium on Security
  and Privacy}}.
\newblock


\bibitem[\protect\citeauthoryear{Hunger, Kazdagli, Rawat, Dimakis, Vishwanath,
  and Tiwari}{Hunger et~al\mbox{.}}{2015}]%
        {hunger2015understanding}
\bibfield{author}{\bibinfo{person}{Casen Hunger}, \bibinfo{person}{Mikhail
  Kazdagli}, \bibinfo{person}{Ankit Rawat}, \bibinfo{person}{Alex Dimakis},
  \bibinfo{person}{Sriram Vishwanath}, {and} \bibinfo{person}{Mohit Tiwari}.}
  \bibinfo{year}{2015}\natexlab{}.
\newblock \showarticletitle{Understanding contention-based channels and using
  them for defense}. In \bibinfo{booktitle}{\emph{International Symposium on
  High Performance Computer Architecture}}.
\newblock


\bibitem[\protect\citeauthoryear{Hussain, Al-Haiqi, Zaidan, Zaidan, Kiah,
  Anuar, and Abdulnabi}{Hussain et~al\mbox{.}}{2016}]%
        {HuAlZa:16}
\bibfield{author}{\bibinfo{person}{Muzammil Hussain}, \bibinfo{person}{Ahmed
  Al-Haiqi}, \bibinfo{person}{AA Zaidan}, \bibinfo{person}{BB Zaidan},
  \bibinfo{person}{ML~Mat Kiah}, \bibinfo{person}{Nor~Badrul Anuar}, {and}
  \bibinfo{person}{Mohamed Abdulnabi}.} \bibinfo{year}{2016}\natexlab{}.
\newblock \showarticletitle{The rise of keyloggers on smartphones: A survey and
  insight into motion-based tap inference attacks}.
\newblock \bibinfo{journal}{\emph{Pervasive and Mobile Computing}}
  (\bibinfo{year}{2016}).
\newblock


\bibitem[\protect\citeauthoryear{Inci, Gulmezoglu, Irazoqui, Eisenbarth, and
  Sunar}{Inci et~al\mbox{.}}{2016}]%
        {InGuIr:16}
\bibfield{author}{\bibinfo{person}{Mehmet~Sinan Inci}, \bibinfo{person}{Berk
  Gulmezoglu}, \bibinfo{person}{Gorka Irazoqui}, \bibinfo{person}{Thomas
  Eisenbarth}, {and} \bibinfo{person}{Berk Sunar}.}
  \bibinfo{year}{2016}\natexlab{}.
\newblock \showarticletitle{Cache attacks enable bulk key recovery on the
  cloud}. In \bibinfo{booktitle}{\emph{International Conference on
  Cryptographic Hardware and Embedded Systems}}.
\newblock


\bibitem[\protect\citeauthoryear{Irazoqui, Cong, Guo, Khattri, Kanuparthi,
  Eisenbarth, and Sunar}{Irazoqui et~al\mbox{.}}{2017}]%
        {IrCoGu:17}
\bibfield{author}{\bibinfo{person}{Gorka Irazoqui}, \bibinfo{person}{Kai Cong},
  \bibinfo{person}{Xiaofei Guo}, \bibinfo{person}{Hareesh Khattri},
  \bibinfo{person}{Arun Kanuparthi}, \bibinfo{person}{Thomas Eisenbarth}, {and}
  \bibinfo{person}{Berk Sunar}.} \bibinfo{year}{2017}\natexlab{}.
\newblock \showarticletitle{Did we learn from LLC Side Channel Attacks? A Cache
  Leakage Detection Tool for Crypto Libraries}.
\newblock \bibinfo{journal}{\emph{arXiv preprint arXiv:1709.01552}}
  (\bibinfo{year}{2017}).
\newblock


\bibitem[\protect\citeauthoryear{Irazoqui, Eisenbarth, and Sunar}{Irazoqui
  et~al\mbox{.}}{2015a}]%
        {irazoqui2015s}
\bibfield{author}{\bibinfo{person}{Gorka Irazoqui}, \bibinfo{person}{Thomas
  Eisenbarth}, {and} \bibinfo{person}{Berk Sunar}.}
  \bibinfo{year}{2015}\natexlab{a}.
\newblock \showarticletitle{S $\$$ A: A shared cache attack that works across
  cores and defies VM sandboxing--and its application to AES}. In
  \bibinfo{booktitle}{\emph{IEEE Symposium on Security and Privacy}}.
\newblock


\bibitem[\protect\citeauthoryear{Irazoqui, Eisenbarth, and Sunar}{Irazoqui
  et~al\mbox{.}}{2016}]%
        {irazoqui2016mascat}
\bibfield{author}{\bibinfo{person}{Gorka Irazoqui}, \bibinfo{person}{Thomas
  Eisenbarth}, {and} \bibinfo{person}{Berk Sunar}.}
  \bibinfo{year}{2016}\natexlab{}.
\newblock \showarticletitle{MASCAT: Stopping Microarchitectural Attacks Before
  Execution.}
\newblock \bibinfo{journal}{\emph{IACR Cryptol. ePrint Arch.}}
  (\bibinfo{year}{2016}).
\newblock


\bibitem[\protect\citeauthoryear{Irazoqui, Inci, Eisenbarth, and
  Sunar}{Irazoqui et~al\mbox{.}}{2014}]%
        {IrInEi:14}
\bibfield{author}{\bibinfo{person}{Gorka Irazoqui},
  \bibinfo{person}{Mehmet~Sinan Inci}, \bibinfo{person}{Thomas Eisenbarth},
  {and} \bibinfo{person}{Berk Sunar}.} \bibinfo{year}{2014}\natexlab{}.
\newblock \showarticletitle{Wait a minute! A fast, Cross-VM attack on AES}. In
  \bibinfo{booktitle}{\emph{International Workshop on Recent Advances in
  Intrusion Detection}}.
\newblock


\bibitem[\protect\citeauthoryear{Irazoqui, Inci, Eisenbarth, and
  Sunar}{Irazoqui et~al\mbox{.}}{2015b}]%
        {IrLnEi:15}
\bibfield{author}{\bibinfo{person}{Gorka Irazoqui},
  \bibinfo{person}{Mehmet~Sinan Inci}, \bibinfo{person}{Thomas Eisenbarth},
  {and} \bibinfo{person}{Berk Sunar}.} \bibinfo{year}{2015}\natexlab{b}.
\newblock \showarticletitle{Lucky 13 strikes back}. In
  \bibinfo{booktitle}{\emph{ACM Symposium on Information, Computer and
  Communications Security}}.
\newblock


\bibitem[\protect\citeauthoryear{Islam, Moghimi, Bruhns, Krebbel, Gulmezoglu,
  Eisenbarth, and Sunar}{Islam et~al\mbox{.}}{2019}]%
        {SaAhId:19}
\bibfield{author}{\bibinfo{person}{Saad Islam}, \bibinfo{person}{Ahmad
  Moghimi}, \bibinfo{person}{Ida Bruhns}, \bibinfo{person}{Moritz Krebbel},
  \bibinfo{person}{Berk Gulmezoglu}, \bibinfo{person}{Thomas Eisenbarth}, {and}
  \bibinfo{person}{Berk Sunar}.} \bibinfo{year}{2019}\natexlab{}.
\newblock \showarticletitle{SPOILER: Speculative Load Hazards Boost Rowhammer
  and Cache Attacks}. In \bibinfo{booktitle}{\emph{USENIX Security Symposium}}.
\newblock


\bibitem[\protect\citeauthoryear{Jang, Lee, and Kim}{Jang
  et~al\mbox{.}}{2016}]%
        {JaLeKi:16}
\bibfield{author}{\bibinfo{person}{Yeongjin Jang}, \bibinfo{person}{Sangho
  Lee}, {and} \bibinfo{person}{Taesoo Kim}.} \bibinfo{year}{2016}\natexlab{}.
\newblock \showarticletitle{Breaking kernel address space layout randomization
  with intel tsx}. In \bibinfo{booktitle}{\emph{ACM Conference on Computer and
  Communications Security}}.
\newblock


\bibitem[\protect\citeauthoryear{Joye and Yen}{Joye and Yen}{2002}]%
        {JoYe:02}
\bibfield{author}{\bibinfo{person}{Marc Joye} {and} \bibinfo{person}{Sung-Ming
  Yen}.} \bibinfo{year}{2002}\natexlab{}.
\newblock \showarticletitle{The Montgomery powering ladder}. In
  \bibinfo{booktitle}{\emph{International Workshop on Cryptographic Hardware
  and Embedded Systems}}.
\newblock


\bibitem[\protect\citeauthoryear{Karatsuba and Ofman}{Karatsuba and
  Ofman}{1962}]%
        {KaOf:62}
\bibfield{author}{\bibinfo{person}{Anatolii~Alekseevich Karatsuba} {and}
  \bibinfo{person}{Yu~P Ofman}.} \bibinfo{year}{1962}\natexlab{}.
\newblock \showarticletitle{Multiplication of many-digital numbers by automatic
  computers}. In \bibinfo{booktitle}{\emph{Doklady Akademii Nauk}}.
\newblock


\bibitem[\protect\citeauthoryear{Katz, Menezes, Van~Oorschot, and
  Vanstone}{Katz et~al\mbox{.}}{1996}]%
        {KaMeVa:96}
\bibfield{author}{\bibinfo{person}{Jonathan Katz}, \bibinfo{person}{Alfred~J
  Menezes}, \bibinfo{person}{Paul~C Van~Oorschot}, {and}
  \bibinfo{person}{Scott~A Vanstone}.} \bibinfo{year}{1996}\natexlab{}.
\newblock \bibinfo{booktitle}{\emph{Handbook of applied cryptography}}.
\newblock \bibinfo{publisher}{CRC press}.
\newblock


\bibitem[\protect\citeauthoryear{Kaufmann, Pelletier, Vaudenay, and
  Villegas}{Kaufmann et~al\mbox{.}}{2016}]%
        {KaPeVa:16}
\bibfield{author}{\bibinfo{person}{Thierry Kaufmann},
  \bibinfo{person}{Herv{\'e} Pelletier}, \bibinfo{person}{Serge Vaudenay},
  {and} \bibinfo{person}{Karine Villegas}.} \bibinfo{year}{2016}\natexlab{}.
\newblock \showarticletitle{When constant-time source yields variable-time
  binary: Exploiting curve25519-donna built with msvc 2015}. In
  \bibinfo{booktitle}{\emph{International Conference on Cryptology and Network
  Security}}.
\newblock


\bibitem[\protect\citeauthoryear{Kayaalp, Abu-Ghazaleh, Ponomarev, and
  Jaleel}{Kayaalp et~al\mbox{.}}{2016}]%
        {kayaalp2016high}
\bibfield{author}{\bibinfo{person}{Mehmet Kayaalp}, \bibinfo{person}{Nael
  Abu-Ghazaleh}, \bibinfo{person}{Dmitry Ponomarev}, {and}
  \bibinfo{person}{Aamer Jaleel}.} \bibinfo{year}{2016}\natexlab{}.
\newblock \showarticletitle{A high-resolution side-channel attack on last-level
  cache}. In \bibinfo{booktitle}{\emph{Annual Design Automation Conference}}.
\newblock


\bibitem[\protect\citeauthoryear{Keerthi, Roy, Rebeiro, Hazra, and
  Bhunia}{Keerthi et~al\mbox{.}}{2020}]%
        {keerthi2020feds}
\bibfield{author}{\bibinfo{person}{K Keerthi}, \bibinfo{person}{Indrani Roy},
  \bibinfo{person}{Chester Rebeiro}, \bibinfo{person}{Aritra Hazra}, {and}
  \bibinfo{person}{Swarup Bhunia}.} \bibinfo{year}{2020}\natexlab{}.
\newblock \showarticletitle{FEDS: Comprehensive Fault Attack Exploitability
  Detection for Software Implementations of Block Ciphers}.
\newblock \bibinfo{journal}{\emph{IACR Transactions on Cryptographic Hardware
  and Embedded Systems}} (\bibinfo{year}{2020}).
\newblock


\bibitem[\protect\citeauthoryear{Keramidas, Antonopoulos, Serpanos, and
  Kaxiras}{Keramidas et~al\mbox{.}}{2008}]%
        {keramidas2008non}
\bibfield{author}{\bibinfo{person}{Georgios Keramidas},
  \bibinfo{person}{Alexandros Antonopoulos}, \bibinfo{person}{Dimitrios~N
  Serpanos}, {and} \bibinfo{person}{Stefanos Kaxiras}.}
  \bibinfo{year}{2008}\natexlab{}.
\newblock \showarticletitle{Non deterministic caches: A simple and effective
  defense against side channel attacks}.
\newblock \bibinfo{journal}{\emph{Design Automation for Embedded Systems}}
  (\bibinfo{year}{2008}).
\newblock


\bibitem[\protect\citeauthoryear{Kim, Patel, Yaglikci, Hassan, Azizi, Orosa,
  and Mutlu}{Kim et~al\mbox{.}}{2020}]%
        {kim2020revisiting}
\bibfield{author}{\bibinfo{person}{Jeremie~S Kim}, \bibinfo{person}{Minesh
  Patel}, \bibinfo{person}{A~Giray Yaglikci}, \bibinfo{person}{Hasan Hassan},
  \bibinfo{person}{Roknoddin Azizi}, \bibinfo{person}{Lois Orosa}, {and}
  \bibinfo{person}{Onur Mutlu}.} \bibinfo{year}{2020}\natexlab{}.
\newblock \showarticletitle{Revisiting RowHammer: An Experimental Analysis of
  Modern DRAM Devices and Mitigation Techniques}.
\newblock \bibinfo{journal}{\emph{arXiv preprint arXiv:2005.13121}}
  (\bibinfo{year}{2020}).
\newblock


\bibitem[\protect\citeauthoryear{Kim, Peinado, and Mainar-Ruiz}{Kim
  et~al\mbox{.}}{2012}]%
        {KiPeMa:12}
\bibfield{author}{\bibinfo{person}{Taesoo Kim}, \bibinfo{person}{Marcus
  Peinado}, {and} \bibinfo{person}{Gloria Mainar-Ruiz}.}
  \bibinfo{year}{2012}\natexlab{}.
\newblock \showarticletitle{STEALTHMEM: System-level Protection Against
  Cache-based Side Channel Attacks in the Cloud}. In
  \bibinfo{booktitle}{\emph{USENIX Conf. on Security Symposium}}.
\newblock


\bibitem[\protect\citeauthoryear{Kim, Daly, Kim, Fallin, Lee, Lee, Wilkerson,
  Lai, and Mutlu}{Kim et~al\mbox{.}}{2014}]%
        {kim2014flipping}
\bibfield{author}{\bibinfo{person}{Yoongu Kim}, \bibinfo{person}{Ross Daly},
  \bibinfo{person}{Jeremie Kim}, \bibinfo{person}{Chris Fallin},
  \bibinfo{person}{Ji~Hye Lee}, \bibinfo{person}{Donghyuk Lee},
  \bibinfo{person}{Chris Wilkerson}, \bibinfo{person}{Konrad Lai}, {and}
  \bibinfo{person}{Onur Mutlu}.} \bibinfo{year}{2014}\natexlab{}.
\newblock \showarticletitle{Flipping bits in memory without accessing them: An
  experimental study of DRAM disturbance errors}.
\newblock \bibinfo{journal}{\emph{ACM SIGARCH Computer Architecture News}}
  (\bibinfo{year}{2014}).
\newblock


\bibitem[\protect\citeauthoryear{Kiriansky, Lebedev, Amarasinghe, Devadas, and
  Emer}{Kiriansky et~al\mbox{.}}{2018}]%
        {kiriansky2018dawg}
\bibfield{author}{\bibinfo{person}{Vladimir Kiriansky}, \bibinfo{person}{Ilia
  Lebedev}, \bibinfo{person}{Saman Amarasinghe}, \bibinfo{person}{Srinivas
  Devadas}, {and} \bibinfo{person}{Joel Emer}.}
  \bibinfo{year}{2018}\natexlab{}.
\newblock \showarticletitle{DAWG: A defense against cache timing attacks in
  speculative execution processors}. In \bibinfo{booktitle}{\emph{IEEE/ACM
  International Symposium on Microarchitecture}}.
\newblock


\bibitem[\protect\citeauthoryear{Kocher, Horn, Fogh, , Genkin, Gruss, Haas,
  Hamburg, Lipp, Mangard, Prescher, Schwarz, and Yarom}{Kocher
  et~al\mbox{.}}{2019}]%
        {KoHoFo:19}
\bibfield{author}{\bibinfo{person}{Paul Kocher}, \bibinfo{person}{Jann Horn},
  \bibinfo{person}{Anders Fogh}, \bibinfo{person}{}, \bibinfo{person}{Daniel
  Genkin}, \bibinfo{person}{Daniel Gruss}, \bibinfo{person}{Werner Haas},
  \bibinfo{person}{Mike Hamburg}, \bibinfo{person}{Moritz Lipp},
  \bibinfo{person}{Stefan Mangard}, \bibinfo{person}{Thomas Prescher},
  \bibinfo{person}{Michael Schwarz}, {and} \bibinfo{person}{Yuval Yarom}.}
  \bibinfo{year}{2019}\natexlab{}.
\newblock \showarticletitle{Spectre Attacks: Exploiting Speculative Execution}.
  In \bibinfo{booktitle}{\emph{IEEE Symposium on Security and Privacy}}.
\newblock


\bibitem[\protect\citeauthoryear{Kocher}{Kocher}{1996}]%
        {Ko:96}
\bibfield{author}{\bibinfo{person}{Paul~C Kocher}.}
  \bibinfo{year}{1996}\natexlab{}.
\newblock \showarticletitle{Timing attacks on implementations of
  Diffie-Hellman, RSA, DSS, and other systems}. In
  \bibinfo{booktitle}{\emph{Annual International Cryptology Conference}}.
\newblock


\bibitem[\protect\citeauthoryear{K{\"o}pf, Mauborgne, and Ochoa}{K{\"o}pf
  et~al\mbox{.}}{2012}]%
        {kopf2012automatic}
\bibfield{author}{\bibinfo{person}{Boris K{\"o}pf}, \bibinfo{person}{Laurent
  Mauborgne}, {and} \bibinfo{person}{Mart{\'\i}n Ochoa}.}
  \bibinfo{year}{2012}\natexlab{}.
\newblock \showarticletitle{Automatic quantification of cache side-channels}.
  In \bibinfo{booktitle}{\emph{International Conference on Computer Aided
  Verification}}.
\newblock


\bibitem[\protect\citeauthoryear{Kwong, Genkin, Gruss, and Yarom}{Kwong
  et~al\mbox{.}}{2020}]%
        {kwong2020rambleed}
\bibfield{author}{\bibinfo{person}{Andrew Kwong}, \bibinfo{person}{Daniel
  Genkin}, \bibinfo{person}{Daniel Gruss}, {and} \bibinfo{person}{Yuval
  Yarom}.} \bibinfo{year}{2020}\natexlab{}.
\newblock \showarticletitle{Rambleed: Reading bits in memory without accessing
  them}. In \bibinfo{booktitle}{\emph{IEEE Symposium on Security and Privacy}}.
\newblock


\bibitem[\protect\citeauthoryear{Langley, Hamburg, and Turner}{Langley
  et~al\mbox{.}}{2016}]%
        {LaHaTu:16}
\bibfield{author}{\bibinfo{person}{Adam Langley}, \bibinfo{person}{Mike
  Hamburg}, {and} \bibinfo{person}{Sean Turner}.}
  \bibinfo{year}{2016}\natexlab{}.
\newblock \bibinfo{booktitle}{\emph{Elliptic curves for security}}.
\newblock \bibinfo{type}{{T}echnical {R}eport}.
\newblock


\bibitem[\protect\citeauthoryear{Leander, Zenner, and Hawkes}{Leander
  et~al\mbox{.}}{2009}]%
        {leander2009cache}
\bibfield{author}{\bibinfo{person}{Gregor Leander}, \bibinfo{person}{Erik
  Zenner}, {and} \bibinfo{person}{Philip Hawkes}.}
  \bibinfo{year}{2009}\natexlab{}.
\newblock \showarticletitle{Cache timing analysis of LFSR-based stream
  ciphers}. In \bibinfo{booktitle}{\emph{IMA International Conference on
  Cryptography and Coding}}.
\newblock


\bibitem[\protect\citeauthoryear{Lee, Shih, Gera, Kim, Kim, and Peinado}{Lee
  et~al\mbox{.}}{2017}]%
        {lee2017inferring}
\bibfield{author}{\bibinfo{person}{Sangho Lee}, \bibinfo{person}{Ming-Wei
  Shih}, \bibinfo{person}{Prasun Gera}, \bibinfo{person}{Taesoo Kim},
  \bibinfo{person}{Hyesoon Kim}, {and} \bibinfo{person}{Marcus Peinado}.}
  \bibinfo{year}{2017}\natexlab{}.
\newblock \showarticletitle{Inferring fine-grained control flow inside SGX
  enclaves with branch shadowing}. In \bibinfo{booktitle}{\emph{USENIX Security
  Symposium}}.
\newblock


\bibitem[\protect\citeauthoryear{Li, Gao, and Reiter}{Li et~al\mbox{.}}{2014}]%
        {LiGaRe:14}
\bibfield{author}{\bibinfo{person}{Peng Li}, \bibinfo{person}{Debin Gao}, {and}
  \bibinfo{person}{Michael~K. Reiter}.} \bibinfo{year}{2014}\natexlab{}.
\newblock \showarticletitle{StopWatch: A Cloud Architecture for Timing Channel
  Mitigation}.
\newblock \bibinfo{journal}{\emph{ACM Trans. Inf. Syst. Secur.}}
  (\bibinfo{year}{2014}).
\newblock


\bibitem[\protect\citeauthoryear{Lipp, Gruss, Spreitzer, Maurice, and
  Mangard}{Lipp et~al\mbox{.}}{2016}]%
        {lipp2016armageddon}
\bibfield{author}{\bibinfo{person}{Moritz Lipp}, \bibinfo{person}{Daniel
  Gruss}, \bibinfo{person}{Raphael Spreitzer}, \bibinfo{person}{Cl{\'e}mentine
  Maurice}, {and} \bibinfo{person}{Stefan Mangard}.}
  \bibinfo{year}{2016}\natexlab{}.
\newblock \showarticletitle{ARMageddon: Cache attacks on mobile devices}. In
  \bibinfo{booktitle}{\emph{USENIX Security Symposium}}.
\newblock


\bibitem[\protect\citeauthoryear{Lipp, Ha{\v{z}}i{\'c}, Schwarz, Perais,
  Maurice, and Gruss}{Lipp et~al\mbox{.}}{2020}]%
        {lipp2020take}
\bibfield{author}{\bibinfo{person}{Moritz Lipp}, \bibinfo{person}{Vedad
  Ha{\v{z}}i{\'c}}, \bibinfo{person}{Michael Schwarz}, \bibinfo{person}{Arthur
  Perais}, \bibinfo{person}{Cl{\'e}mentine Maurice}, {and}
  \bibinfo{person}{Daniel Gruss}.} \bibinfo{year}{2020}\natexlab{}.
\newblock \showarticletitle{Take A Way: Exploring the Security Implications of
  AMD's Cache Way Predictors}. In \bibinfo{booktitle}{\emph{ACM Asia Conference
  on Computer and Communications Security}}.
\newblock


\bibitem[\protect\citeauthoryear{Lipp, Schwarz, Gruss, Prescher, Haas, Fogh,
  Horn, Mangard, Kocher, Genkin, Yarom, and Hamburg}{Lipp
  et~al\mbox{.}}{2018}]%
        {MoMiDa:18}
\bibfield{author}{\bibinfo{person}{Moritz Lipp}, \bibinfo{person}{Michael
  Schwarz}, \bibinfo{person}{Daniel Gruss}, \bibinfo{person}{Thomas Prescher},
  \bibinfo{person}{Werner Haas}, \bibinfo{person}{Anders Fogh},
  \bibinfo{person}{Jann Horn}, \bibinfo{person}{Stefan Mangard},
  \bibinfo{person}{Paul Kocher}, \bibinfo{person}{Daniel Genkin},
  \bibinfo{person}{Yuval Yarom}, {and} \bibinfo{person}{Mike Hamburg}.}
  \bibinfo{year}{2018}\natexlab{}.
\newblock \showarticletitle{Meltdown: Reading Kernel Memory from User Space}.
  In \bibinfo{booktitle}{\emph{USENIX Security Symposium}}.
\newblock


\bibitem[\protect\citeauthoryear{Liu and Lee}{Liu and Lee}{2014}]%
        {LiLe:14}
\bibfield{author}{\bibinfo{person}{Fangfei Liu} {and} \bibinfo{person}{Ruby~B.
  Lee}.} \bibinfo{year}{2014}\natexlab{}.
\newblock \showarticletitle{Random Fill Cache Architecture}. In
  \bibinfo{booktitle}{\emph{IEEE/ACM Intl. Symp. on Microarchitecture}}.
\newblock


\bibitem[\protect\citeauthoryear{Liu, Wu, Mai, and Lee}{Liu
  et~al\mbox{.}}{2016}]%
        {liu2016newcache}
\bibfield{author}{\bibinfo{person}{Fangfei Liu}, \bibinfo{person}{Hao Wu},
  \bibinfo{person}{Kenneth Mai}, {and} \bibinfo{person}{Ruby~B Lee}.}
  \bibinfo{year}{2016}\natexlab{}.
\newblock \showarticletitle{Newcache: Secure cache architecture thwarting cache
  side-channel attacks}.
\newblock \bibinfo{journal}{\emph{IEEE Micro}} (\bibinfo{year}{2016}).
\newblock


\bibitem[\protect\citeauthoryear{Liu, Yarom, Ge, Heiser, and Lee}{Liu
  et~al\mbox{.}}{2015}]%
        {LiYaGe:15}
\bibfield{author}{\bibinfo{person}{Fangfei Liu}, \bibinfo{person}{Yuval Yarom},
  \bibinfo{person}{Qian Ge}, \bibinfo{person}{Gernot Heiser}, {and}
  \bibinfo{person}{Ruby~B Lee}.} \bibinfo{year}{2015}\natexlab{}.
\newblock \showarticletitle{Last-level cache side-channel attacks are
  practical}. In \bibinfo{booktitle}{\emph{IEEE Symposium on Security and
  Privacy}}.
\newblock


\bibitem[\protect\citeauthoryear{Lou, Zhang, Chua, Liang, Cheng, and Zhou}{Lou
  et~al\mbox{.}}{2019}]%
        {lou2019understanding}
\bibfield{author}{\bibinfo{person}{Xiaoxuan Lou}, \bibinfo{person}{Fan Zhang},
  \bibinfo{person}{Zheng~Leong Chua}, \bibinfo{person}{Zhenkai Liang},
  \bibinfo{person}{Yueqiang Cheng}, {and} \bibinfo{person}{Yajin Zhou}.}
  \bibinfo{year}{2019}\natexlab{}.
\newblock \showarticletitle{Understanding Rowhammer attacks through the lens of
  a unified reference framework}.
\newblock \bibinfo{journal}{\emph{arXiv preprint arXiv:1901.03538}}
  (\bibinfo{year}{2019}).
\newblock


\bibitem[\protect\citeauthoryear{Lyubashevsky, Peikert, and Regev}{Lyubashevsky
  et~al\mbox{.}}{2010}]%
        {LyPeRe:10}
\bibfield{author}{\bibinfo{person}{Vadim Lyubashevsky}, \bibinfo{person}{Chris
  Peikert}, {and} \bibinfo{person}{Oded Regev}.}
  \bibinfo{year}{2010}\natexlab{}.
\newblock \showarticletitle{On ideal lattices and learning with errors over
  rings}. In \bibinfo{booktitle}{\emph{Annual International Conference on the
  Theory and Applications of Cryptographic Techniques}}.
\newblock


\bibitem[\protect\citeauthoryear{Martin, Demme, and Sethumadhavan}{Martin
  et~al\mbox{.}}{2012}]%
        {martin2012timewarp}
\bibfield{author}{\bibinfo{person}{Robert Martin}, \bibinfo{person}{John
  Demme}, {and} \bibinfo{person}{Simha Sethumadhavan}.}
  \bibinfo{year}{2012}\natexlab{}.
\newblock \showarticletitle{TimeWarp: rethinking timekeeping and performance
  monitoring mechanisms to mitigate side-channel attacks}. In
  \bibinfo{booktitle}{\emph{Annual International Symposium on Computer
  Architecture}}.
\newblock


\bibitem[\protect\citeauthoryear{Miller}{Miller}{1985}]%
        {Mi:85}
\bibfield{author}{\bibinfo{person}{Victor~S Miller}.}
  \bibinfo{year}{1985}\natexlab{}.
\newblock \showarticletitle{Use of elliptic curves in cryptography}. In
  \bibinfo{booktitle}{\emph{Conference on the theory and application of
  cryptographic techniques}}.
\newblock


\bibitem[\protect\citeauthoryear{Moghimi, Irazoqui, and Eisenbarth}{Moghimi
  et~al\mbox{.}}{2017}]%
        {moghimi2017cachezoom}
\bibfield{author}{\bibinfo{person}{Ahmad Moghimi}, \bibinfo{person}{Gorka
  Irazoqui}, {and} \bibinfo{person}{Thomas Eisenbarth}.}
  \bibinfo{year}{2017}\natexlab{}.
\newblock \showarticletitle{Cachezoom: How SGX amplifies the power of cache
  attacks}. In \bibinfo{booktitle}{\emph{International Conference on
  Cryptographic Hardware and Embedded Systems}}.
\newblock


\bibitem[\protect\citeauthoryear{Moghimi, Wichelmann, Eisenbarth, and
  Sunar}{Moghimi et~al\mbox{.}}{2019}]%
        {moghimi2019memjam}
\bibfield{author}{\bibinfo{person}{Ahmad Moghimi}, \bibinfo{person}{Jan
  Wichelmann}, \bibinfo{person}{Thomas Eisenbarth}, {and} \bibinfo{person}{Berk
  Sunar}.} \bibinfo{year}{2019}\natexlab{}.
\newblock \showarticletitle{Memjam: A false dependency attack against
  constant-time crypto implementations}.
\newblock \bibinfo{journal}{\emph{International Journal of Parallel
  Programming}} (\bibinfo{year}{2019}).
\newblock


\bibitem[\protect\citeauthoryear{Moghimi, Sunar, Eisenbarth, and
  Heninger}{Moghimi et~al\mbox{.}}{2020a}]%
        {moghimi2020tpm}
\bibfield{author}{\bibinfo{person}{Daniel Moghimi}, \bibinfo{person}{Berk
  Sunar}, \bibinfo{person}{Thomas Eisenbarth}, {and} \bibinfo{person}{Nadia
  Heninger}.} \bibinfo{year}{2020}\natexlab{a}.
\newblock \showarticletitle{TPM-FAIL:$\{$TPM$\}$ meets Timing and Lattice
  Attacks}. In \bibinfo{booktitle}{\emph{USENIX Security Symposium}}.
\newblock


\bibitem[\protect\citeauthoryear{Moghimi, Van~Bulck, Heninger, Piessens, and
  Sunar}{Moghimi et~al\mbox{.}}{2020b}]%
        {moghimi2020copycat}
\bibfield{author}{\bibinfo{person}{Daniel Moghimi}, \bibinfo{person}{Jo
  Van~Bulck}, \bibinfo{person}{Nadia Heninger}, \bibinfo{person}{Frank
  Piessens}, {and} \bibinfo{person}{Berk Sunar}.}
  \bibinfo{year}{2020}\natexlab{b}.
\newblock \showarticletitle{CopyCat: Controlled Instruction-Level Attacks on
  Enclaves}. In \bibinfo{booktitle}{\emph{29th $\{$USENIX$\}$ Security
  Symposium ($\{$USENIX$\}$ Security 20)}}. \bibinfo{pages}{469--486}.
\newblock


\bibitem[\protect\citeauthoryear{M{\"o}ller}{M{\"o}ller}{2012}]%
        {Mo:12}
\bibfield{author}{\bibinfo{person}{Bodo M{\"o}ller}.}
  \bibinfo{year}{2012}\natexlab{}.
\newblock \bibinfo{title}{Security of CBC Ciphersuites in SSL/TLS: Problems and
  Countermeasures}.
\newblock
  \bibinfo{howpublished}{\url{http://www.openssl.org/~bodo/tls-cbc.txt}}.
\newblock


\bibitem[\protect\citeauthoryear{Molnar, Piotrowski, Schultz, and
  Wagner}{Molnar et~al\mbox{.}}{2005}]%
        {MoPiSc:05}
\bibfield{author}{\bibinfo{person}{David Molnar}, \bibinfo{person}{Matt
  Piotrowski}, \bibinfo{person}{David Schultz}, {and} \bibinfo{person}{David
  Wagner}.} \bibinfo{year}{2005}\natexlab{}.
\newblock \showarticletitle{The program counter security model: Automatic
  detection and removal of control-flow side channel attacks}. In
  \bibinfo{booktitle}{\emph{International Conference on Information Security
  and Cryptology}}.
\newblock


\bibitem[\protect\citeauthoryear{Monaco}{Monaco}{2018}]%
        {Mo:18}
\bibfield{author}{\bibinfo{person}{John Monaco}.}
  \bibinfo{year}{2018}\natexlab{}.
\newblock \showarticletitle{SoK: Keylogging Side Channels}. In
  \bibinfo{booktitle}{\emph{IEEE Symposium on Security and Privacy}}.
\newblock


\bibitem[\protect\citeauthoryear{Montgomery}{Montgomery}{1987}]%
        {Mo:87}
\bibfield{author}{\bibinfo{person}{Peter~L Montgomery}.}
  \bibinfo{year}{1987}\natexlab{}.
\newblock \showarticletitle{Speeding the Pollard and elliptic curve methods of
  factorization}.
\newblock \bibinfo{journal}{\emph{Mathematics of computation}}
  (\bibinfo{year}{1987}).
\newblock


\bibitem[\protect\citeauthoryear{Mushtaq, Akram, Bhatti, Chaudhry, Lapotre, and
  Gogniat}{Mushtaq et~al\mbox{.}}{2018}]%
        {mushtaq2018nights}
\bibfield{author}{\bibinfo{person}{Maria Mushtaq}, \bibinfo{person}{Ayaz
  Akram}, \bibinfo{person}{Muhammad~Khurram Bhatti}, \bibinfo{person}{Maham
  Chaudhry}, \bibinfo{person}{Vianney Lapotre}, {and} \bibinfo{person}{Guy
  Gogniat}.} \bibinfo{year}{2018}\natexlab{}.
\newblock \showarticletitle{Nights-watch: A cache-based side-channel intrusion
  detector using hardware performance counters}. In
  \bibinfo{booktitle}{\emph{International Workshop on Hardware and
  Architectural Support for Security and Privacy}}.
\newblock


\bibitem[\protect\citeauthoryear{Mutlu and Kim}{Mutlu and Kim}{2019}]%
        {mutlu2019rowhammer}
\bibfield{author}{\bibinfo{person}{Onur Mutlu} {and} \bibinfo{person}{Jeremie~S
  Kim}.} \bibinfo{year}{2019}\natexlab{}.
\newblock \showarticletitle{RowHammer: A retrospective}.
\newblock \bibinfo{journal}{\emph{IEEE Transactions on Computer-Aided Design of
  Integrated Circuits and Systems}} (\bibinfo{year}{2019}).
\newblock


\bibitem[\protect\citeauthoryear{Nahapetian}{Nahapetian}{2016}]%
        {Na:16}
\bibfield{author}{\bibinfo{person}{Ani Nahapetian}.}
  \bibinfo{year}{2016}\natexlab{}.
\newblock \showarticletitle{Side-channel attacks on mobile and wearable
  systems}. In \bibinfo{booktitle}{\emph{IEEE Annual Consumer Communications \&
  Networking Conference}}. IEEE.
\newblock


\bibitem[\protect\citeauthoryear{Neve and Seifert}{Neve and Seifert}{2006}]%
        {neve2006advances}
\bibfield{author}{\bibinfo{person}{Michael Neve} {and}
  \bibinfo{person}{Jean-Pierre Seifert}.} \bibinfo{year}{2006}\natexlab{}.
\newblock \showarticletitle{Advances on access-driven cache attacks on AES}. In
  \bibinfo{booktitle}{\emph{International Workshop on Selected Areas in
  Cryptography}}.
\newblock


\bibitem[\protect\citeauthoryear{Neve, Seifert, and Wang}{Neve
  et~al\mbox{.}}{2006}]%
        {neve2006cache}
\bibfield{author}{\bibinfo{person}{Michael Neve}, \bibinfo{person}{Jean-Pierre
  Seifert}, {and} \bibinfo{person}{Zhenghong Wang}.}
  \bibinfo{year}{2006}\natexlab{}.
\newblock \showarticletitle{Cache time-behavior analysis on AES}.
\newblock \bibinfo{journal}{\emph{Selected Area of Cryptology}}
  (\bibinfo{year}{2006}).
\newblock


\bibitem[\protect\citeauthoryear{Osvik, Shamir, and Tromer}{Osvik
  et~al\mbox{.}}{2006}]%
        {OsShTr:06}
\bibfield{author}{\bibinfo{person}{Dag~Arne Osvik}, \bibinfo{person}{Adi
  Shamir}, {and} \bibinfo{person}{Eran Tromer}.}
  \bibinfo{year}{2006}\natexlab{}.
\newblock \showarticletitle{Cache attacks and countermeasures: the case of
  AES}. In \bibinfo{booktitle}{\emph{Cryptographers' Track at the RSA
  Conference}}.
\newblock


\bibitem[\protect\citeauthoryear{O’Hanlon and Tonge}{O’Hanlon and
  Tonge}{2005}]%
        {o2005investigation}
\bibfield{author}{\bibinfo{person}{Mair{\'e}ad O’Hanlon} {and}
  \bibinfo{person}{Anthony Tonge}.} \bibinfo{year}{2005}\natexlab{}.
\newblock \showarticletitle{Investigation of cache timing attacks on AES}.
\newblock \bibinfo{journal}{\emph{School of Computing, Dublin City University}}
  (\bibinfo{year}{2005}).
\newblock


\bibitem[\protect\citeauthoryear{Payer}{Payer}{2016}]%
        {payer2016hexpads}
\bibfield{author}{\bibinfo{person}{Mathias Payer}.}
  \bibinfo{year}{2016}\natexlab{}.
\newblock \showarticletitle{HexPADS: a platform to detect “stealth”
  attacks}. In \bibinfo{booktitle}{\emph{International Symposium on Engineering
  Secure Software and Systems}}.
\newblock


\bibitem[\protect\citeauthoryear{Peikert}{Peikert}{2010}]%
        {Pe:10}
\bibfield{author}{\bibinfo{person}{Chris Peikert}.}
  \bibinfo{year}{2010}\natexlab{}.
\newblock \showarticletitle{An efficient and parallel Gaussian sampler for
  lattices}. In \bibinfo{booktitle}{\emph{Annual Cryptology Conference}}.
\newblock


\bibitem[\protect\citeauthoryear{Percival}{Percival}{2005}]%
        {Pe:05}
\bibfield{author}{\bibinfo{person}{Colin Percival}.}
  \bibinfo{year}{2005}\natexlab{}.
\newblock \bibinfo{title}{Cache missing for fun and profit}.
\newblock
\newblock


\bibitem[\protect\citeauthoryear{Pereida~Garc{\'\i}a, Brumley, and
  Yarom}{Pereida~Garc{\'\i}a et~al\mbox{.}}{2016}]%
        {GaBrYa:16}
\bibfield{author}{\bibinfo{person}{Cesar Pereida~Garc{\'\i}a},
  \bibinfo{person}{Billy~Bob Brumley}, {and} \bibinfo{person}{Yuval Yarom}.}
  \bibinfo{year}{2016}\natexlab{}.
\newblock \showarticletitle{Make sure DSA signing exponentiations really are
  constant-time}. In \bibinfo{booktitle}{\emph{ACM Conference on Computer and
  Communications Security}}.
\newblock


\bibitem[\protect\citeauthoryear{Pessl, Bruinderink, and Yarom}{Pessl
  et~al\mbox{.}}{2017}]%
        {PeBrYa:17}
\bibfield{author}{\bibinfo{person}{Peter Pessl}, \bibinfo{person}{Leon~Groot
  Bruinderink}, {and} \bibinfo{person}{Yuval Yarom}.}
  \bibinfo{year}{2017}\natexlab{}.
\newblock \showarticletitle{To BLISS-B or not to be: Attacking strongSwan's
  Implementation of Post-Quantum Signatures}. In \bibinfo{booktitle}{\emph{ACM
  Conference on Computer and Communications Security}}.
\newblock


\bibitem[\protect\citeauthoryear{Pessl, Gruss, Maurice, Schwarz, and
  Mangard}{Pessl et~al\mbox{.}}{2016}]%
        {pessl2016drama}
\bibfield{author}{\bibinfo{person}{Peter Pessl}, \bibinfo{person}{Daniel
  Gruss}, \bibinfo{person}{Cl{\'e}mentine Maurice}, \bibinfo{person}{Michael
  Schwarz}, {and} \bibinfo{person}{Stefan Mangard}.}
  \bibinfo{year}{2016}\natexlab{}.
\newblock \showarticletitle{DRAMA: Exploiting DRAM addressing for cross-cpu
  attacks}. In \bibinfo{booktitle}{\emph{USENIX Security Symposium}}.
\newblock


\bibitem[\protect\citeauthoryear{Purnal, Giner, Gruss, and Verbauwhede}{Purnal
  et~al\mbox{.}}{2020}]%
        {purnal2020systematic}
\bibfield{author}{\bibinfo{person}{Antoon Purnal}, \bibinfo{person}{Lukas
  Giner}, \bibinfo{person}{Daniel Gruss}, {and} \bibinfo{person}{Ingrid
  Verbauwhede}.} \bibinfo{year}{2020}\natexlab{}.
\newblock \showarticletitle{Systematic analysis of randomization-based
  protected cache architectures}. In \bibinfo{booktitle}{\emph{42th IEEE
  Symposium on Security and Privacy}}, Vol.~\bibinfo{volume}{5}.
  \bibinfo{pages}{2021}.
\newblock


\bibitem[\protect\citeauthoryear{Qureshi}{Qureshi}{2018}]%
        {qureshi2018ceaser}
\bibfield{author}{\bibinfo{person}{Moinuddin~K Qureshi}.}
  \bibinfo{year}{2018}\natexlab{}.
\newblock \showarticletitle{Ceaser: Mitigating conflict-based cache attacks via
  encrypted-address and remapping}. In \bibinfo{booktitle}{\emph{Annual
  IEEE/ACM International Symposium on Microarchitecture}}.
\newblock


\bibitem[\protect\citeauthoryear{Qureshi}{Qureshi}{2019}]%
        {qureshi2019new}
\bibfield{author}{\bibinfo{person}{Moinuddin~K Qureshi}.}
  \bibinfo{year}{2019}\natexlab{}.
\newblock \showarticletitle{New attacks and defense for encrypted-address
  cache}. In \bibinfo{booktitle}{\emph{ACM/IEEE Annual International Symposium
  on Computer Architecture}}.
\newblock


\bibitem[\protect\citeauthoryear{Rebeiro, Mukhopadhyay, Takahashi, and
  Fukunaga}{Rebeiro et~al\mbox{.}}{2009}]%
        {rebeiro2009cache}
\bibfield{author}{\bibinfo{person}{Chester Rebeiro}, \bibinfo{person}{Debdeep
  Mukhopadhyay}, \bibinfo{person}{Junko Takahashi}, {and}
  \bibinfo{person}{Toshinori Fukunaga}.} \bibinfo{year}{2009}\natexlab{}.
\newblock \showarticletitle{Cache timing attacks on Clefia}. In
  \bibinfo{booktitle}{\emph{International conference on cryptology in India}}.
\newblock


\bibitem[\protect\citeauthoryear{Reparaz, Balasch, and Verbauwhede}{Reparaz
  et~al\mbox{.}}{2017}]%
        {ReBaVe:17}
\bibfield{author}{\bibinfo{person}{Oscar Reparaz}, \bibinfo{person}{Josep
  Balasch}, {and} \bibinfo{person}{Ingrid Verbauwhede}.}
  \bibinfo{year}{2017}\natexlab{}.
\newblock \showarticletitle{Dude, is my code constant time?}. In
  \bibinfo{booktitle}{\emph{IEEE Design, Automation \& Test in Europe
  Conference \& Exhibition (DATE)}}.
\newblock


\bibitem[\protect\citeauthoryear{Rivest, Shamir, and Adleman}{Rivest
  et~al\mbox{.}}{1978}]%
        {RiShAd:78}
\bibfield{author}{\bibinfo{person}{R.~L. Rivest}, \bibinfo{person}{A. Shamir},
  {and} \bibinfo{person}{L. Adleman}.} \bibinfo{year}{1978}\natexlab{}.
\newblock \showarticletitle{A Method for Obtaining Digital Signatures and
  Public-key Cryptosystems}.
\newblock \bibinfo{journal}{\emph{Commun. ACM}} (\bibinfo{year}{1978}).
\newblock


\bibitem[\protect\citeauthoryear{Rodrigues, Quint{\~a}o~Pereira, and
  Aranha}{Rodrigues et~al\mbox{.}}{2016}]%
        {RoQuAr:16}
\bibfield{author}{\bibinfo{person}{Bruno Rodrigues},
  \bibinfo{person}{Fernando~Magno Quint{\~a}o~Pereira}, {and}
  \bibinfo{person}{Diego~F Aranha}.} \bibinfo{year}{2016}\natexlab{}.
\newblock \showarticletitle{Sparse representation of implicit flows with
  applications to side-channel detection}. In \bibinfo{booktitle}{\emph{ACM
  International Conference on Compiler Construction}}.
\newblock


\bibitem[\protect\citeauthoryear{Ronen, Gillham, Genkin, Shamir, Wong, and
  Yarom}{Ronen et~al\mbox{.}}{2019}]%
        {RoGiGe:19}
\bibfield{author}{\bibinfo{person}{Eyal Ronen}, \bibinfo{person}{Robert
  Gillham}, \bibinfo{person}{Daniel Genkin}, \bibinfo{person}{Adi Shamir},
  \bibinfo{person}{David Wong}, {and} \bibinfo{person}{Yuval Yarom}.}
  \bibinfo{year}{2019}\natexlab{}.
\newblock \showarticletitle{The 9 Lives of Bleichenbacher’s CAT: New Cache
  ATtacks on TLS Implementations}. In \bibinfo{booktitle}{\emph{IEEE Symposium
  on Security and Privacy}}.
\newblock


\bibitem[\protect\citeauthoryear{Ronen, Paterson, and Shamir}{Ronen
  et~al\mbox{.}}{2018}]%
        {RoPaSh:18}
\bibfield{author}{\bibinfo{person}{Eyal Ronen}, \bibinfo{person}{Kenneth~G
  Paterson}, {and} \bibinfo{person}{Adi Shamir}.}
  \bibinfo{year}{2018}\natexlab{}.
\newblock \showarticletitle{Pseudo constant time implementations of TLS are
  only pseudo secure}. In \bibinfo{booktitle}{\emph{ACM Conference on Computer
  and Communications Security}}.
\newblock


\bibitem[\protect\citeauthoryear{Roy, Rebeiro, Hazra, and Bhunia}{Roy
  et~al\mbox{.}}{2019}]%
        {roy2019safari}
\bibfield{author}{\bibinfo{person}{Indrani Roy}, \bibinfo{person}{Chester
  Rebeiro}, \bibinfo{person}{Aritra Hazra}, {and} \bibinfo{person}{Swarup
  Bhunia}.} \bibinfo{year}{2019}\natexlab{}.
\newblock \showarticletitle{Safari: Automatic synthesis of fault-attack
  resistant block cipher implementations}.
\newblock \bibinfo{journal}{\emph{IEEE Transactions on Computer-Aided Design of
  Integrated Circuits and Systems}} (\bibinfo{year}{2019}).
\newblock


\bibitem[\protect\citeauthoryear{Saileshwar and Qureshi}{Saileshwar and
  Qureshi}{2019}]%
        {saileshwar2019lookout}
\bibfield{author}{\bibinfo{person}{Gururaj Saileshwar} {and}
  \bibinfo{person}{Moinuddin~K Qureshi}.} \bibinfo{year}{2019}\natexlab{}.
\newblock \showarticletitle{Lookout for zombies: Mitigating flush+ reload
  attack on shared caches by monitoring invalidated lines}.
\newblock \bibinfo{journal}{\emph{arXiv preprint arXiv:1906.02362}}
  (\bibinfo{year}{2019}).
\newblock


\bibitem[\protect\citeauthoryear{Sanchez and Kozyrakis}{Sanchez and
  Kozyrakis}{2011}]%
        {sanchez2011vantage}
\bibfield{author}{\bibinfo{person}{Daniel Sanchez} {and}
  \bibinfo{person}{Christos Kozyrakis}.} \bibinfo{year}{2011}\natexlab{}.
\newblock \showarticletitle{Vantage: scalable and efficient fine-grain cache
  partitioning}. In \bibinfo{booktitle}{\emph{Annual International Symposium on
  Computer Architecture}}.
\newblock


\bibitem[\protect\citeauthoryear{Sari, Demir, and Kucuk}{Sari
  et~al\mbox{.}}{2019}]%
        {sari2019fairsdp}
\bibfield{author}{\bibinfo{person}{Sercan Sari}, \bibinfo{person}{Onur Demir},
  {and} \bibinfo{person}{Gurhan Kucuk}.} \bibinfo{year}{2019}\natexlab{}.
\newblock \showarticletitle{FairSDP: Fair and Secure Dynamic Cache
  Partitioning}. In \bibinfo{booktitle}{\emph{International Conference on
  Computer Science and Engineering}}.
\newblock


\bibitem[\protect\citeauthoryear{Schwarz and Gruss}{Schwarz and Gruss}{2020}]%
        {schwarz2020trusted}
\bibfield{author}{\bibinfo{person}{Michael Schwarz} {and}
  \bibinfo{person}{Daniel Gruss}.} \bibinfo{year}{2020}\natexlab{}.
\newblock \showarticletitle{How Trusted Execution Environments Fuel Research on
  Microarchitectural Attacks}.
\newblock \bibinfo{journal}{\emph{IEEE Security \& Privacy}}
  (\bibinfo{year}{2020}).
\newblock


\bibitem[\protect\citeauthoryear{Schwarz, Lackner, and Gruss}{Schwarz
  et~al\mbox{.}}{2019a}]%
        {schwarz2019javascript}
\bibfield{author}{\bibinfo{person}{Michael Schwarz}, \bibinfo{person}{Florian
  Lackner}, {and} \bibinfo{person}{Daniel Gruss}.}
  \bibinfo{year}{2019}\natexlab{a}.
\newblock \showarticletitle{JavaScript Template Attacks: Automatically
  Inferring Host Information for Targeted Exploits}. In
  \bibinfo{booktitle}{\emph{Network and Distributed System Security
  Symposium}}.
\newblock


\bibitem[\protect\citeauthoryear{Schwarz, Lipp, and Gruss}{Schwarz
  et~al\mbox{.}}{2018}]%
        {schwarz2018javascript}
\bibfield{author}{\bibinfo{person}{Michael Schwarz}, \bibinfo{person}{Moritz
  Lipp}, {and} \bibinfo{person}{Daniel Gruss}.}
  \bibinfo{year}{2018}\natexlab{}.
\newblock \showarticletitle{JavaScript Zero: Real JavaScript and Zero
  Side-Channel Attacks}. In \bibinfo{booktitle}{\emph{Network and Distributed
  System Security Symposium}}.
\newblock


\bibitem[\protect\citeauthoryear{Schwarz, Schwarzl, Lipp, Masters, and
  Gruss}{Schwarz et~al\mbox{.}}{2019b}]%
        {schwarz2019netspectre}
\bibfield{author}{\bibinfo{person}{Michael Schwarz}, \bibinfo{person}{Martin
  Schwarzl}, \bibinfo{person}{Moritz Lipp}, \bibinfo{person}{Jon Masters},
  {and} \bibinfo{person}{Daniel Gruss}.} \bibinfo{year}{2019}\natexlab{b}.
\newblock \showarticletitle{Netspectre: Read arbitrary memory over network}. In
  \bibinfo{booktitle}{\emph{European Symposium on Research in Computer
  Security}}.
\newblock


\bibitem[\protect\citeauthoryear{Schwarz, Weiser, Gruss, Maurice, and
  Mangard}{Schwarz et~al\mbox{.}}{2017}]%
        {schwarz2017malware}
\bibfield{author}{\bibinfo{person}{Michael Schwarz}, \bibinfo{person}{Samuel
  Weiser}, \bibinfo{person}{Daniel Gruss}, \bibinfo{person}{Cl{\'e}mentine
  Maurice}, {and} \bibinfo{person}{Stefan Mangard}.}
  \bibinfo{year}{2017}\natexlab{}.
\newblock \showarticletitle{Malware guard extension: Using SGX to conceal cache
  attacks}. In \bibinfo{booktitle}{\emph{International Conference on Detection
  of Intrusions and Malware, and Vulnerability Assessment}}.
\newblock


\bibitem[\protect\citeauthoryear{Shi, Song, Chen, and Zang}{Shi
  et~al\mbox{.}}{2011}]%
        {ShSoCh:11}
\bibfield{author}{\bibinfo{person}{Jicheng Shi}, \bibinfo{person}{Xiang Song},
  \bibinfo{person}{Haibo Chen}, {and} \bibinfo{person}{Binyu Zang}.}
  \bibinfo{year}{2011}\natexlab{}.
\newblock \showarticletitle{Limiting cache-based side-channel in multi-tenant
  cloud using dynamic page coloring}. In \bibinfo{booktitle}{\emph{IEEE/IFIP
  Intl. Conf. on Dependable Systems and Networks Workshops}}.
\newblock


\bibitem[\protect\citeauthoryear{Shih, Lee, Kim, and Peinado}{Shih
  et~al\mbox{.}}{2017}]%
        {shih2017t}
\bibfield{author}{\bibinfo{person}{Ming-Wei Shih}, \bibinfo{person}{Sangho
  Lee}, \bibinfo{person}{Taesoo Kim}, {and} \bibinfo{person}{Marcus Peinado}.}
  \bibinfo{year}{2017}\natexlab{}.
\newblock \showarticletitle{T-SGX: Eradicating Controlled-Channel Attacks
  Against Enclave Programs.}. In \bibinfo{booktitle}{\emph{Network and
  Distributed System Security Symposium}}.
\newblock


\bibitem[\protect\citeauthoryear{Shin, Kim, Kwon, Jeong, and Hur}{Shin
  et~al\mbox{.}}{2018}]%
        {ShKiKw:18}
\bibfield{author}{\bibinfo{person}{Youngjoo Shin}, \bibinfo{person}{Hyung~Chan
  Kim}, \bibinfo{person}{Dokeun Kwon}, \bibinfo{person}{Ji~Hoon Jeong}, {and}
  \bibinfo{person}{Junbeom Hur}.} \bibinfo{year}{2018}\natexlab{}.
\newblock \showarticletitle{Unveiling Hardware-based Data Prefetcher, a Hidden
  Source of Information Leakage}. In \bibinfo{booktitle}{\emph{ACM Conference
  on Computer and Communications Security}}.
\newblock


\bibitem[\protect\citeauthoryear{Shinde, Chua, Narayanan, and Saxena}{Shinde
  et~al\mbox{.}}{2016}]%
        {shinde2016preventing}
\bibfield{author}{\bibinfo{person}{Shweta Shinde}, \bibinfo{person}{Zheng~Leong
  Chua}, \bibinfo{person}{Viswesh Narayanan}, {and} \bibinfo{person}{Prateek
  Saxena}.} \bibinfo{year}{2016}\natexlab{}.
\newblock \showarticletitle{Preventing page faults from telling your secrets}.
  In \bibinfo{booktitle}{\emph{ACM on Asia Conference on Computer and
  Communications Security}}.
\newblock


\bibitem[\protect\citeauthoryear{Silverman and Whyte}{Silverman and
  Whyte}{2007}]%
        {SiWh:07}
\bibfield{author}{\bibinfo{person}{Joseph~H Silverman} {and}
  \bibinfo{person}{William Whyte}.} \bibinfo{year}{2007}\natexlab{}.
\newblock \showarticletitle{Timing attacks on NTRUEncrypt via variation in the
  number of hash calls}. In \bibinfo{booktitle}{\emph{Cryptographers’ Track
  at the RSA Conference}}.
\newblock


\bibitem[\protect\citeauthoryear{Song, Wagner, and Tian}{Song
  et~al\mbox{.}}{2001}]%
        {SoWaTi:01}
\bibfield{author}{\bibinfo{person}{Dawn~Xiaodong Song},
  \bibinfo{person}{David~A Wagner}, {and} \bibinfo{person}{Xuqing Tian}.}
  \bibinfo{year}{2001}\natexlab{}.
\newblock \showarticletitle{Timing analysis of keystrokes and timing attacks on
  ssh.}. In \bibinfo{booktitle}{\emph{USENIX Security Symposium}}.
\newblock


\bibitem[\protect\citeauthoryear{Sprabery, Evchenko, Raj, Bobba, Mohan, and
  Campbell}{Sprabery et~al\mbox{.}}{2018}]%
        {sprabery2018scheduling}
\bibfield{author}{\bibinfo{person}{Read Sprabery}, \bibinfo{person}{Konstantin
  Evchenko}, \bibinfo{person}{Abhilash Raj}, \bibinfo{person}{Rakesh~B Bobba},
  \bibinfo{person}{Sibin Mohan}, {and} \bibinfo{person}{Roy Campbell}.}
  \bibinfo{year}{2018}\natexlab{}.
\newblock \showarticletitle{Scheduling, isolation, and cache allocation: A
  side-channel defense}. In \bibinfo{booktitle}{\emph{IEEE International
  Conference on Cloud Engineering}}.
\newblock


\bibitem[\protect\citeauthoryear{Spreitzer, Moonsamy, Korak, and
  Mangard}{Spreitzer et~al\mbox{.}}{2018a}]%
        {SpMoKo:18}
\bibfield{author}{\bibinfo{person}{R. Spreitzer}, \bibinfo{person}{V.
  Moonsamy}, \bibinfo{person}{T. Korak}, {and} \bibinfo{person}{S. Mangard}.}
  \bibinfo{year}{2018}\natexlab{a}.
\newblock \showarticletitle{Systematic Classification of Side-Channel Attacks:
  A Case Study for Mobile Devices}.
\newblock \bibinfo{journal}{\emph{IEEE Communications Surveys Tutorials}}
  (\bibinfo{year}{2018}).
\newblock


\bibitem[\protect\citeauthoryear{Spreitzer, Palfinger, and Mangard}{Spreitzer
  et~al\mbox{.}}{2018b}]%
        {SpPaMa:18}
\bibfield{author}{\bibinfo{person}{Raphael Spreitzer}, \bibinfo{person}{Gerald
  Palfinger}, {and} \bibinfo{person}{Stefan Mangard}.}
  \bibinfo{year}{2018}\natexlab{b}.
\newblock \showarticletitle{SCAnDroid: Automated Side-Channel Analysis of
  Android APIs}. In \bibinfo{booktitle}{\emph{ACM Conference on Security \&
  Privacy in Wireless and Mobile Networks}}.
\newblock


\bibitem[\protect\citeauthoryear{Srivastava, Slpsk, Roy, Rebeiro, Hazra, and
  Bhunia}{Srivastava et~al\mbox{.}}{2020}]%
        {srivastava2020solomon}
\bibfield{author}{\bibinfo{person}{Milind Srivastava},
  \bibinfo{person}{Patanjali Slpsk}, \bibinfo{person}{Indrani Roy},
  \bibinfo{person}{Chester Rebeiro}, \bibinfo{person}{Aritra Hazra}, {and}
  \bibinfo{person}{Swarup Bhunia}.} \bibinfo{year}{2020}\natexlab{}.
\newblock \showarticletitle{SOLOMON: An automated framework for detecting fault
  attack vulnerabilities in hardware}. In \bibinfo{booktitle}{\emph{Design,
  Automation \& Test in Europe Conference \& Exhibition}}.
\newblock


\bibitem[\protect\citeauthoryear{Strackx and Piessens}{Strackx and
  Piessens}{2017}]%
        {strackx2017heisenberg}
\bibfield{author}{\bibinfo{person}{Raoul Strackx} {and} \bibinfo{person}{Frank
  Piessens}.} \bibinfo{year}{2017}\natexlab{}.
\newblock \showarticletitle{The Heisenberg defense: Proactively defending SGX
  enclaves against page-table-based side-channel attacks}.
\newblock \bibinfo{journal}{\emph{arXiv preprint arXiv:1712.08519}}
  (\bibinfo{year}{2017}).
\newblock


\bibitem[\protect\citeauthoryear{Szefer}{Szefer}{2018}]%
        {Sz:18}
\bibfield{author}{\bibinfo{person}{Jakub Szefer}.}
  \bibinfo{year}{2018}\natexlab{}.
\newblock \showarticletitle{Survey of Microarchitectural Side and Covert
  Channels, Attacks, and Defenses}.
\newblock \bibinfo{journal}{\emph{Journal of Hardware and Systems Security}}
  (\bibinfo{year}{2018}).
\newblock


\bibitem[\protect\citeauthoryear{Tan, Zeng, Bu, and Ren}{Tan
  et~al\mbox{.}}{2020}]%
        {tanphantomcache}
\bibfield{author}{\bibinfo{person}{Qinhan Tan}, \bibinfo{person}{Zhihua Zeng},
  \bibinfo{person}{Kai Bu}, {and} \bibinfo{person}{Kui Ren}.}
  \bibinfo{year}{2020}\natexlab{}.
\newblock \showarticletitle{PhantomCache: Obfuscating cache conflicts with
  localized randomization}. In \bibinfo{booktitle}{\emph{Network and
  Distributed System Security Symposium}}.
\newblock


\bibitem[\protect\citeauthoryear{Tibouchi and Wallet}{Tibouchi and
  Wallet}{2019}]%
        {TiWa:19}
\bibfield{author}{\bibinfo{person}{Mehdi Tibouchi} {and}
  \bibinfo{person}{Alexandre Wallet}.} \bibinfo{year}{2019}\natexlab{}.
\newblock \bibinfo{booktitle}{\emph{One Bit is All It Takes: A Devastating
  Timing Attack on BLISS’s Non-Constant Time Sign Flips}}.
\newblock \bibinfo{type}{{T}echnical {R}eport}.
  \bibinfo{institution}{Cryptology ePrint Archive, Report 2019/898}.
\newblock


\bibitem[\protect\citeauthoryear{Trilla, Hernandez, Abella, and Cazorla}{Trilla
  et~al\mbox{.}}{2018}]%
        {trilla2018cache}
\bibfield{author}{\bibinfo{person}{David Trilla}, \bibinfo{person}{Carles
  Hernandez}, \bibinfo{person}{Jaume Abella}, {and}
  \bibinfo{person}{Francisco~J Cazorla}.} \bibinfo{year}{2018}\natexlab{}.
\newblock \showarticletitle{Cache side-channel attacks and time-predictability
  in high-performance critical real-time systems}. In
  \bibinfo{booktitle}{\emph{Annual Design Automation Conference}}.
\newblock


\bibitem[\protect\citeauthoryear{Trippel, Lustig, and Martonosi}{Trippel
  et~al\mbox{.}}{2018}]%
        {TrLuMa:18}
\bibfield{author}{\bibinfo{person}{Caroline Trippel}, \bibinfo{person}{Daniel
  Lustig}, {and} \bibinfo{person}{Margaret Martonosi}.}
  \bibinfo{year}{2018}\natexlab{}.
\newblock \showarticletitle{Checkmate: Automated synthesis of hardware exploits
  and security litmus tests}. In \bibinfo{booktitle}{\emph{Annual IEEE/ACM
  International Symposium on Microarchitecture}}.
\newblock


\bibitem[\protect\citeauthoryear{Tsunoo}{Tsunoo}{2002}]%
        {tsunoo2002crypt}
\bibfield{author}{\bibinfo{person}{Yukiyasu Tsunoo}.}
  \bibinfo{year}{2002}\natexlab{}.
\newblock \showarticletitle{Crypt-analysis of block ciphers implemented on
  computers with cache}.
\newblock \bibinfo{journal}{\emph{Proc. ISITA2002, Oct.}}
  (\bibinfo{year}{2002}).
\newblock


\bibitem[\protect\citeauthoryear{Tsunoo, Saito, Suzaki, Shigeri, and
  Miyauchi}{Tsunoo et~al\mbox{.}}{2003}]%
        {tsunoo2003cryptanalysis}
\bibfield{author}{\bibinfo{person}{Yukiyasu Tsunoo}, \bibinfo{person}{Teruo
  Saito}, \bibinfo{person}{Tomoyasu Suzaki}, \bibinfo{person}{Maki Shigeri},
  {and} \bibinfo{person}{Hiroshi Miyauchi}.} \bibinfo{year}{2003}\natexlab{}.
\newblock \showarticletitle{Cryptanalysis of DES implemented on computers with
  cache}. In \bibinfo{booktitle}{\emph{International Workshop on Cryptographic
  Hardware and Embedded Systems}}.
\newblock


\bibitem[\protect\citeauthoryear{Tunstall}{Tunstall}{2017}]%
        {Tu:17}
\bibfield{author}{\bibinfo{person}{Michael Tunstall}.}
  \bibinfo{year}{2017}\natexlab{}.
\newblock \showarticletitle{Smart card security}.
\newblock In \bibinfo{booktitle}{\emph{Smart Cards, Tokens, Security and
  Applications}}.
\newblock


\bibitem[\protect\citeauthoryear{Ullrich, Zseby, Fabini, and Weippl}{Ullrich
  et~al\mbox{.}}{2017}]%
        {UlZsFa:17}
\bibfield{author}{\bibinfo{person}{Johanna Ullrich}, \bibinfo{person}{Tanja
  Zseby}, \bibinfo{person}{Joachim Fabini}, {and} \bibinfo{person}{Edgar
  Weippl}.} \bibinfo{year}{2017}\natexlab{}.
\newblock \showarticletitle{Network-based secret communication in clouds: A
  survey}.
\newblock \bibinfo{journal}{\emph{IEEE Communications Surveys \& Tutorials}}
  (\bibinfo{year}{2017}).
\newblock


\bibitem[\protect\citeauthoryear{Van~Bulck, Piessens, and Strackx}{Van~Bulck
  et~al\mbox{.}}{2017a}]%
        {van2017sgx}
\bibfield{author}{\bibinfo{person}{Jo Van~Bulck}, \bibinfo{person}{Frank
  Piessens}, {and} \bibinfo{person}{Raoul Strackx}.}
  \bibinfo{year}{2017}\natexlab{a}.
\newblock \showarticletitle{SGX-Step: A practical attack framework for precise
  enclave execution control}. In \bibinfo{booktitle}{\emph{Workshop on System
  Software for Trusted Execution}}.
\newblock


\bibitem[\protect\citeauthoryear{Van~Bulck, Weichbrodt, Kapitza, Piessens, and
  Strackx}{Van~Bulck et~al\mbox{.}}{2017b}]%
        {van2017telling}
\bibfield{author}{\bibinfo{person}{Jo Van~Bulck}, \bibinfo{person}{Nico
  Weichbrodt}, \bibinfo{person}{R{\"u}diger Kapitza}, \bibinfo{person}{Frank
  Piessens}, {and} \bibinfo{person}{Raoul Strackx}.}
  \bibinfo{year}{2017}\natexlab{b}.
\newblock \showarticletitle{Telling your secrets without page faults: Stealthy
  page table-based attacks on enclaved execution}. In
  \bibinfo{booktitle}{\emph{USENIX Security Symposium}}.
\newblock


\bibitem[\protect\citeauthoryear{van~de Pol, Smart, and Yarom}{van~de Pol
  et~al\mbox{.}}{2015}]%
        {VaSmYa:15}
\bibfield{author}{\bibinfo{person}{Joop van~de Pol}, \bibinfo{person}{Nigel~P
  Smart}, {and} \bibinfo{person}{Yuval Yarom}.}
  \bibinfo{year}{2015}\natexlab{}.
\newblock \showarticletitle{Just a little bit more}. In
  \bibinfo{booktitle}{\emph{Cryptographers' Track at the RSA Conference}}.
\newblock


\bibitem[\protect\citeauthoryear{Varadarajan, Ristenpart, and
  Swift}{Varadarajan et~al\mbox{.}}{2014}]%
        {varadarajan2014scheduler}
\bibfield{author}{\bibinfo{person}{Venkatanathan Varadarajan},
  \bibinfo{person}{Thomas Ristenpart}, {and} \bibinfo{person}{Michael Swift}.}
  \bibinfo{year}{2014}\natexlab{}.
\newblock \showarticletitle{Scheduler-based defenses against cross-VM
  side-channels}. In \bibinfo{booktitle}{\emph{USENIX Security Symposium}}.
\newblock


\bibitem[\protect\citeauthoryear{Vattikonda, Das, and Shacham}{Vattikonda
  et~al\mbox{.}}{2011}]%
        {VaDaSh:11}
\bibfield{author}{\bibinfo{person}{Bhanu~C. Vattikonda},
  \bibinfo{person}{Sambit Das}, {and} \bibinfo{person}{Hovav Shacham}.}
  \bibinfo{year}{2011}\natexlab{}.
\newblock \showarticletitle{Eliminating Fine Grained Timers in Xen}. In
  \bibinfo{booktitle}{\emph{ACM Workshop on Cloud Computing Security}}.
\newblock


\bibitem[\protect\citeauthoryear{Vaudenay}{Vaudenay}{2002}]%
        {Va:02}
\bibfield{author}{\bibinfo{person}{Serge Vaudenay}.}
  \bibinfo{year}{2002}\natexlab{}.
\newblock \showarticletitle{Security Flaws Induced by CBC
  Padding—Applications to SSL, IPSEC, WTLS...}. In
  \bibinfo{booktitle}{\emph{International Conference on the Theory and
  Applications of Cryptographic Techniques}}.
\newblock


\bibitem[\protect\citeauthoryear{Wang, Zhu, Wang, and Meng}{Wang
  et~al\mbox{.}}{2020}]%
        {wang2020analyzing}
\bibfield{author}{\bibinfo{person}{Limin Wang}, \bibinfo{person}{Ziyuan Zhu},
  \bibinfo{person}{Zhanpeng Wang}, {and} \bibinfo{person}{Dan Meng}.}
  \bibinfo{year}{2020}\natexlab{}.
\newblock \showarticletitle{Analyzing The Security of The Cache Side Channel
  Defences With Attack Graphs}. In \bibinfo{booktitle}{\emph{Asia and South
  Pacific Design Automation Conference}}.
\newblock


\bibitem[\protect\citeauthoryear{Wang and Chen}{Wang and Chen}{2014}]%
        {wang2014futility}
\bibfield{author}{\bibinfo{person}{Ruisheng Wang} {and}
  \bibinfo{person}{Lizhong Chen}.} \bibinfo{year}{2014}\natexlab{}.
\newblock \showarticletitle{Futility scaling: High-associativity cache
  partitioning}. In \bibinfo{booktitle}{\emph{Annual IEEE/ACM International
  Symposium on Microarchitecture}}.
\newblock


\bibitem[\protect\citeauthoryear{Wang, Bao, Liu, Wang, Zhang, and Wu}{Wang
  et~al\mbox{.}}{2019}]%
        {WaBaLi:19}
\bibfield{author}{\bibinfo{person}{Shuai Wang}, \bibinfo{person}{Yuyan Bao},
  \bibinfo{person}{Xiao Liu}, \bibinfo{person}{Pei Wang},
  \bibinfo{person}{Danfeng Zhang}, {and} \bibinfo{person}{Dinghao Wu}.}
  \bibinfo{year}{2019}\natexlab{}.
\newblock \showarticletitle{Identifying Cache-Based Side Channels through
  Secret-Augmented Abstract Interpretation}. In
  \bibinfo{booktitle}{\emph{USENIX Security Symposium}}.
\newblock


\bibitem[\protect\citeauthoryear{Wang, Wang, Liu, Zhang, and Wu}{Wang
  et~al\mbox{.}}{2017b}]%
        {WaWaLi:17}
\bibfield{author}{\bibinfo{person}{Shuai Wang}, \bibinfo{person}{Pei Wang},
  \bibinfo{person}{Xiao Liu}, \bibinfo{person}{Danfeng Zhang}, {and}
  \bibinfo{person}{Dinghao Wu}.} \bibinfo{year}{2017}\natexlab{b}.
\newblock \showarticletitle{Cached: Identifying cache-based timing channels in
  production software}. In \bibinfo{booktitle}{\emph{USENIX Security
  Symposium}}.
\newblock


\bibitem[\protect\citeauthoryear{Wang, Chen, Pan, Zhang, Wang, Bindschaedler,
  Tang, and Gunter}{Wang et~al\mbox{.}}{2017a}]%
        {wang2017leaky}
\bibfield{author}{\bibinfo{person}{Wenhao Wang}, \bibinfo{person}{Guoxing
  Chen}, \bibinfo{person}{Xiaorui Pan}, \bibinfo{person}{Yinqian Zhang},
  \bibinfo{person}{XiaoFeng Wang}, \bibinfo{person}{Vincent Bindschaedler},
  \bibinfo{person}{Haixu Tang}, {and} \bibinfo{person}{Carl~A Gunter}.}
  \bibinfo{year}{2017}\natexlab{a}.
\newblock \showarticletitle{Leaky cauldron on the dark land: Understanding
  memory side-channel hazards in SGX}. In \bibinfo{booktitle}{\emph{ACM SIGSAC
  Conference on Computer and Communications Security}}.
\newblock


\bibitem[\protect\citeauthoryear{Wang, Ferraiuolo, and Suh}{Wang
  et~al\mbox{.}}{2014}]%
        {wang2014timing}
\bibfield{author}{\bibinfo{person}{Yao Wang}, \bibinfo{person}{Andrew
  Ferraiuolo}, {and} \bibinfo{person}{G~Edward Suh}.}
  \bibinfo{year}{2014}\natexlab{}.
\newblock \showarticletitle{Timing channel protection for a shared memory
  controller}. In \bibinfo{booktitle}{\emph{IEEE International Symposium on
  High Performance Computer Architecture}}.
\newblock


\bibitem[\protect\citeauthoryear{Wang, Ferraiuolo, Zhang, Myers, and Suh}{Wang
  et~al\mbox{.}}{2016}]%
        {wang2016secdcp}
\bibfield{author}{\bibinfo{person}{Yao Wang}, \bibinfo{person}{Andrew
  Ferraiuolo}, \bibinfo{person}{Danfeng Zhang}, \bibinfo{person}{Andrew~C
  Myers}, {and} \bibinfo{person}{G~Edward Suh}.}
  \bibinfo{year}{2016}\natexlab{}.
\newblock \showarticletitle{SecDCP: secure dynamic cache partitioning for
  efficient timing channel protection}. In \bibinfo{booktitle}{\emph{Annual
  Design Automation Conference}}.
\newblock


\bibitem[\protect\citeauthoryear{Wang and Suh}{Wang and Suh}{2012}]%
        {wang2012efficient}
\bibfield{author}{\bibinfo{person}{Yao Wang} {and} \bibinfo{person}{G~Edward
  Suh}.} \bibinfo{year}{2012}\natexlab{}.
\newblock \showarticletitle{Efficient timing channel protection for on-chip
  networks}. In \bibinfo{booktitle}{\emph{IEEE/ACM Sixth International
  Symposium on Networks-on-Chip}}.
\newblock


\bibitem[\protect\citeauthoryear{Wang, Wu, and Suh}{Wang
  et~al\mbox{.}}{2017c}]%
        {wang2017secure}
\bibfield{author}{\bibinfo{person}{Yao Wang}, \bibinfo{person}{Benjamin Wu},
  {and} \bibinfo{person}{G~Edward Suh}.} \bibinfo{year}{2017}\natexlab{c}.
\newblock \showarticletitle{Secure dynamic memory scheduling against timing
  channel attacks}. In \bibinfo{booktitle}{\emph{IEEE International Symposium
  on High Performance Computer Architecture}}.
\newblock


\bibitem[\protect\citeauthoryear{Wang and Lee}{Wang and Lee}{2006}]%
        {wang2006covert}
\bibfield{author}{\bibinfo{person}{Zhenghong Wang} {and}
  \bibinfo{person}{Ruby~B Lee}.} \bibinfo{year}{2006}\natexlab{}.
\newblock \showarticletitle{Covert and side channels due to processor
  architecture}. In \bibinfo{booktitle}{\emph{Annual Computer Security
  Applications Conference}}.
\newblock


\bibitem[\protect\citeauthoryear{Wang and Lee}{Wang and Lee}{2007}]%
        {WaLe:07}
\bibfield{author}{\bibinfo{person}{Zhenghong Wang} {and}
  \bibinfo{person}{Ruby~B. Lee}.} \bibinfo{year}{2007}\natexlab{}.
\newblock \showarticletitle{New Cache Designs for Thwarting Software
  Cache-based Side Channel Attacks}. In \bibinfo{booktitle}{\emph{ACM Intl.
  Symp. on Computer Architecture}}.
\newblock


\bibitem[\protect\citeauthoryear{Wang and Lee}{Wang and Lee}{2008}]%
        {wang2008novel}
\bibfield{author}{\bibinfo{person}{Zhenghong Wang} {and}
  \bibinfo{person}{Ruby~B Lee}.} \bibinfo{year}{2008}\natexlab{}.
\newblock \showarticletitle{A novel cache architecture with enhanced
  performance and security}. In \bibinfo{booktitle}{\emph{IEEE/ACM
  International Symposium on Microarchitecture}}.
\newblock


\bibitem[\protect\citeauthoryear{Weiser, Spreitzer, and Bodner}{Weiser
  et~al\mbox{.}}{2018a}]%
        {WeSpBo:18}
\bibfield{author}{\bibinfo{person}{Samuel Weiser}, \bibinfo{person}{Raphael
  Spreitzer}, {and} \bibinfo{person}{Lukas Bodner}.}
  \bibinfo{year}{2018}\natexlab{a}.
\newblock \showarticletitle{Single Trace Attack Against RSA Key Generation in
  Intel SGX SSL}. In \bibinfo{booktitle}{\emph{Asia Conference on Computer and
  Communications Security}}.
\newblock


\bibitem[\protect\citeauthoryear{Weiser, Zankl, Spreitzer, Miller, Mangard, and
  Sigl}{Weiser et~al\mbox{.}}{2018b}]%
        {WeZaSp:18}
\bibfield{author}{\bibinfo{person}{Samuel Weiser}, \bibinfo{person}{Andreas
  Zankl}, \bibinfo{person}{Raphael Spreitzer}, \bibinfo{person}{Katja Miller},
  \bibinfo{person}{Stefan Mangard}, {and} \bibinfo{person}{Georg Sigl}.}
  \bibinfo{year}{2018}\natexlab{b}.
\newblock \showarticletitle{DATA--differential address trace analysis: finding
  address-based side-channels in binaries}. In \bibinfo{booktitle}{\emph{USENIX
  Security Symposium}}.
\newblock


\bibitem[\protect\citeauthoryear{Werner, Unterluggauer, Giner, Schwarz, Gruss,
  and Mangard}{Werner et~al\mbox{.}}{2019}]%
        {werner2019scattercache}
\bibfield{author}{\bibinfo{person}{Mario Werner}, \bibinfo{person}{Thomas
  Unterluggauer}, \bibinfo{person}{Lukas Giner}, \bibinfo{person}{Michael
  Schwarz}, \bibinfo{person}{Daniel Gruss}, {and} \bibinfo{person}{Stefan
  Mangard}.} \bibinfo{year}{2019}\natexlab{}.
\newblock \showarticletitle{ScatterCache: thwarting cache attacks via cache set
  randomization}. In \bibinfo{booktitle}{\emph{USENIX Security Symposium}}.
\newblock


\bibitem[\protect\citeauthoryear{Wichelmann, Moghimi, Eisenbarth, and
  Sunar}{Wichelmann et~al\mbox{.}}{2018}]%
        {WiMoEi:18}
\bibfield{author}{\bibinfo{person}{Jan Wichelmann}, \bibinfo{person}{Ahmad
  Moghimi}, \bibinfo{person}{Thomas Eisenbarth}, {and} \bibinfo{person}{Berk
  Sunar}.} \bibinfo{year}{2018}\natexlab{}.
\newblock \showarticletitle{MicroWalk: A Framework for Finding Side Channels in
  Binaries}. In \bibinfo{booktitle}{\emph{ACM Annual Computer Security
  Applications Conference}}.
\newblock


\bibitem[\protect\citeauthoryear{Wu, Guo, Schaumont, and Wang}{Wu
  et~al\mbox{.}}{2018}]%
        {wu2018eliminating}
\bibfield{author}{\bibinfo{person}{Meng Wu}, \bibinfo{person}{Shengjian Guo},
  \bibinfo{person}{Patrick Schaumont}, {and} \bibinfo{person}{Chao Wang}.}
  \bibinfo{year}{2018}\natexlab{}.
\newblock \showarticletitle{Eliminating timing side-channel leaks using program
  repair}. In \bibinfo{booktitle}{\emph{ACM SIGSOFT International Symposium on
  Software Testing and Analysis}}.
\newblock


\bibitem[\protect\citeauthoryear{Xiao, Li, Chen, and Zhang}{Xiao
  et~al\mbox{.}}{2017}]%
        {XiLiCh:17}
\bibfield{author}{\bibinfo{person}{Yuan Xiao}, \bibinfo{person}{Mengyuan Li},
  \bibinfo{person}{Sanchuan Chen}, {and} \bibinfo{person}{Yinqian Zhang}.}
  \bibinfo{year}{2017}\natexlab{}.
\newblock \showarticletitle{Stacco: Differentially analyzing side-channel
  traces for detecting SSL/TLS vulnerabilities in secure enclaves}. In
  \bibinfo{booktitle}{\emph{ACM Conference on Computer and Communications
  Security}}.
\newblock


\bibitem[\protect\citeauthoryear{Xiong and Szefer}{Xiong and Szefer}{2020}]%
        {xiong2020leaking}
\bibfield{author}{\bibinfo{person}{Wenjie Xiong} {and} \bibinfo{person}{Jakub
  Szefer}.} \bibinfo{year}{2020}\natexlab{}.
\newblock \showarticletitle{Leaking information through cache LRU states}. In
  \bibinfo{booktitle}{\emph{IEEE International Symposium on High Performance
  Computer Architecture}}.
\newblock


\bibitem[\protect\citeauthoryear{Xu, Cui, and Peinado}{Xu
  et~al\mbox{.}}{2015}]%
        {xu2015controlled}
\bibfield{author}{\bibinfo{person}{Yuanzhong Xu}, \bibinfo{person}{Weidong
  Cui}, {and} \bibinfo{person}{Marcus Peinado}.}
  \bibinfo{year}{2015}\natexlab{}.
\newblock \showarticletitle{Controlled-channel attacks: Deterministic side
  channels for untrusted operating systems}. In \bibinfo{booktitle}{\emph{IEEE
  Symposium on Security and Privacy}}.
\newblock


\bibitem[\protect\citeauthoryear{Yan, Gopireddy, Shull, and Torrellas}{Yan
  et~al\mbox{.}}{2017}]%
        {yan2017secure}
\bibfield{author}{\bibinfo{person}{Mengjia Yan}, \bibinfo{person}{Bhargava
  Gopireddy}, \bibinfo{person}{Thomas Shull}, {and} \bibinfo{person}{Josep
  Torrellas}.} \bibinfo{year}{2017}\natexlab{}.
\newblock \showarticletitle{Secure hierarchy-aware cache replacement policy
  (SHARP): Defending against cache-based side channel attacks}. In
  \bibinfo{booktitle}{\emph{ACM/IEEE Annual International Symposium on Computer
  Architecture}}.
\newblock


\bibitem[\protect\citeauthoryear{Yarom and Benger}{Yarom and Benger}{2014}]%
        {YaBe:14}
\bibfield{author}{\bibinfo{person}{Yuval Yarom} {and} \bibinfo{person}{Naomi
  Benger}.} \bibinfo{year}{2014}\natexlab{}.
\newblock \showarticletitle{Recovering OpenSSL ECDSA Nonces Using the FLUSH+
  RELOAD Cache Side-channel Attack.}
\newblock \bibinfo{journal}{\emph{IACR Cryptology ePrint Archive}}
  \bibinfo{volume}{2014} (\bibinfo{year}{2014}).
\newblock


\bibitem[\protect\citeauthoryear{Yarom and Falkner}{Yarom and Falkner}{2014}]%
        {YaFa:14}
\bibfield{author}{\bibinfo{person}{Yuval Yarom} {and} \bibinfo{person}{Katrina
  Falkner}.} \bibinfo{year}{2014}\natexlab{}.
\newblock \showarticletitle{FLUSH+ RELOAD: A High Resolution, Low Noise, L3
  Cache Side-Channel Attack.}. In \bibinfo{booktitle}{\emph{USENIX Security
  Symposium}}.
\newblock


\bibitem[\protect\citeauthoryear{Yarom, Genkin, and Heninger}{Yarom
  et~al\mbox{.}}{2017}]%
        {YaGeHe:17}
\bibfield{author}{\bibinfo{person}{Yuval Yarom}, \bibinfo{person}{Daniel
  Genkin}, {and} \bibinfo{person}{Nadia Heninger}.}
  \bibinfo{year}{2017}\natexlab{}.
\newblock \showarticletitle{CacheBleed: a timing attack on OpenSSL
  constant-time RSA}.
\newblock \bibinfo{journal}{\emph{Journal of Cryptographic Engineering}}
  (\bibinfo{year}{2017}).
\newblock


\bibitem[\protect\citeauthoryear{Zander, Armitage, and Branch}{Zander
  et~al\mbox{.}}{2007}]%
        {ZaArBr:07}
\bibfield{author}{\bibinfo{person}{Sebastian Zander},
  \bibinfo{person}{Grenville Armitage}, {and} \bibinfo{person}{Philip Branch}.}
  \bibinfo{year}{2007}\natexlab{}.
\newblock \showarticletitle{A survey of covert channels and countermeasures in
  computer network protocols}.
\newblock \bibinfo{journal}{\emph{IEEE Communications Surveys \& Tutorials}}
  (\bibinfo{year}{2007}).
\newblock


\bibitem[\protect\citeauthoryear{Zankl, Heyszl, and Sigl}{Zankl
  et~al\mbox{.}}{2016}]%
        {ZaHeSi:16}
\bibfield{author}{\bibinfo{person}{Andreas Zankl}, \bibinfo{person}{Johann
  Heyszl}, {and} \bibinfo{person}{Georg Sigl}.}
  \bibinfo{year}{2016}\natexlab{}.
\newblock \showarticletitle{Automated detection of instruction cache leaks in
  modular exponentiation software}. In \bibinfo{booktitle}{\emph{International
  Conference on Smart Card Research and Advanced Applications}}.
\newblock


\bibitem[\protect\citeauthoryear{Zenner}{Zenner}{2008}]%
        {zenner2008cache}
\bibfield{author}{\bibinfo{person}{Erik Zenner}.}
  \bibinfo{year}{2008}\natexlab{}.
\newblock \showarticletitle{A cache timing analysis of HC-256}. In
  \bibinfo{booktitle}{\emph{International Workshop on Selected Areas in
  Cryptography}}.
\newblock


\bibitem[\protect\citeauthoryear{Zenner}{Zenner}{2009}]%
        {zenner2009cache}
\bibfield{author}{\bibinfo{person}{Erik Zenner}.}
  \bibinfo{year}{2009}\natexlab{}.
\newblock \showarticletitle{Cache timing analysis of eStream finalists}. In
  \bibinfo{booktitle}{\emph{Dagstuhl Seminar Proceedings}}. Schloss
  Dagstuhl-Leibniz-Zentrum f{\"u}r Informatik.
\newblock


\bibitem[\protect\citeauthoryear{Zhang, Askarov, and Myers}{Zhang
  et~al\mbox{.}}{2012a}]%
        {zhang2012language}
\bibfield{author}{\bibinfo{person}{Danfeng Zhang}, \bibinfo{person}{Aslan
  Askarov}, {and} \bibinfo{person}{Andrew~C Myers}.}
  \bibinfo{year}{2012}\natexlab{a}.
\newblock \showarticletitle{Language-based control and mitigation of timing
  channels}. In \bibinfo{booktitle}{\emph{ACM SIGPLAN conference on Programming
  Language Design and Implementation}}.
\newblock


\bibitem[\protect\citeauthoryear{Zhang, Wang, Suh, and Myers}{Zhang
  et~al\mbox{.}}{2015}]%
        {zhang2015hardware}
\bibfield{author}{\bibinfo{person}{Danfeng Zhang}, \bibinfo{person}{Yao Wang},
  \bibinfo{person}{G~Edward Suh}, {and} \bibinfo{person}{Andrew~C Myers}.}
  \bibinfo{year}{2015}\natexlab{}.
\newblock \showarticletitle{A hardware design language for timing-sensitive
  information-flow security}.
\newblock \bibinfo{journal}{\emph{Acm Sigplan Notices}} (\bibinfo{year}{2015}).
\newblock


\bibitem[\protect\citeauthoryear{Zhang, Lou, Zhao, Bhasin, He, Ding, Qureshi,
  and Ren}{Zhang et~al\mbox{.}}{2018a}]%
        {zhang2018persistent}
\bibfield{author}{\bibinfo{person}{Fan Zhang}, \bibinfo{person}{Xiaoxuan Lou},
  \bibinfo{person}{Xinjie Zhao}, \bibinfo{person}{Shivam Bhasin},
  \bibinfo{person}{Wei He}, \bibinfo{person}{Ruyi Ding},
  \bibinfo{person}{Samiya Qureshi}, {and} \bibinfo{person}{Kui Ren}.}
  \bibinfo{year}{2018}\natexlab{a}.
\newblock \showarticletitle{Persistent fault analysis on block ciphers}.
\newblock \bibinfo{journal}{\emph{IACR Transactions on Cryptographic Hardware
  and Embedded Systems}} (\bibinfo{year}{2018}).
\newblock


\bibitem[\protect\citeauthoryear{Zhang and Lee}{Zhang and Lee}{2014}]%
        {zhang2014new}
\bibfield{author}{\bibinfo{person}{Tianwei Zhang} {and} \bibinfo{person}{Ruby~B
  Lee}.} \bibinfo{year}{2014}\natexlab{}.
\newblock \showarticletitle{New models of cache architectures characterizing
  information leakage from cache side channels}. In
  \bibinfo{booktitle}{\emph{Annual computer security applications conference}}.
\newblock


\bibitem[\protect\citeauthoryear{Zhang, Liu, Chen, and Lee}{Zhang
  et~al\mbox{.}}{2013}]%
        {zhang2013side}
\bibfield{author}{\bibinfo{person}{Tianwei Zhang}, \bibinfo{person}{Fangfei
  Liu}, \bibinfo{person}{Si Chen}, {and} \bibinfo{person}{Ruby~B Lee}.}
  \bibinfo{year}{2013}\natexlab{}.
\newblock \showarticletitle{Side channel vulnerability metrics: the promise and
  the pitfalls}. In \bibinfo{booktitle}{\emph{International Workshop on
  Hardware and Architectural Support for Security and Privacy}}.
\newblock


\bibitem[\protect\citeauthoryear{Zhang, Zhang, and Lee}{Zhang
  et~al\mbox{.}}{2016}]%
        {ZhZhLe:16}
\bibfield{author}{\bibinfo{person}{Tianwei Zhang}, \bibinfo{person}{Yinqian
  Zhang}, {and} \bibinfo{person}{Ruby~B Lee}.} \bibinfo{year}{2016}\natexlab{}.
\newblock \showarticletitle{Cloudradar: A real-time side-channel attack
  detection system in clouds}. In \bibinfo{booktitle}{\emph{Intl. Symp. on
  Research in Attacks, Intrusions, and Defenses}}.
\newblock


\bibitem[\protect\citeauthoryear{Zhang, Zhang, and Lee}{Zhang
  et~al\mbox{.}}{2018b}]%
        {zhang2018analyzing}
\bibfield{author}{\bibinfo{person}{Tianwei Zhang}, \bibinfo{person}{Yinqian
  Zhang}, {and} \bibinfo{person}{Ruby~B Lee}.}
  \bibinfo{year}{2018}\natexlab{b}.
\newblock \showarticletitle{Analyzing cache side channels using deep neural
  networks}. In \bibinfo{booktitle}{\emph{Annual Computer Security Applications
  Conference}}.
\newblock


\bibitem[\protect\citeauthoryear{Zhang, Juels, Reiter, and Ristenpart}{Zhang
  et~al\mbox{.}}{2012b}]%
        {ZhJuRe:12}
\bibfield{author}{\bibinfo{person}{Yinqian Zhang}, \bibinfo{person}{Ari Juels},
  \bibinfo{person}{Michael~K Reiter}, {and} \bibinfo{person}{Thomas
  Ristenpart}.} \bibinfo{year}{2012}\natexlab{b}.
\newblock \showarticletitle{Cross-VM side channels and their use to extract
  private keys}. In \bibinfo{booktitle}{\emph{ACM conference on Computer and
  communications security}}.
\newblock


\bibitem[\protect\citeauthoryear{Zhang, Juels, Reiter, and Ristenpart}{Zhang
  et~al\mbox{.}}{2014}]%
        {ZhJuRe:14}
\bibfield{author}{\bibinfo{person}{Yinqian Zhang}, \bibinfo{person}{Ari Juels},
  \bibinfo{person}{Michael~K Reiter}, {and} \bibinfo{person}{Thomas
  Ristenpart}.} \bibinfo{year}{2014}\natexlab{}.
\newblock \showarticletitle{Cross-tenant side-channel attacks in PaaS clouds}.
  In \bibinfo{booktitle}{\emph{ACM Conference on Computer and Communications
  Security}}.
\newblock


\bibitem[\protect\citeauthoryear{Zhang and Reiter}{Zhang and Reiter}{2013}]%
        {ZhRe:13}
\bibfield{author}{\bibinfo{person}{Yinqian Zhang} {and}
  \bibinfo{person}{Michael~K. Reiter}.} \bibinfo{year}{2013}\natexlab{}.
\newblock \showarticletitle{D\"{u}Ppel: Retrofitting Commodity Operating
  Systems to Mitigate Cache Side Channels in the Cloud}. In
  \bibinfo{booktitle}{\emph{ACM Conf. on Computer and Communications
  Security}}.
\newblock


\bibitem[\protect\citeauthoryear{Zhao, Wang, and Zheng}{Zhao
  et~al\mbox{.}}{2009}]%
        {zhao2009cache}
\bibfield{author}{\bibinfo{person}{Xin-jie Zhao}, \bibinfo{person}{Tao Wang},
  {and} \bibinfo{person}{Yuanyuan Zheng}.} \bibinfo{year}{2009}\natexlab{}.
\newblock \showarticletitle{Cache Timing Attacks on Camellia Block Cipher.}
\newblock \bibinfo{journal}{\emph{IACR Cryptol. ePrint Arch.}}
  (\bibinfo{year}{2009}).
\newblock


\bibitem[\protect\citeauthoryear{Zhou, Wagh, Mittal, and Wentzlaff}{Zhou
  et~al\mbox{.}}{2017}]%
        {zhou2017camouflage}
\bibfield{author}{\bibinfo{person}{Yanqi Zhou}, \bibinfo{person}{Sameer Wagh},
  \bibinfo{person}{Prateek Mittal}, {and} \bibinfo{person}{David Wentzlaff}.}
  \bibinfo{year}{2017}\natexlab{}.
\newblock \showarticletitle{Camouflage: Memory traffic shaping to mitigate
  timing attacks}. In \bibinfo{booktitle}{\emph{IEEE International Symposium on
  High Performance Computer Architecture}}.
\newblock


\bibitem[\protect\citeauthoryear{Zhou, Reiter, and Zhang}{Zhou
  et~al\mbox{.}}{2016}]%
        {zhou2016software}
\bibfield{author}{\bibinfo{person}{Ziqiao Zhou}, \bibinfo{person}{Michael~K
  Reiter}, {and} \bibinfo{person}{Yinqian Zhang}.}
  \bibinfo{year}{2016}\natexlab{}.
\newblock \showarticletitle{A software approach to defeating side channels in
  last-level caches}. In \bibinfo{booktitle}{\emph{ACM SIGSAC Conference on
  Computer and Communications Security}}.
\newblock


\end{thebibliography}

\end{document}